\documentclass{aa} % for a 2 col version                     
\usepackage{longtable}  \usepackage{float}                   
\usepackage{graphicx}           \usepackage{array}           
\usepackage{txfonts}  \usepackage{color}                     
\begin{document}                                             
\title{Ultra-deep 31.0-50.3 GHz spectral survey of IRC+10216 
\thanks{Based on observations carried out with the Yebes 40m 
radiotelescope at Yebes Observatory, operated by the Spanish 
Geographic Institute (IGN, Ministerio de Transportes,        
Movilidad y Agenda Urbana).}}                                
\titlerunning{IRC+10216 Q-band survey}                       
\authorrunning{J. R. Pardo et al.}                           
%\subtitle{  }                                               
\author{J.~R.~Pardo\inst{1},                                 
 J.~Cernicharo\inst{1}, B. Tercero\inst{2}, C.~Cabezas       
\inst{1}, C.~Berm\'udez\inst{1}, M. Ag\'undez\inst{1}        
, J.~D.~Gallego\inst{2}, F. Tercero\inst{2},                 
M.~G\'omez-Garrido\inst{2}, P. de Vicente\inst{2},           
J.~A. L\'opez-P\'erez\inst{2}                           }    
\institute{Consejo Superior de Investigaciones               
  Cientif\'{\i}cas, Instituto de F\'{\i}sica Fundamental,    
  Serrano 123, 28006 Madrid, Spain \\                        
  \email{jr.pardo@csic.es}                                   
  \and                                                       
  Instituto Geogr\'afico Nacional, Centro de Desarrollos     
  Tecnol\'ogicos, Observatorio de Yebes, Apartado 148,       
  19080 Yebes, Spain    }                                    
\date{Received September 20, 2021; accepted October 29, 2021}
\abstract                                                    
{IRC+10216, the carbon-rich envelope of the asymptotic giant 
branch (AGB) star CW Leo, is one of the richest molecular    
sources in the sky. Available spectral surveys below 51 GHz  
are more than 25 years old and new work is needed.}          
{Characterizing the rich molecular content of this source,   
specially for heavy species, requires to carry out very      
sensitive spectral surveys at low frequencies. In particular 
in this work we have achieved an rms in the range 0.2-0.6 mK 
per MHz.} % methods heading (mandatory)  {Very               
{long Q-band (31.0-50.3 GHz) single dish integrations were   
carried out with the Yebes 40m telescope using specifically  
built receivers. State of the art line catalogs are used     
for line identification.} % results heading (mandatory)      
 {A total of                                                 
 652
 spectral features                                           
 corresponding to  713 transitions from 
  81
 species (we count as different the isomers, isotopologues   
 and ortho/para species) are present in the data. Only       
  57
 unidentified lines remain with signal to noise ratios       
$\geqslant$3. Some new species and/or vibrational modes have 
been discovered for the first time with this survey.}        
{This IRC+10216 spectral survey is, by far, the most         
sensitive one carried out to this date in the Q-band. It     
 therefore provides the most complete view of IRC+10216 from 
 31.0 to 50.3 GHz, giving unique information on its molecular
 content, specially for heavy species. Rotational diagrams   
 built from the data provide valuable information on the     
 physical conditions and chemical content of this            
 circumstellar envelope.}                                    
\keywords{line:identification --- radio lines: stars ---     
 stars: AGB and post-AGB --- stars: carbon ---               
  stars: individual (IRC+10216) --- surveys }                
\maketitle                                                   
\section{Introduction}                                       
The combination of proximity (123\,$\pm$\,14 pc from the Sun;
 \citealt{Gro2012}) and high mass-loss rate                  
(\.M\,=\,(2-4)\,$\times$\,10$^{-5}$ M$_\odot$ yr$^{-1}$;     
 \citealt{Cer15,Gue2018})                                    
 makes IRC+10216 the main stellar target for molecule hunting
 in space \citep{Cer00}. In this context,                    
 the largest work at millimeter wavelengths (4.3-1.0 mm) has 
been carried out with the IRAM-30m telescope. Accumulated    
data achieved during more than three decades are now in their
 final phase of analysis for publication                     
 (Cernicharo et al. 2021, in prep.). In order to complement  
 those data at the longer end of the millimeter              
 range (6.0-9.7 mm, or 31.0 to 50.3 GHz in frequency), which 
 is specially important for the detection of heavy molecules,
  we decided to undergo an extremely deep integration in the 
 so-called Q-band towards IRC+10216 and several other stellar
 sources known for their molecular richness. For this work it
 was important to have access to a large antenna in order to 
 keep angular resolutions similar to those provided by the   
 IRAM-30m (but here at wavelengths 1.4 to 10 times larger),  
 state of the art receivers with very broad instantaneous    
band to reduce telescope time for a given rms goal, and large
 spectral resolutions to allow for degradation if desired.   
                                                             
The importance of sensitive surveys for understanding        
the rich molecular complexity found in the Universe,         
eventually leading to the appearance of life, is             
unquestionable and has made a large progress in recent years 
due to two key factors. On one hand the rapid and large      
increase in the instantaneous bands of receivers, allowing to
accumulate large amounts of integration times over large     
frequency bands because short-spaced frequency tunings are   
no longer necessary. On the other hand, large interferometers
 now in operation such as ALMA or NOEMA, allow to explore the
sources with rich molecular content at narrower              
angular scales thus providing a new molecular view of them   
since some molecular species, specially for stellar sources, 
 appear only in the innermost regions of their envelopes.    
                                                             
\begin{figure*}[ht] \begin{center}                           
\includegraphics[width=\textwidth]{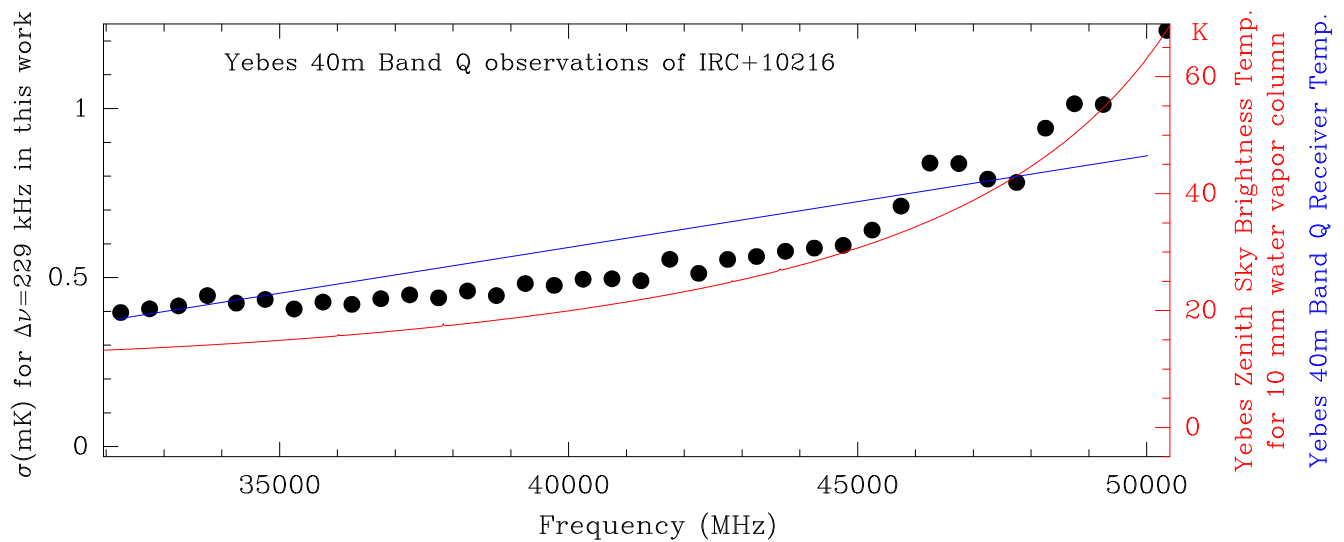}\end{center}  
\caption{Noise level in the spectra presented in this work at
their final resolution (see section \ref{sect:analysis}) and 
typical atmospheric brightness temperature and average       
receiver temperature across the band.} \label{fg:rms}        
\end{figure*}                                                
The spectral region below the atmospheric 60 GHz O$_2$       
spin-rotation band generally lacks coverage                  
by the most important millimeterwave single dish telescopes  
or interferometers. In addition, due to wavelength,  large   
 antennas are needed to keep a reasonably coupling to stellar
 sources. As a consequence, available spectral surveys of    
 IRC+10216 in the Q-band are more than 25 years old          
 \citep{Kaw95}. This is the reason that has motivated our    
 work, in which we present a big improvement with respect to 
 the past. The step forward is not only due to technical     
reasons related to the observational capabilities themselves,
  but also to a continuous effort on extending the           
 spectroscopic databases with the latest laboratory and      
 theoretical results.                                        
 The observations are described in section                   
\ref{sect:obs}. The systematic procedure                     
for data reduction, line identification, assignment and      
fitting is described in section                              
\ref{sect:analysis}. We present the results in section       
\ref{sect:results}, although an extended series of figures   
and tables, with the full results, has been left for appendix
 \ref{sect:appendix}. The procedure for line identification  
 and handling this large amount of information is described  
in section \ref{sect:lineiden}.                              
 A summary stressing the legacy importance of the results    
 obtained in this work, in the framework of our knowledge of 
 the physical conditions and chemistry of IRC+10216, is      
 finally given in section \ref{sect:summary}.                
\section{Observations}    \label{sect:obs}                   
This IRC+10216 31.0-50.3 GHz spectral survey was carried out 
in several runs from May 2019 to February 2020, and in April 
 2021, with the 40 meter antenna                             
of Yebes Observatory (IGN, Spain), hereafter                 
Yebes-40m, after several improvements in its equipment       
funded by the Nanocosmos                                     
project\footnote{\texttt{https://nanocosmos.iff.csic.es/}}.  
This large antenna provides now a main beam efficiency from  
0.6 at 31 GHz to 0.43 at 50 GHz, and a beam size in the range
 36-56$"$ for those frequencies. Although the experimental   
 setup is described in                                       
detail in a separate technical paper \citep{Tercero2021},    
 the most relevant information for this work is the use of a 
 receiver consisting of two HEMT cold amplifiers covering the
 31.0-50.3 GHz band with horizontal and vertical             
polarizations. The receiver temperature ranges from 22 K     
at 31 GHz to 42 K at 50 GHz. The backends are 16$\times$2.5  
GHz fast Fourier transform spectrometers (FFTS), with a      
primary spectral resolution of 38.1 kHz (later smoothed in   
the data reduction for this work), providing the whole       
 coverage of the observed band.                              
                                                             
 The observing mode was position switching with an off       
 position at 300$"$ in azimuth. The reference position used  
 for IRC+10216 has been J2000 RA 09:47:57.36 Dec +13:16:44.4.
 The HPBW of the antenna at 39.2 GHz is 45".                 
Pointing corrections were obtained by observing              
 the SiO masers of R\,Leo, and errors were always            
 within 2-3$"$. The intensity scale of the first-calibrated  
 is antenna temperature (T$_A^*$) corrected for atmospheric  
 absorption using the ATM package                            
 \citep{Cernicharo1985,Pardo2001}. A conversion to main beam 
temperature (T$_{MB}$) is done in order to generate the      
rotational diagrams of Figure \ref{fg:rotdiag}. Calibration  
uncertainties are estimated to be within 10~\%.              
 More details on the observational procedure can be found in 
 \cite{Pardo2020}                                            
 %The resulting spectra were                                 
%smoothed to a resolution of 0.229 MHz, for which the        
%sensitivity of the final spectra ranges from 0.4 to 1 mK    
%the surveyed frequency range.                               

\begin{figure*}[h]                                           
\includegraphics[width=\textwidth]{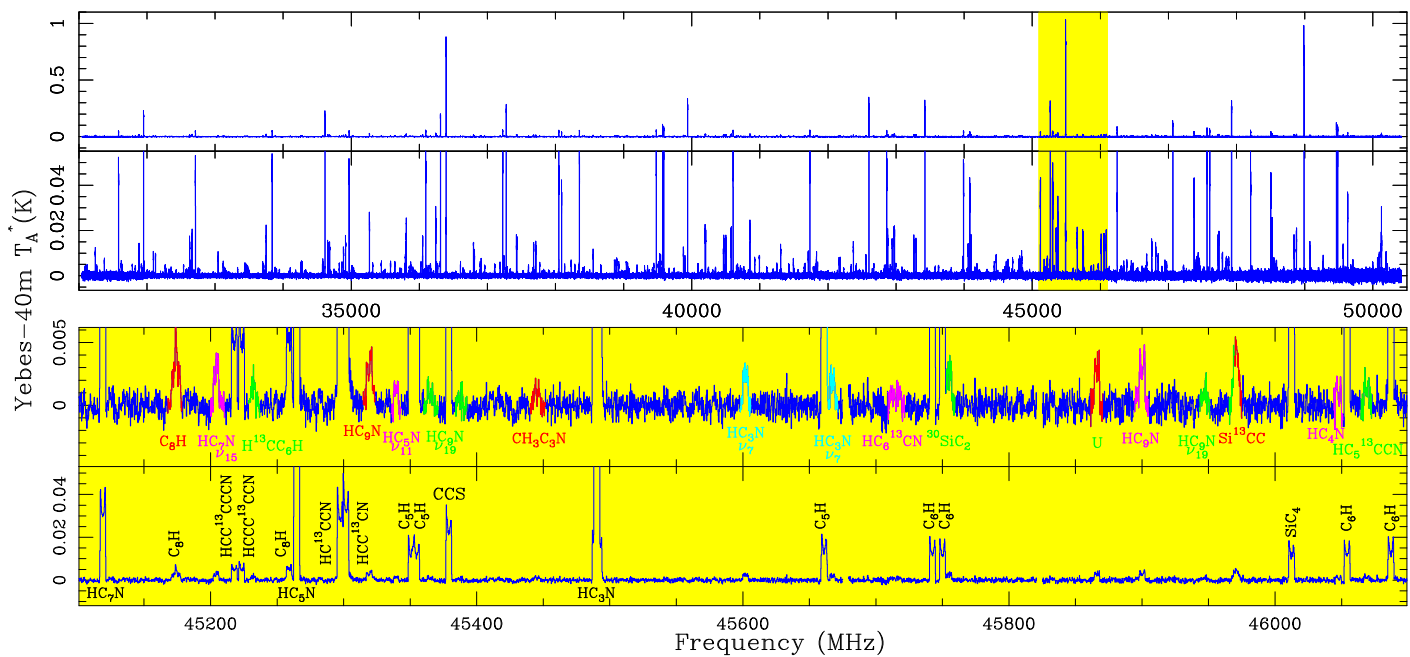}      
\caption{Overall view of the data presented in this paper    
with a zoom around 45.6 GHz. Several weak spectral features  
($\sim$1-3 mK), revealed in this work, are shown in different
 colors in one of the panels.} \label{fg:zooms} \end{figure*}

\section{Data reduction and analysis}\label{sect:analysis}   
 The GILDAS                                                  
 package\footnote{\texttt{http://www.iram.fr/IRAMFR/GILDAS}} 
 was used to add up the raw spectra provided by the telescope
 calibration software, to fit individual baselines in the    
 different FFTS sections, and to stitch all these sections   
 into a final spectrum using the appropriate weights         
 according to system temperatures and integration times of   
 each individual scan. If necessary, additional baselines    
 have been fitted to narrow sections of the data, but these  
 interventions have been kept to the strict minimum.         
 The resulting spectrum was                                  
 smoothed to a resolution of 0.229 MHz, for which the        
 sensitivity ranges from $\sim$ 0.4 to 1.2 mK in the surveyed
 frequency range (see Figure \ref{fg:rms}). The average      
 data noise every 500 MHz shown in the latter figure is the  
 basis for a first automatic procedure that checks which     
 observed spectral features are above the 3$\sigma$ level.   
 Obviously, a second check is done by hand to spot features  
 that were missed in the first step.                         
 \linebreak Once the detected spectral features are          
 well established, the next step has been to assign them to  
 molecular lines. For this task we rely upon the latest      
 version of the MADEX line catalog \citep{Cer12} with up to  
 6150 molecular species an 23.5 million transitions. We      
 proceed in several steps. First, as part of the automatic   
 procedures, there is a cross-check between the data and the 
catalog  to assign as many lines as possible. Second, all the
 assignments are checked by hand in order to eliminate errors
 and to disentangle blended lines from different species,    
 lines spread due to hyperfine structure, etc. A third step  
consists of assigning lines not present in the catalog to new
 species, isotopologues or vibrational levels. This step has 
 resulted in the discovery for the first time in space of the
 $\nu_{19}$ vibrational mode of HC$_9$N \citep{Pardo2020} and
 the Mg-containing carbon chains MgC$_5$N and MgC$_6$H       
 \citep{Pardo2021}. This careful                             
 work has allowed to identify                                
93 $\%$ of all lines 
 present in the data. In addition, features at               
 the 2-3 $\sigma$ level with a linewidth compatible with the 
 $\pm$14.5 kms$^{-1}$ typical of the source have also been   
 individually checked by hand, and this has resulted on extra
  assignments, mainly to $^{13}$C substitutions of HC$_7$N.  
                                                             
 The overall summary of the identification work described in 
 this section is the following:                              
 652
 spectral features belonging to                              
 713
 identified lines and                                        
  57
 unidentified lines (U-lines). An overview is given in Table 
 \ref{tb:molecules}; the details on the results of individual
 fits to all lines, using the SHELL method implemented in    
the GILDAS software, can be found in Table \ref{tb:id_qband};
 and the remaining U-lines are listed in Table \ref{ulines}. 
 The final spectra presented in this paper (Figures          
\ref{fg:zooms}, and \ref{fig01} to \ref{fig39}) improve by a 
factor of $\approx$10 the sensitivity of those presented in  
the previous work conducted with the Nobeyama 45m telescope  
\citep{Kaw95} in a similar frequency range. In total         
they label 132 lines and report 33 U-lines. Among the latter,
 22 of them could be labeled with the spectroscopic data     
 available nowadays (they correspond to C$_8$H, C$_6$H$^-$   
 C$_5$N$^-$, C$_6$H, HC$_9$N and HC$_7$N $\nu_{15}$), 6      
are spurious features as they are not present in our data,   
and 5 are still U-lines for us in 2021. The strongest        
observed feature (at 14$\sigma$ in our data) that still      
remains unidentified stands at 41712 MHz.                    
\section{Results}    \label{sect:results}                    
The detected lines can be divided into different families of 
 carriers. We present the results for each one of these      
 separately.                                                 
 \subsection{Cyanopolyynes} This molecular family,           
 HC$_{2n+1}$N, dominates the millimeter wave spectrum of     
 this source, as it happens with other C-rich evolved stars  
 such as CRL 618. In IRC+10216, up to                        
         304
 lines (                                                     
43$\%$ of all detected lines in the Q-band)                    
belong to this family, which comprises HCN, HC$_3$N, HC$_5$N,
 HC$_7$N, and HC$_9$N in their main species, isotopic        
 substitutions, vibrationally excited states, together with a
 few isomers. The upper panel in Figures \ref{fg:freqdiag}   
and \ref{fg:rotdiag} provide a  quick view of all lines from 
 this family in this work. The difference with respect to    
 \cite{Kaw95} is quite large since they cannot detect        
 isotopic substitutions of                                   
 HC$_5$N nor the vibrational states of HC$_5$N, HC$_7$N      
 and HC$_9$N. Sometimes, $\nu_{11}$ lines of HC$_5$N are     
 present in their data but labeled as U probably due to lack 
 of information in line catalogs available in 1995.          
                                                             
The alternative single and triple C-bond is, in fact, the    
dominant chemical feature in this source since most radicals,
 anions, and metal-bearing species, are carbon chains of this
 type.

 Concerning HCN, although there are 
 no lines  from its main rotational ladder in the 
  Q-band, three $l$-doubling transitions from the 
 $\nu_2$=1 vibrational state are present in the data 
 (J=12-12,13-13,14-14).
  The upper energy 
  levels of these transitions lie between 1300 and 
  1500 K and, therefore, the signal detected arises 
  from much closer to the photosphere of CW Leo than 
  most lines in this work. They are narrower too. 
  Although this is the first Q-band detection in 
  IRC+10216 of HCN $l$-doubling lines, higher-J 
  ones have been previously reported \citep{Ce11}.
 \subsection{Carbon chain Radicals}                          
 The list of radicals detected that are related with the     
 cyanopolyynes is also long: C$_3$H, C$_4$H, C$_5$H, C$_6$H, 
 C$_7$H, C$_8$H, C$_3$N ($^{13}$CCCN, C$^{13}$CCN,           
 CC$^{13}$CN) and C$_5$N. They are responsible for           
 a total of                                                  
         167
 lines or                                                    
          23
$\%$ of all detected and labeled spectral lines in this work.
 Their open structure makes that several                     
 configurations are close in energy,                         
 and hyperfine structure is also common for them.            
 C$_3$H presents 6 hyperfine components (two of them blended)
 in its ground vibrational state, from 32617 to 32668 MHz.   
 Another set of hyperfine components from 44857 to 44939 GHz,
 corresponding to the $\nu_4$ vibrational mode (at 42.4 K)   
 does not show up in the data.                               
 C$_4$H presents a total of 9 doublets corresponding to its  
 ground, $\nu_7$ and 2$\nu_7$ vibrational states.            
 C$_5$H                                                      
        presents 4 doublets for each one of the lowest       
 states: $^2\Pi_{3/2}$ (not resolved) and                    
 $^2\Pi_{1/2}$ (resolved). C$_6$H                            
 is detected as resolved doublets from its $^2\Pi_{3/2}$     
and $^2\Pi_{1/2}$ states, and other lines from the vibronic  
 states $^2\Sigma^-$, $^2\Delta_{3/2}$, and $^2\Delta_{5/2}$ 
 belonging to the $\nu_{11}$ vibrational state are           
 also detected. The total number of lines from C$_6$H is     
 56
, the largest number in this family.                         
 Finally, C$_7$H and C$_8$H also display doublets in         
 the electronic states $^2\Pi_{3/2}$ and $^2\Pi_{1/2}$, with 
 only the last ones resolved in both cases, although         
 the lowest energy state is $^2\Pi_{1/2}$ for C$_7$H and     
 $^2\Pi_{3/2}$ for C$_8$H.

 \begin{table*}[h]
 \caption{Molecules and total number of spectral
 features (N) and lines (L) detected in this work for 
  each one of them. $\mu$ is the molecule's electric 
 dipole moment in Debyes (1 Debye 
 = 3.33564$\cdot$10$^{-30}$ Coulomb$\cdot$meter).}
 \label{tb:molecules}
 \begin{tabular}{|l|c|c|p{45mm}||l|c|c|p{45mm}|}
 \hline 
 \footnotesize{Molecule} &  \footnotesize{$\mu$} & \footnotesize{N/L} & \footnotesize{References for spectroscopy} & \footnotesize{Molecule} &  \footnotesize{$\mu$} & \footnotesize{N/L} & \footnotesize{References for spectroscopy} \\ 
 \hline
\scriptsize{HCN                 }&   \scriptsize{   2.985}&   \scriptsize{ 3/ 3}&   \scriptsize{ \cite{Zel2003}, \cite{Ebenstein1984}                          }& 
\scriptsize{HC$_3$N             }&   \scriptsize{   3.732}&   \scriptsize{ 8/10}&   \scriptsize{ \cite{Mbosei2000}, \cite{Thorwirth2000}                       }\\
\scriptsize{H$^{13}$CCCN        }&   \scriptsize{   3.732}&   \scriptsize{ 2/ 2}&   \scriptsize{ \cite{Creswell1977}, \cite{Mallins1978}, \cite{Thorwirth2001} }& 
\scriptsize{HC$^{13}$CCN        }&   \scriptsize{   3.732}&   \scriptsize{ 2/ 2}&   \scriptsize{ Same as H$^{13}$CCCN.                                         }\\
\scriptsize{HCC$^{13}$CN        }&   \scriptsize{   3.732}&   \scriptsize{ 2/ 2}&   \scriptsize{ Same as H$^{13}$CCCN.                                         }& 
\scriptsize{HCCNC               }&   \scriptsize{   2.957}&   \scriptsize{ 2/ 2}&   \scriptsize{ \cite{Guarnieri1992}, \cite{Gripp2000}                        }\\
\scriptsize{HCCN$^{13}$C        }&   \scriptsize{   2.957}&   \scriptsize{ 2/ 2}&   \scriptsize{ Same as HCCNC                                                 }& 
\scriptsize{HNCCC               }&   \scriptsize{   5.665}&   \scriptsize{ 2/ 2}&   \scriptsize{ \cite{Kawaguchi1992}, \cite{Hirahara1993}, \cite{Bots1992}    }\\
\scriptsize{HC$_5$N             }&   \scriptsize{   4.330}&   \scriptsize{20/22}&   \scriptsize{ \cite{Bizzo2004a}, \cite{Kroto1976}                           }& 
\scriptsize{H$^{13}$CCCCCN      }&   \scriptsize{   4.330}&   \scriptsize{ 8/ 8}&   \scriptsize{ Same as HC$_5$N.                                              }\\
\scriptsize{HC$^{13}$CCCCN      }&   \scriptsize{   4.330}&   \scriptsize{ 8/ 8}&   \scriptsize{ Same as HC$_5$N.                                              }& 
\scriptsize{HCC$^{13}$CCCN      }&   \scriptsize{   4.330}&   \scriptsize{ 7/ 7}&   \scriptsize{ Same as HC$_5$N.                                              }\\
\scriptsize{HCCC$^{13}$CCN      }&   \scriptsize{   4.330}&   \scriptsize{ 7/ 7}&   \scriptsize{ Same as HC$_5$N.                                              }& 
\scriptsize{HCCCC$^{13}$CN      }&   \scriptsize{   4.330}&   \scriptsize{ 8/ 8}&   \scriptsize{ Same as HC$_5$N.                                              }\\
\scriptsize{HC$_7$N             }&   \scriptsize{   4.820}&   \scriptsize{72/74}&   \scriptsize{ \cite{McCarthy2000}, \cite{Bizzo2004b}, \cite{Bots1997}       }& 
\scriptsize{H$^{13}$CCCCCCCN    }&   \scriptsize{   4.820}&   \scriptsize{10/11}&   \scriptsize{  \cite{McCarthy2000},  \cite{Bots1997}                        }\\
\scriptsize{HC$^{13}$CCCCCCN    }&   \scriptsize{   4.820}&   \scriptsize{ 8/ 8}&   \scriptsize{ Same as H$^{13}$CCCCCCCN                                      }& 
\scriptsize{HCC$^{13}$CCCCCN    }&   \scriptsize{   4.820}&   \scriptsize{10/10}&   \scriptsize{ Same as H$^{13}$CCCCCCCN                                      }\\
\scriptsize{HCCC$^{13}$CCCCN    }&   \scriptsize{   4.820}&   \scriptsize{10/10}&   \scriptsize{ Same as H$^{13}$CCCCCCCN                                      }& 
\scriptsize{HCCCC$^{13}$CCCN    }&   \scriptsize{   4.820}&   \scriptsize{ 6/ 6}&   \scriptsize{ Same as H$^{13}$CCCCCCCN                                      }\\
\scriptsize{HCCCCC$^{13}$CCN    }&   \scriptsize{   4.820}&   \scriptsize{11/11}&   \scriptsize{ Same as H$^{13}$CCCCCCCN                                      }& 
\scriptsize{HCCCCCC$^{13}$CN    }&   \scriptsize{   4.820}&   \scriptsize{ 9/ 9}&   \scriptsize{ Same as H$^{13}$CCCCCCCN                                      }\\
\scriptsize{HC$_9$N             }&   \scriptsize{   5.200}&   \scriptsize{80/80}&   \scriptsize{ \cite{McCarthy2000}, \cite{Bots1997}, \cite{Pardo2020}        }& 
\scriptsize{HCCN                }&   \scriptsize{   3.600}&   \scriptsize{ 1/ 1}&   \scriptsize{ \cite{Allen2001}, \cite{Aoki1993}                             }\\
\scriptsize{l-HC$_4$N           }&   \scriptsize{   4.330}&   \scriptsize{ 6/ 6}&   \scriptsize{ \cite{Tang1999}, \cite{Ikuta2000}                             }& 
\scriptsize{o-H$_2$C$_4$        }&   \scriptsize{   4.457}&   \scriptsize{ 6/ 6}&   \scriptsize{ \cite{Killian1990}, \cite{Oswald1995}                         }\\
\scriptsize{p-H$_2$C$_4$        }&   \scriptsize{   4.457}&   \scriptsize{ 4/ 5}&   \scriptsize{ \cite{Killian1990}, \cite{Oswald1995}                         }& 
\scriptsize{o-H$_2$C$_3$        }&   \scriptsize{   4.100}&   \scriptsize{ 2/ 2}&   \scriptsize{ \cite{Vr90}                                                   }\\
\scriptsize{p-H$_2$C$_3$        }&   \scriptsize{   4.100}&   \scriptsize{ 1/ 1}&   \scriptsize{ \cite{Vr90}                                                   }& 
\scriptsize{C$_3$H              }&   \scriptsize{   3.550}&   \scriptsize{ 6/ 6}&   \scriptsize{ \cite{Yama90}, \cite{Gott85}, \cite{Kaif04}, \cite{Woon1995}  }\\
\scriptsize{C$_4$H              }&   \scriptsize{   2.100}&   \scriptsize{18/19}&   \scriptsize{ \cite{Gue1982}, \cite{Gott83}, \cite{Chen95}, \cite{Oyam2020} }& 
\scriptsize{C$_5$H              }&   \scriptsize{   4.881}&   \scriptsize{12/16}&   \scriptsize{ \cite{McCa1999}, \cite{Woon1995}                              }\\
\scriptsize{C$_6$H              }&   \scriptsize{   5.540}&   \scriptsize{56/56}&   \scriptsize{ \cite{Linn1999}, \cite{Woon1995}                              }& 
\scriptsize{C$_7$H              }&   \scriptsize{   5.945}&   \scriptsize{23/33}&   \scriptsize{ \cite{McCa1997}, \cite{Woon1995}                              }\\
\scriptsize{C$_8$H              }&   \scriptsize{   6.500}&   \scriptsize{31/46}&   \scriptsize{ \cite{McCa1996}, \cite{McCa1999}                              }& 
\scriptsize{C$_3$N              }&   \scriptsize{   2.850}&   \scriptsize{ 4/ 4}&   \scriptsize{ \cite{McCa1995}, \cite{Gue1982}, \cite{Gott83}                }\\
\scriptsize{CC$^{13}$CN         }&   \scriptsize{   2.850}&   \scriptsize{ 4/ 6}&   \scriptsize{ \cite{McCa1995}, \cite{McCa2003}                              }& 
\scriptsize{C$^{13}$CCN         }&   \scriptsize{   2.850}&   \scriptsize{ 4/ 4}&   \scriptsize{ Same as CC$^{13}$CN                                           }\\
\scriptsize{$^{13}$CCCN         }&   \scriptsize{   2.850}&   \scriptsize{ 5/ 6}&   \scriptsize{ Same as CC$^{13}$CN                                           }& 
\scriptsize{C$_5$N              }&   \scriptsize{   3.385}&   \scriptsize{ 4/ 4}&   \scriptsize{ \cite{Kasai1997}, \cite{Bots1996}                             }\\
\scriptsize{C$_5$N$^-$          }&   \scriptsize{   5.200}&   \scriptsize{ 6/ 6}&   \scriptsize{ \cite{Cerni2008}                                              }& 
\scriptsize{C$_6$H$^-$          }&   \scriptsize{   8.200}&   \scriptsize{ 7/ 8}&   \scriptsize{ \cite{McCarthy2007}, \cite{Blanksby0000}                      }\\
\scriptsize{C$_8$H$^-$          }&   \scriptsize{  10.400}&   \scriptsize{13/13}&   \scriptsize{ \cite{Gupta2007}, \cite{Blanksby0000}                         }& 
\scriptsize{c-C$_3$H$_2$ (ortho)}&   \scriptsize{   3.430}&   \scriptsize{ 2/ 2}&   \scriptsize{ \cite{Th85}, \cite{Bo86}, \cite{Vr87}, \cite{Sp12}, \cite{K7} }\\
\scriptsize{c-C$_3$H$_2$ (para) }&   \scriptsize{   3.430}&   \scriptsize{ 2/ 2}&   \scriptsize{ Same as c-C$_3$H$_2$ (ortho)                                  }& 
\scriptsize{SiS                 }&   \scriptsize{   1.735}&   \scriptsize{ 1/ 1}&   \scriptsize{ \cite{Mueller2007}, \cite{Tiemann1972}, \cite{Sanz03}         }\\
\scriptsize{$^{29}$SiS          }&   \scriptsize{   1.735}&   \scriptsize{ 1/ 1}&   \scriptsize{ Same as $^{(28)}$SiS.                                         }& 
\scriptsize{$^{30}$SiS          }&   \scriptsize{   1.735}&   \scriptsize{ 1/ 1}&   \scriptsize{ Same as $^{(28)}$SiS.                                         }\\
\scriptsize{Si$^{33}$S          }&   \scriptsize{   1.735}&   \scriptsize{ 1/ 1}&   \scriptsize{ Same as $^{(28)}$SiS.                                         }& 
\scriptsize{Si$^{34}$S          }&   \scriptsize{   1.735}&   \scriptsize{ 1/ 1}&   \scriptsize{ Same as $^{(28)}$SiS.                                         }\\
\scriptsize{SiO                 }&   \scriptsize{   3.098}&   \scriptsize{ 1/ 1}&   \scriptsize{ \cite{M77}, \cite{M91}, \cite{Sanz03}, \cite{C98}, \cite{R70} }& 
\scriptsize{$^{29}$SiO          }&   \scriptsize{   3.098}&   \scriptsize{ 1/ 1}&   \scriptsize{ Same as $^{(28)}$SiO.                                         }\\
\scriptsize{$^{30}$SiO          }&   \scriptsize{   3.098}&   \scriptsize{ 1/ 1}&   \scriptsize{ Same as $^{(28)}$SiO.                                         }& 
\scriptsize{Si$^{17}$O          }&   \scriptsize{   3.098}&   \scriptsize{ 1/ 1}&   \scriptsize{ Same as $^{(28)}$SiO.                                         }\\
\scriptsize{SiC$_2$             }&   \scriptsize{   2.393}&   \scriptsize{ 2/ 2}&   \scriptsize{ \cite{Got1989}, \cite{Sue1989}                                }& 
\scriptsize{Si$^{13}$CC         }&   \scriptsize{   2.393}&   \scriptsize{ 2/ 2}&   \scriptsize{ \cite{Cer1991}                                                }\\
\scriptsize{$^{29}$SiC$_2$      }&   \scriptsize{   2.393}&   \scriptsize{ 1/ 1}&   \scriptsize{ \cite{Sue1989}, \cite{Cer1991}, \cite{Kok1991}                }& 
\scriptsize{$^{30}$SiC$_2$      }&   \scriptsize{   2.393}&   \scriptsize{ 2/ 2}&   \scriptsize{ Same as $^{29}$SiC$_2$                                        }\\
\scriptsize{SiC$_4$             }&   \scriptsize{   6.420}&   \scriptsize{ 6/ 6}&   \scriptsize{ \cite{Ohi1989}, \cite{Gor2000}                                }& 
\scriptsize{$^{29}$SiC$_4$      }&   \scriptsize{   6.420}&   \scriptsize{ 5/ 5}&   \scriptsize{ Same as $^{29}$SiC$_4$                                        }\\
\scriptsize{SiC$_6$             }&   \scriptsize{   8.220}&   \scriptsize{ 7/ 7}&   \scriptsize{ \cite{Gor2000}                                                }& 
\scriptsize{CS                  }&   \scriptsize{   1.958}&   \scriptsize{ 1/ 1}&   \scriptsize{ \cite{Got2003}, \cite{W68}                                    }\\
\scriptsize{$^{13}$CS           }&   \scriptsize{   1.958}&   \scriptsize{ 1/ 1}&   \scriptsize{ Same as CS.                                                   }& 
\scriptsize{C$^{33}$S           }&   \scriptsize{   1.958}&   \scriptsize{ 1/ 1}&   \scriptsize{ Same as CS.                                                   }\\
\scriptsize{C$^{34}$S           }&   \scriptsize{   1.958}&   \scriptsize{ 1/ 1}&   \scriptsize{ Same as CS.                                                   }& 
\scriptsize{C$^{36}$S           }&   \scriptsize{   1.958}&   \scriptsize{ 1/ 1}&   \scriptsize{ Same as CS.                                                   }\\
\scriptsize{CCS                 }&   \scriptsize{   2.855}&   \scriptsize{ 4/ 4}&   \scriptsize{ \cite{Sai1987}, \cite{Yama1990}, \cite{Lo92}, \cite{Lee97}    }& 
\scriptsize{C$_3$S              }&   \scriptsize{   3.704}&   \scriptsize{ 3/ 3}&   \scriptsize{ \cite{Ya87}, \cite{Ta95}, \cite{Lo92}, \cite{O92}, \cite{S94} }\\
\scriptsize{NaCN                }&   \scriptsize{   8.850}&   \scriptsize{ 7/ 8}&   \scriptsize{ \cite{VV84}, \cite{HZ11}                                      }& 
\scriptsize{MgCN                }&   \scriptsize{   5.150}&   \scriptsize{ 2/ 2}&   \scriptsize{ \cite{An94}, \cite{Hi02}                                      }\\
\scriptsize{MgNC                }&   \scriptsize{   5.308}&   \scriptsize{ 4/ 4}&   \scriptsize{ \cite{Ka93}, \cite{SB01}                                      }& 
\scriptsize{MgC$_3$N            }&   \scriptsize{   6.250}&   \scriptsize{14/14}&   \scriptsize{ \cite{Cerni2019}                                              }\\
\scriptsize{MgC$_5$N            }&   \scriptsize{   7.300}&   \scriptsize{20/30}&   \scriptsize{ \cite{Pardo2021}                                              }& 
\scriptsize{MgC$_4$H            }&   \scriptsize{   2.100}&   \scriptsize{10/10}&   \scriptsize{ \cite{Cerni2019}                                              }\\
\scriptsize{MgC$_6$H            }&   \scriptsize{   2.500}&   \scriptsize{23/30}&   \scriptsize{ \cite{Pardo2021}                                              }& 
\scriptsize{CH$_3$CN            }&   \scriptsize{   3.925}&   \scriptsize{ 2/ 3}&   \scriptsize{ \cite{Pav90}, \cite{C06}, \cite{S04}, \cite{M09}, \cite{A93}  }\\
\scriptsize{CH$_3$C$_3$N        }&   \scriptsize{   4.750}&   \scriptsize{ 5/ 5}&   \scriptsize{ \cite{Moi1982}, \cite{Bes1983}, \cite{Bes1984}                }& 
\scriptsize{KCl                 }&   \scriptsize{  10.270}&   \scriptsize{ 2/ 2}&   \scriptsize{ \cite{Clo}, \cite{Caris2004}                                  }\\
\scriptsize{NaCl                }&   \scriptsize{   9.002}&   \scriptsize{ 1/ 1}&   \scriptsize{ \cite{Clo}, \cite{Caris2002}, \cite{Timp2012}                 }& 
\scriptsize{Na$^{37}$Cl         }&   \scriptsize{   9.002}&   \scriptsize{ 1/ 1}&   \scriptsize{ Same as NaCl.                                                 }\\
\scriptsize{AlCl                }&   \scriptsize{   1.500}&   \scriptsize{ 1/ 1}&   \scriptsize{ \cite{WG72}, \cite{Li65}, \cite{He93}                         }& 
 \scriptsize{U Lines} & & \scriptsize{  57} & \\
 \hline
 \end{tabular}
 \end{table*}
 \clearpage

 The total number of spectral features 
 of C$_7$H and C$_8$H detected is 
 23 and  31, respectively, with a counterpart of
 33 and  46, respectively, rotational transitions.

 Other radicals detected in this family
  are C$_3$N (two doublets at 
 39571/39590 and 49466/49485 MHz), the $^{13}$C 
 isotopic substitutions of it (marginal), and C$_5$N 
 through two extremely weak doublets at 
 39280/39291 and 42086/42097 MHz.
 For a graphic summary of all lines from carbon chain        
 radicals, see the second panel of Figures \ref{fg:freqdiag} 
 and \ref{fg:rotdiag}.                                       
 \subsection{Anions}  Up to date, IRC+10216 is the           
 only circumstellar environnment where anions have           
 been detected. The survey presented in this paper provided  
a good opportunity for detecting heavy linear anions and the 
                 results have confirmed those expectations:  
C$_5$N$^-$, C$_6$H$^-$, and C$_8$H$^-$ are detected through  
  6,   8 and  13
  lines, respectively.  The lighter anion C$_4$H$^-$         
 could be present but its two transitions at 37239.41        
 and 46549.16 MHz are very close in frequency with lines     
of HCC$^{13}$CCCN and HCCC$^{13}$CCN, for the first one, and 
HC$_9$N $\nu_{19}$, for the second one, to the point of      
 making impossible to confirm its detection. The same happens
 to the C$_3$N$^-$ line at 48515.87 (blended with a strong   
 C$_6$H line). The other line of C$_3$N$^-$ at 38812.79 MHz  
 is not seen,                 a result that is compatible    
 with its low abundance revealed from its discovery in this  
 source at higher frequencies \citep{Tha2008}.               
  C$_5$N$^-$ and C$_6$H$^-$ are expected (and confirmed) to  
  display the strongest lines among anions in this survey.   
 A graphic summary of all lines from anions  can             
 be found on the third panel of Figures \ref{fg:freqdiag}    
 and \ref{fg:rotdiag}.                                       
                                                             
 \subsection{Si-bearing species} \label{sct:si}              
 The only oxygen-bearing species detected in our line survey 
 is SiO. The J=1-0 transition is detected for                
 the main isotopologue as well as for $^{29}$SiO and         
 $^{30}$SiO. On the other hand Si$^{17}$O at 41794.658 MHz   
 and Si$^{18}$O at 40352.771 could be there but their        
 detection is still marginal at the actual  noise level.     
                                                             
 Another Si bearing molecule with only one rotational        
 transition in this range (J=1-0) is SiS. It has been        
 detected for its main isotopologue as well as for           
 $^{29}$SiS, $^{30}$SiS, Si$^{33}$S and Si$^{34}$S.          
                                                             
 The cyclic C$_{2v}$ molecule SiC$_2$ has two                
 strong enough lines for detection in the surveyed frequency 
 range. Except $^{29}$SiC$_2$ (1 line detected), all other   
 substitutions (SiC$_2$, $^{30}$SiC$_2$ and $^{13}$CSiC)     
 exhibit two detected lines here.                            
                                                             
The linear molecule SiC$_4$ has 6 rotational transitions in  
 the surveyed frequency range (from J=11-10 to J=16-15). The 
 multiple possible isotopic substitutions                    
 make it a candidate for a large number of lines in this     
 source but only $^{29}$SiC$_4$ can be confirmed as detected 
 here with five of its individual six rotational transitions 
 present in the Q-band range just marginally seen.           
                                                             
 Finally, the detection of SiC$_6$ is marginal, although     
 possible, through several transitions from                  
 J=26$\rightarrow$25 to J=38$\rightarrow$37 at the 1-2 mK    
 level (sometimes blended with lines from other species).    
 If confirmed, this would be the first time that this        
 Si-bearing carbon chain is seen in space.                   
                                                             
 \subsection{S-bearing species}                              
 Apart from SiS (see section \ref{sct:si}) other S-bearing   
 molecules detected are CS, CCS and C$_3$S. The diatomic     
 molecule CS is detected in 5 isotopic substitutions, the    
 main one and $^{13}$CS, C$^{33}$S, C$^{34}$S and C$^{36}$S. 
 The linear species C$_3$S has 3 transitions in the surveyed 
 frequency range (J=6-5,7-6,8-7), all three detected, and the
 $^3\Sigma$ species CCS displays also four detected lines.   
 The isotopic substitutions of CCS and C$_3$S are not seen.  
                                                             
  \subsection{Metal chlorides and cyanides}                  
 These interesting molecules appear in most cases  with just 
 one or two lines (MgCN, KCl, NaCl, Na$^{37}$Cl and AlCl)    
 given their molecular weight and the fact that they are     
 mostly diatomic or linear. NaCN, due to its bent structure, 
 has more rotational lines, 7 of them detected in this work. 
  Two doublets have been detected corresponding to the       
 isomer MgNC with the first one (35793/35809 MHz) within     
  a blended complex  of lines from other molecules.          
                                                             
 Related to the cyanides, we should mention the detection of 
 CH$_3$CN (1 line), CH$_3$C$_3$N (5 lines), and the radicals 
 MgC$_3$N and MgC$_5$N. The last two species, together with  
 MgC$_4$H and MgC$_6$H, have been discovered in space for the
 first time in IRC+10216 \citep{Cerni2019, Pardo2021}.       
                                                             
\subsection{Cyclic molecules}                Besides SiC$_2$ 
(section\ref{sct:si}), the only cyclic molecule we can report
 as detected here is c-C$_3$H$_2$ with just 2 lines for both 
the ortho and para configuration of H spins. Only one line   
(the one involving the lowest energy levels within the       
Q-band) for each species is strong and nicely detected. The  
other line, at 42231.25 MHz and 42139.19 GHz for the para and
 ortho species respectively, is almost 10 times weaker. In   
addition, the 42231.25 MHz is blended with another line from 
an unidentified species. For these reasons, a proper analysis
 using rotational diagrams needs other lines observed, for   
example, with the IRAM-30m telescope.  %This task remains    
%therefore for follow-up.}                                   
    
  It is interesting to mention that IRC+10216 
  remains poor in cyclic molecules. A similar 
  Q-band survey with the same telescope on TMC-1 
  has discovered several cyclic molecules not 
  seen in the envelope of CW Leo, such as indene 
  and benzyne \citep{Ce21a, Ce21b}.

 \subsection{Latest findings}\label{sect:newfindigs}         
 As the line identification goes on and the number of        
 unidentified features reduces, it becomes easier to find    
 harmonical relations in other subsets of U-lines. The great 
 progress made in line catalogs in recent years, based on    
 both laboratory work and ab initio calculations is obvious. 
 Nevertheless, spectral surveys of space objects such as this
 one are also part of this progress as some species are      
 directly discovered in them thanks to these subsets of      
 harmonically related U-lines. This is the case, for example,
 of the $\nu_{19}$ vibrational level of HC$_9$N, for which   
 a series of 26 doublets, harmonically related with integer  
 quantum numbers ranging from J$_{up}$ = 54 to 80, were      
 found among the U-lines in an early stage of analysis       
 \citep{Pardo2020}. A fit of the observed central frequencies
 and line intensities allowed us to derive the rotational    
 constants and, therefore, incorporate HC$_9$N $\nu_{19}$ to 
 MADEX and the identifications to this paper.                
                                                             
Similarly to HC$_9$N $\nu_{19}$, another harmonically related
 series of doublets was      discovered later                
   resulting on the first space detection of  MgC$_{5}$N.    
 This discovery follows the previous one by \cite{Cerni2019} 
 of MgC$_3$N and and MgC$_4$H, which motivated the search for
 MgC$_6$H in in IRC+10126. However, given the line           
 intensities of MgC$_4$H, the expected ones for              
 MgC$_6$H  would be quite in the limit                       
 for detection. Extra observations were carried out in 2021  
 in order to confirm the detection of MgC$_6$H, and finally  
 both species were presented in \cite{Pardo2021}.

\begin{table*}[t]
  \caption{Results from rotational diagram fits in this work.
    $T_{rot}$: Rotational temperature, $N_{col}$: Column density, $Z_{rot}$: Partition function.
    The resulting fits are plotted in Figure \ref{fg:rotdiag},
    on top of the data points used for those fits
    (extracted from the individual line fits as listed in Table \ref{tb:id_qband}).
    Species with unreliable fits, due to a very low number of data points or extremely
      high data disperison in them, are not shown. These species, and others with a very low number of
      detected lines here, will be revisited in future
      publications of similar 4.3 to 1.0 mm surveys carried out with the IRAM 30m telescope.
  }
\label{tb:linefits}
\setlength\extrarowheight{-4.0pt}
\begin{center}
\begin{tabular}{cccccl}
\normalsize{Molecule}          & \normalsize{\# lines}  &\normalsize{ $T_{rot}$(K)   }&\normalsize{ $N_{col}$(cm$^{-2}$)   }&\normalsize{ $Z_{rot}$ }&\normalsize{  Comments }\\
\hline
\normalsize{  HC$_5$N                              }&\normalsize{  7         }&\normalsize{  $10.1\pm 0.6$              }&\normalsize{  $(4.2\pm 0.4)\times 10^{14}$    }&\normalsize{  158.80              }&\normalsize{  Reliable fit   }\\
\normalsize{ HC$_5$N $v_{11}=1$                   }&\normalsize{ 12        }&\normalsize{ $6.2\pm 1.8$               }&\normalsize{ $(7\pm 5)\times 10^{12}$        }&\normalsize{ 195.20              }&\normalsize{ Large data dispersion  }\\
% \normalsize{ HC$_5$N $v_{11}=2$                   }&\normalsize{ 3         }&\normalsize{ ---                        }&\normalsize{ ---                             }&\normalsize{ ---                 }&\normalsize{ $T_{rot}<0$~K  }\\
\normalsize{  H$^{13}$CCCCCN                       }&\normalsize{  8         }&\normalsize{  $10.0\pm 1.7$              }&\normalsize{  $(9.7\pm 2.7)\times 10^{12}$    }&\normalsize{  161.52              }&\normalsize{  Reliable fit  }\\
\normalsize{  HC$^{13}$CCCCN                       }&\normalsize{  8         }&\normalsize{  $14.4\pm 2.6$              }&\normalsize{  $(8.2\pm 1.7)\times 10^{12}$    }&\normalsize{  228.04              }&\normalsize{  Reliable fit   }\\
\normalsize{  HCC$^{13}$CCCN                       }&\normalsize{  7         }&\normalsize{  $10.4\pm  1.3$             }&\normalsize{  $(1.00\pm 0.19)\times 10^{13}$  }&\normalsize{  162.42              }&\normalsize{  Reliable fit  }\\
\normalsize{  HCCC$^{13}$CCN                       }&\normalsize{  7         }&\normalsize{  $10.6\pm  1.7$             }&\normalsize{  $(1.00\pm 0.25)\times 10^{13}$  }&\normalsize{  166.20              }&\normalsize{  Reliable fit   }\\
\normalsize{  HCCCC$^{13}$CN                       }&\normalsize{  8         }&\normalsize{  $9.2\pm 0.6$               }&\normalsize{  $(1.09\pm 0.13)\times 10^{13}$  }&\normalsize{  145.42              }&\normalsize{  Reliable fit  }\\
\normalsize{  HC$_7$N                              }&\normalsize{  17        }&\normalsize{  $15.8\pm 0.4$              }&\normalsize{  $(1.95\pm 0.11)\times 10^{14}$  }&\normalsize{  583.44              }&\normalsize{  Reliable fit  }\\
\normalsize{  HC$_7$N $v_{15}=1$                   }&\normalsize{  31        }&\normalsize{  $21.3\pm 1.7$              }&\normalsize{  $(2.0\pm 0.3)\times 10^{13}$    }&\normalsize{  1572.97             }&\normalsize{  Reliable fit (3 points removed)  }\\
\normalsize{ HC$_7$N $v_{15}=2$                   }&\normalsize{ 26        }&\normalsize{ $12.6\pm 2.5$              }&\normalsize{ $(1.0\pm 0.5)\times 10^{13}$    }&\normalsize{ 1358.51             }&\normalsize{ Large data dispersion   }\\
\normalsize{  H$^{13}$CCCCCCCN                     }&\normalsize{  11        }&\normalsize{  $23\pm 4$                  }&\normalsize{  $(3.0\pm 0.8)\times 10^{12}$    }&\normalsize{  879.75              }&\normalsize{  Reliable fit (1 point removed)   }\\
\normalsize{ HC$^{13}$CCCCCCN                     }&\normalsize{ 8         }&\normalsize{  $15\pm 7$                 }&\normalsize{ $(3.31\pm 3.41)\times 10^{12}$  }&\normalsize{ 559.13              }&\normalsize{ Large data dispersion  }\\
\normalsize{ HCC$^{13}$CCCCCN                     }&\normalsize{ 10        }&\normalsize{ $19\pm 6$                  }&\normalsize{ $(3.5\pm 2.1)\times 10^{12}$    }&\normalsize{ 713.79              }&\normalsize{ Large data dispersion  }\\
\normalsize{  HCCC$^{13}$CCCCN                     }&\normalsize{  10        }&\normalsize{  $23\pm 4$                  }&\normalsize{  $(2.7\pm 0.8)\times 10^{12}$    }&\normalsize{  836.22              }&\normalsize{  Reliable fit (2 points removed)   }\\
\normalsize{  HCCCC$^{13}$CCCN                     }&\normalsize{  6         }&\normalsize{   $16\pm 3$                 }&\normalsize{  $(3.9\pm 1.7)\times 10^{12}$    }&\normalsize{  573.14              }&\normalsize{  Reliable fit   }\\
\normalsize{ HCCCCC$^{13}$CCN                     }&\normalsize{ 11        }&\normalsize{ $22\pm 6$                  }&\normalsize{ $(2.7\pm 1.3)\times 10^{12}$    }&\normalsize{ 824.82              }&\normalsize{ Large data dispersion  }\\
\normalsize{  HCCCCCC$^{13}$CN                     }&\normalsize{  9         }&\normalsize{  $31\pm 10$                }&\normalsize{  $(3.2\pm 1.1)\times 10^{12}$    }&\normalsize{  1156.63             }&\normalsize{  Reliable fit (1 point removed) }\\
\normalsize{  HC$_9$N                              }&\normalsize{  30        }&\normalsize{ $22.9\pm 1.1$              }&\normalsize{  $(4.5\pm 0.6)\times 10^{13}$    }&\normalsize{  1644.03             }&\normalsize{  Reliable fit (1 point removed)  }\\
\normalsize{  HC$_9$N $v_{19}=1$                   }&\normalsize{  50        }&\normalsize{ $28.0\pm 2.2$              }&\normalsize{  $(2.4\pm 0.4)\times 10^{13}$    }&\normalsize{  4009.46             }&\normalsize{  Reliable fit (1 point removed)  }\\
\normalsize{ l-HC$_4$N                            }&\normalsize{ 6         }&\normalsize{ $5.1\pm 2.5$               }&\normalsize{ $(3.9\pm 3.7)\times 10^{12}$    }&\normalsize{ 126.85              }&\normalsize{ Large data dispersion   }\\
%\normalsize{ o-H$_2$C$_4$                         }&\normalsize{ 6         }&\normalsize{ ---                        }&\normalsize{ ---                             }&\normalsize{ ---                 }&\normalsize{ Bad fit. Large data dispersion   }\\
%\normalsize{ p-H$_2$C$_4$                         }&\normalsize{ 5         }&\normalsize{ ---                        }&\normalsize{ ---                             }&\normalsize{ ---                 }&\normalsize{ $T_{rot}<0$~K. Large data dispersion  }\\
%\normalsize{ o-H$_2$C$_3$                         }&\normalsize{ 2         }&\normalsize{ ---                        }&\normalsize{ ---                             }&\normalsize{ ---                 }&\normalsize{ Negligible $E_{rot}$ range   }\\
%\normalsize{ C$_3$H                               }&\normalsize{ 6         }&\normalsize{ ---                        }&\normalsize{ ---                             }&\normalsize{ ---                 }&\normalsize{ Negligible $E_{rot}$ range   }\\
\normalsize{  C$_4$H                               }&\normalsize{  4         }&\normalsize{  $4.9\pm 0.6$               }&\normalsize{  $(1.84\pm 0.25)\times 10^{14}$  }&\normalsize{  86.42               }&\normalsize{  Reliable fit   }\\
\normalsize{ C$_4$H $v_7=1$                       }&\normalsize{ 10        }&\normalsize{ $10\pm 5$                  }&\normalsize{ $(6.6\pm 2.0)\times 10^{13}$    }&\normalsize{ 314.65              }&\normalsize{ Large data dispersion   }\\
\normalsize{  C$_4$H $v_7=2$                       }&\normalsize{  4         }&\normalsize{  $6.4\pm 1.3$               }&\normalsize{  $(2.2\pm 0.4)\times 10^{13}$    }&\normalsize{  113.77              }&\normalsize{  Reliable fit    }\\
\normalsize{  C$_5$H ${}^2\Pi_{1/2}$               }&\normalsize{  8         }&\normalsize{  $5.0\pm 0.4$               }&\normalsize{  $(2.7\pm 0.5)\times 10^{13}$    }&\normalsize{  172.74              }&\normalsize{  Reliable fit   }\\
\normalsize{ C$_5$H ${}^2\Pi_{3/2}$               }&\normalsize{ 8         }&\normalsize{ $5.9\pm 0.8$               }&\normalsize{ $(2.52\pm 2.48)\times 10^{14}$  }&\normalsize{ 10.67               }&\normalsize{ Moderate data dispersion   }\\
\normalsize{  C$_6$H ${}^2\Pi_{3/2}$               }&\normalsize{  14        }&\normalsize{  $11.0\pm 0.8$              }&\normalsize{  $(1.51\pm 0.14)\times 10^{13}$  }&\normalsize{  334.91              }&\normalsize{  Reliable fit   }\\
\normalsize{  C$_6$H ${}^2\Pi_{1/2}$               }&\normalsize{  14        }&\normalsize{  $12.4\pm 0.6$              }&\normalsize{  $(1.23\pm 0.18)\times 10^{13}$  }&\normalsize{  66.83               }&\normalsize{  Reliable fit   }\\
\normalsize{  C$_6$H $v_{11}=1$                    }&\normalsize{  14        }&\normalsize{  $10.4\pm 0.5$              }&\normalsize{  $(9.3\pm 0.7)\times 10^{12}$    }&\normalsize{  310.61              }&\normalsize{  Reliable fit (2 points removed)  }\\
\normalsize{ C$_6$H ${}^2\Delta_{5/2}$ $v_{11}=1$ }&\normalsize{ 6         }&\normalsize{  $13\pm 3$                 }&\normalsize{ $(2.0\pm 0.5)\times 10^{12}$    }&\normalsize{ 201.51              }&\normalsize{ Moderate data dispersion   }\\
\normalsize{ C$_6$H ${}^2\Delta_{3/2}$ $v_{11}=1$ }&\normalsize{ 7         }&\normalsize{  $21\pm 13$                }&\normalsize{ $(1.4\pm 0.7)\times 10^{12}$    }&\normalsize{ 311.75              }&\normalsize{ Large data dispersion   }\\
\normalsize{  C$_7$H ${}^2\Pi_{1/2}$               }&\normalsize{  22        }&\normalsize{   $13.2\pm 1.4$             }&\normalsize{  $(5.1\pm 1.0)\times 10^{12}$    }&\normalsize{  1260.39             }&\normalsize{  Reliable fit (1 point removed)   }\\
\normalsize{ C$_7$H ${}^2\Pi_{3/2}$               }&\normalsize{ 11        }&\normalsize{  $14.7\pm 2.2$             }&\normalsize{ $(3.2\pm 2.0)\times 10^{12}$    }&\normalsize{ 110.99              }&\normalsize{ Moderate data dispersion   }\\
\normalsize{  C$_8$H ${}^2\Pi_{3/2}$               }&\normalsize{  30        }&\normalsize{   $14.8\pm 0.6$             }&\normalsize{   $(8.9\pm 0.9)\times 10^{12}$   }&\normalsize{  2111.08             }&\normalsize{  Reliable fit (1 point removed)  }\\
\normalsize{ C$_8$H ${}^2\Pi_{1/2}$               }&\normalsize{ 16        }&\normalsize{  $8.0\pm 1.4$              }&\normalsize{  $(1.15\pm 1.38)\times 10^{13}$ }&\normalsize{ 35.16               }&\normalsize{ Large data dispersion   }\\
%\normalsize{ C$_3$N                               }&\normalsize{ 4         }&\normalsize{ $3.29\pm 0.19$             }&\normalsize{ $(1.91\pm 0.21)\times 10^{14}$  }&\normalsize{ 85.11               }&\normalsize{Large data dispersion. Short $E_{rot}$ range. Unreliable fit.  }\\
%\normalsize{ CC$^{13}$CN                          }&\normalsize{ 7         }&\normalsize{ $1.6\pm 0.5$               }&\normalsize{ $(9.96\pm 11.30)\times 10^{13}$ }&\normalsize{ 156.19              }&\normalsize{Large data dispersion. Short $E_{rot}$ range. Unreliable fit.  }\\
%\normalsize{ C$^{13}$CCN                          }&\normalsize{ 4         }&\normalsize{ $2.0\pm 0.9$               }&\normalsize{ $(4.50\pm 6.14)\times 10^{13}$  }&\normalsize{ 156.15              }&\normalsize{Large data dispersion. Short $E_{rot}$ range. Unreliable fit.   }\\
%\normalsize{ $^{13}$CCCN                          }&\normalsize{ 6         }&\normalsize{ $6\pm 4$                   }&\normalsize{ $(5.65\pm 3.45)\times 10^{12}$  }&\normalsize{ 324.01              }&\normalsize{Large data dispersion. Short $E_{rot}$ range. Unreliable fit.   }\\
%\normalsize{ C$_5$N                               }&\normalsize{ 4         }&\normalsize{ ---                        }&\normalsize{ ---                             }&\normalsize{ ---                 }&\normalsize{ $T_{rot}<0$~K. Large data dispersion   }\\
\normalsize{  C$_5$N$^-$                           }&\normalsize{  6         }&\normalsize{  $8.8\pm 1.4$               }&\normalsize{  $(4.9\pm 1.6)\times 10^{12}$    }&\normalsize{  132.62              }&\normalsize{  Reliable fit   }\\
\normalsize{  C$_6$H$^-$                           }&\normalsize{  7         }&\normalsize{  $11.4\pm 1.0$              }&\normalsize{  $(4.3\pm 0.6)\times 10^{12}$    }&\normalsize{  173.51              }&\normalsize{  Reliable fit   }\\
\normalsize{ C$_8$H$^-$                           }&\normalsize{ 13        }&\normalsize{ $23\pm 5$                  }&\normalsize{ $(1.1\pm 0.4)\times 10^{12}$    }&\normalsize{ 810.24              }&\normalsize{Reasonable fit. Large data dispersion   }\\
%\normalsize{ p-C$_3$H$_2$                         }&\normalsize{ 2         }&\normalsize{ 7.47                       }&\normalsize{ $1.84\times 10^{13}$            }&\normalsize{ 13.04               }&\normalsize{ Unreliable fit. Only 2 data points   }\\
%\normalsize{ SiC$_2$                              }&\normalsize{ 2         }&\normalsize{ 35.46                      }&\normalsize{ $1.04\times 10^{15}$            }&\normalsize{ 231.45              }&\normalsize{ Unreliable fit. Only 2 data points  }\\
%\normalsize{ Si$^{13}$CC                          }&\normalsize{ 2         }&\normalsize{ 2.42                       }&\normalsize{ $5.46\times 10^{12}$            }&\normalsize{ 11.39               }&\normalsize{ Unreliable fit. Only 2 data points   }\\
\normalsize{  SiC$_4$                              }&\normalsize{  6         }&\normalsize{  $12.9\pm 1.8$              }&\normalsize{  $(6.2\pm 1.0)\times 10^{12}$    }&\normalsize{  175.36              }&\normalsize{  Reliable fit   }\\
\normalsize{ $^{29}$SiC$_4$                       }&\normalsize{ 5         }&\normalsize{ $10\pm 6$                  }&\normalsize{ $(5.73\pm 5.06)\times 10^{11}$  }&\normalsize{ 134.88              }&\normalsize{ Large data dispersion   }\\
\normalsize{  SiC$_6$                              }&\normalsize{  7         }&\normalsize{  $17.3\pm 2.5$              }&\normalsize{  $(1.3\pm 0.4)\times 10^{12}$    }&\normalsize{  588.57              }&\normalsize{  Reliable fit (1 point removed)   }\\
\normalsize{  CCS                                  }&\normalsize{  4         }&\normalsize{  $7.9\pm 1.1$               }&\normalsize{  $(4.0\pm 0.7)\times 10^{13}$    }&\normalsize{  44.48               }&\normalsize{  Reliable fit   }\\
\normalsize{ C$_3$S                               }&\normalsize{ 3         }&\normalsize{ $23.60\pm 96.85$           }&\normalsize{ $(2.58\pm 3.60)\times 10^{13}$  }&\normalsize{ 170.50              }&\normalsize{ Very high data dispersion   }\\
\normalsize{ NaCN                                 }&\normalsize{ 7         }&\normalsize{ $24\pm 13$                 }&\normalsize{ $(8.5\pm 1.7)\times 10^{12}$    }&\normalsize{ 425.76              }&\normalsize{ Large data dispersion (1 point removed)  }\\
%\normalsize{ MgNC                                 }&\normalsize{ 4         }&\normalsize{ $16.31\pm 22.15$           }&\normalsize{ $(9.89\pm 3.86)\times 10^{12}$  }&\normalsize{ 113.60              }&\normalsize{ Unreliable fit. Very high data dispersion   }\\
\normalsize{  MgC$_3$N                             }&\normalsize{  14        }&\normalsize{  $11.1\pm 0.9$              }&\normalsize{  $(5.2\pm 0.6)\times 10^{12}$    }&\normalsize{  336.00              }&\normalsize{  Reliable fit (1 point removed)  }\\
\normalsize{  MgC$_5$N                             }&\normalsize{  30        }&\normalsize{  $15.4\pm 1.8$              }&\normalsize{  $(4.7\pm 1.3)\times 10^{12}$    }&\normalsize{  1109.27             }&\normalsize{  Reliable fit (1 point removed)   }\\
\normalsize{ MgC$_4$H                             }&\normalsize{ 10        }&\normalsize{ $4.7\pm 1.2$               }&\normalsize{ $(1.5\pm 1.3)\times 10^{13}$    }&\normalsize{ 140.43              }&\normalsize{ Large data dispersion  }\\
\normalsize{ MgC$_6$H                             }&\normalsize{ 30         }&\normalsize{ $24.8\pm 8.9$               }&\normalsize{ $(2.0\pm 0.9)\times 10^{13}$    }&\normalsize{ 966.24             }&\normalsize{ Large data dispersion  }\\
%\normalsize{ CH$_3$CN                             }&\normalsize{ 3         }&\normalsize{ ---                        }&\normalsize{ ---                             }&\normalsize{ ---                 }&\normalsize{ Negligible $E_{rot}$ range  }\\
%\normalsize{ CH$_3$C$_3$N                         }&\normalsize{ 5         }&\normalsize{ ---                        }&\normalsize{ ---                             }&\normalsize{ ---                 }&\normalsize{ $T_{rot}<0$~K. Very high data dispersion }\\
%\normalsize{ KCl                                  }&\normalsize{ 2         }&\normalsize{ ---                        }&\normalsize{ ---                             }&\normalsize{ ---                 }&\normalsize{ $T_{rot}<0$~K  }\\
\hline
\end{tabular}
\end{center}
\end{table*}

\begin{figure*}[h]     \begin{center}                        
\includegraphics[width=0.86\textwidth]{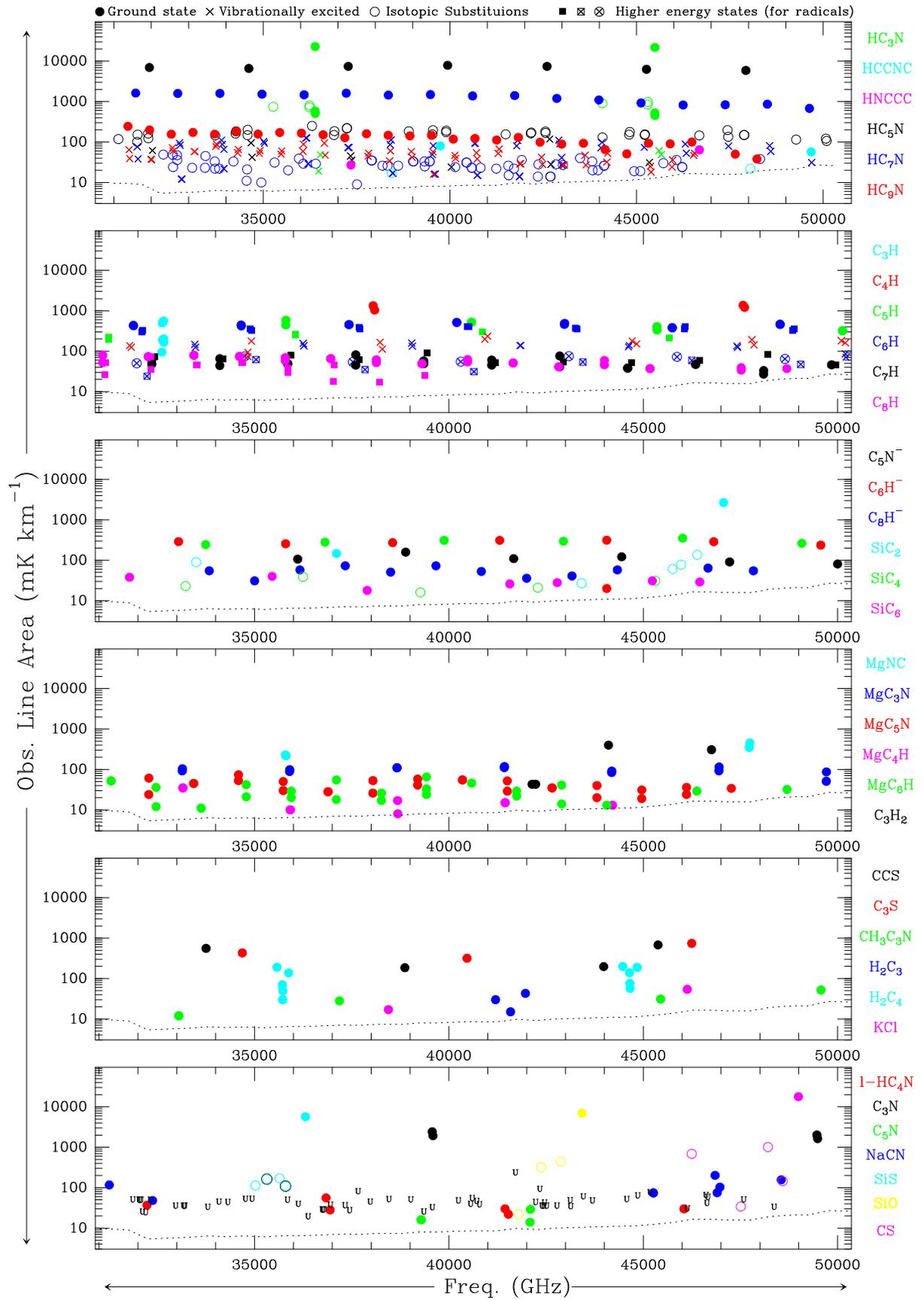}         
\end{center} \caption{Graphic summary of all lines detected  
in this work ($\int$T$_A^*$dv versus frequency). Dotted line 
marks the 1 $\sigma$ limit in the data according to Fig.     
\ref{fg:rms}. The remaining U-lines appear in the bottom     
panel.} \label{fg:freqdiag} \end{figure*}  \clearpage        
\begin{figure*}[h]         \begin{center}                    
\includegraphics[width=0.86\textwidth]{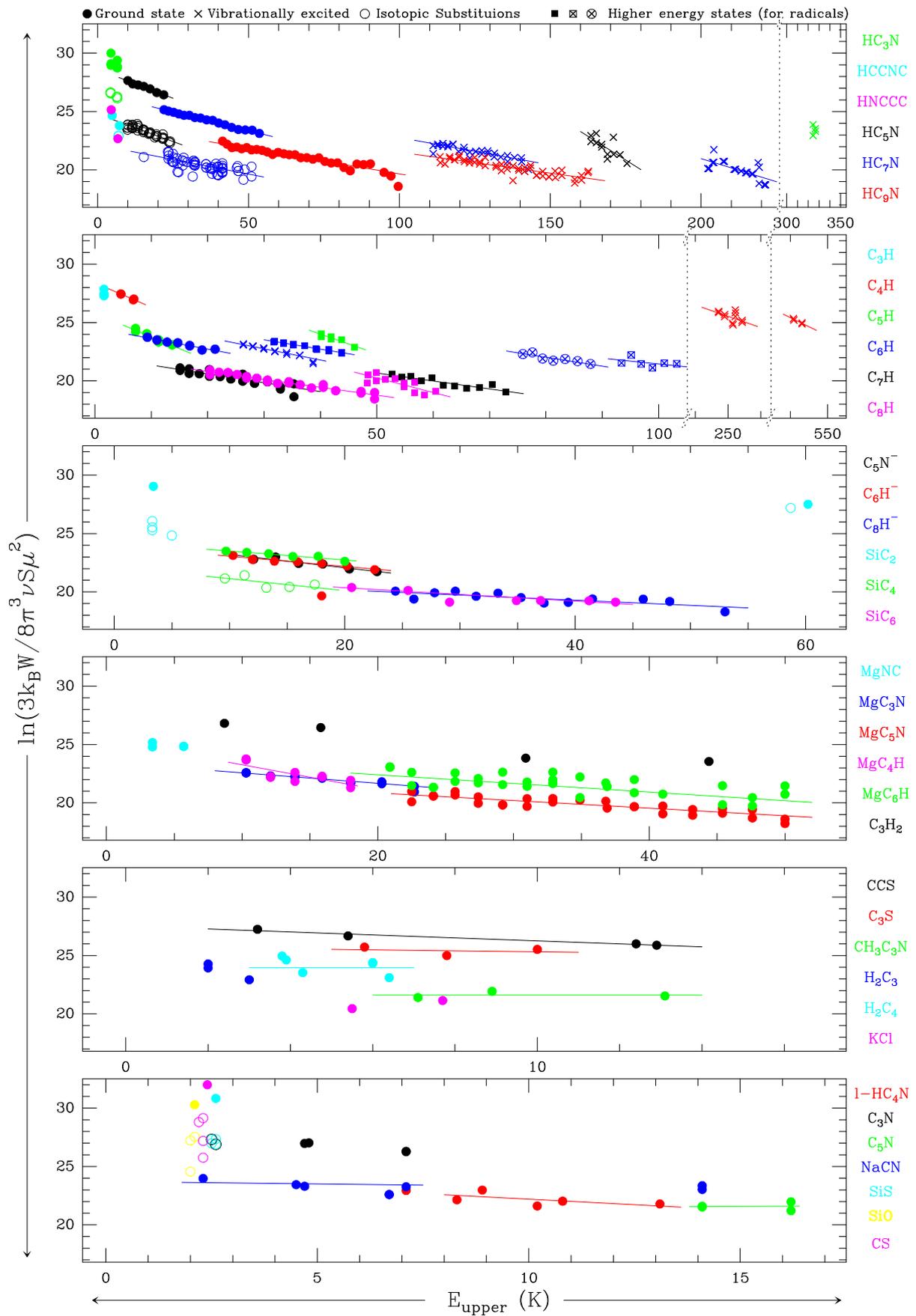}          
\end{center} \caption{A more physical view of most lines     
 detected in this work, in the form of rotational diagrams.  
 See fit results in Table \ref{tb:linefits}.                 
Note that the changing telescope beam across the Q-band is   
taken into account here through W=$\int$T$_{MB}$dv in the    
y-axis, instead of $\int$T$_A^*$dv as in Figure              
\ref{fg:freqdiag}. The values of W for each individual line  
can be found in Table \ref{tb:id_qband}. The fits shown      
correspond to molecules listed in Table \ref{tb:linefits}.}  
\label{fg:rotdiag} \end{figure*}                             
\clearpage                                                   
 \subsection{Most significant U-lines} \label{sct:most-sig}  
                                                             
 A careful, long and painstaking work, has allowed us to     
 reduce to a minimum the number of unidentified features     
 in this work. They are summarized in Table \ref{ulines}.    
                                                             
 Among all the remaining U-Lines, the strongest one,         
 at 14$\sigma$ appears at a frequency of 41711.90 MHz        
 and was already present and not identified by               
 \cite{Kaw95}. MADEX does not %\clearpage                    
 provide any candidate to be the carrier, and the same       
 applies for other catalogs.                                 
                                                             
 After this strong U-line we have to go down to 6$\sigma$    
 for the next unidentified line, standing at a central       
 frequency of 42340 MHz.   %The first one is partially       
 %blended with HCCC$^{13}$CCN J=13-12 and, therefore,        
 %the determination of its central frequency is not easy.    
 %Nevertheless, the extension of the signal in the blue part 
 %of the  HCCC$^{13}$CCN J=13-12 line appears on all three   
%different receiver tunings used for these observations with 
 %band would appear at different signal side band frequencies
 %and can therefore be eliminated from the data, which we did
 %in a few cases. So, there is definitely a line around      
 %34586 but its carrier is not clear yet.                    
  This line  appears as a nice and isolated spectral feature 
 also present at the same position on all receiver tunings   
 (reference central frequencies: 42300.0,                    
  42280.0 and 41370.0 MHz).                                  
 A look at MADEX                                             
 catalog does not provide any plausible                      
 carrier as the only candidate, Ag$^{37}$Cl, does not have   
 any emission at other frequencies in this survey, around    
 35284 and 49396 MHz, and Ag$^{35}$Cl is not found either.   
  %\clearpage                                                
                                                             
 At the 5$\sigma$ level there are just 2 unidentified lines  
 at the following frequencies in round numbers: 35849        
 and 37663 MHz. These two lines, again, appear isolated and  
  well detected on all three different receiver tunings at   
 the same position. Again,                                   
               no answers are found in MADEX about possible  
 carriers. It is worth noting, in any case, that we are      
 talking about features with peak intensities below 3 mK     
 in the T$_A^*$ scale.                                       
 %blending, the central frequency is difficult to know and,  
                                                             
 At the 4$\sigma$ level we have only one unidentified line,  
%a doublet at 33431 and 33433 MHz (both lines together have  
 at 43457.12 MHz. Again, it                                  
 appears     at the same position in all three receiver      
 tunings but no carrier has been found for it yet.           
                                                             
 \subsection{Unidentified lines at 3$\sigma$ level}          
 Apart from the most significant unidentified lines discussed
 in section  \ref{sct:most-sig},                             
  52
more features remain unidentified just at the 3$\sigma$ level
 and with line widths within $\pm$20\% from that typical of  
the source. Most of these features are in fact at $\sim$1 mK 
 level so that some of them could just be artifacts that     
 would disappear on deeper integrations (that we plan to do) 
 or might be baseline effects although we                    
do not have evidence to remove any of the remaining ones, as 
we have done with several other in our careful data reduction
 process. Table \ref{ulines} should therefore be kept as a   
 reference for follow-up in future works.                    
      
\section{Procedure for line identification}                  
\label{sect:lineiden}                                        
                                                             
 The first thing to say in this section is that not all the  
 identifications are 100\% sure, although we are quite       
 confident about most of them. The main doubts arise when (i)
 the frequency coincidence is not within $\sim$1 MHz         
 typically, (ii) the line width is too narrow or too broad   
 with respect to that typical of the source, (iii) there is a
 close blending with a stronger line, and (iv) only one line 
 for a given species is found and we cannot confirm it from  
 other lines detected at higher frequencies with other       
 telescopes. These circumstances are pointed out in our      
 line identification Table \ref{tb:id_qband}, when necessary.
                                                             
 For our data analysis and line identification process, we   
 have defined a reference file containing the data smoothed  
 to a resolution of 229 kHz (6 channel smoothing from the    
 original output), from which we have removed all artifacts  
 (spikes, baseline undulations)                              
 as much as we could. Then we fitted all the spectral        
 features in the data using the method SHELL implemented in  
 GILDAS package, which is adequate for the CW Leo expanding  
 envelope. We left the expansion velocity as a free parameter
 in this fit as an extra check for the line identification   
 process (see below). The fitted line profiles were stored as
 a third column in the reference file so that we can         
 incorporate them to the figures for the convenience of the  
 reader. We created a separate file containing the MADEX     
 catalog entries that we identified as corresponding to the  
 observed spectral features, and another file storing the    
 SHELL method fitting parameters. The first file             
 (irc10216\_qband.dat) is used to create figures .           
 The other two files (irc10216\_qband.lines and              
 irc10216\_qband.parameters) are used as inputs for a        
 fortran script, specially designed for this paper, that     
 automatically creates Tables \ref{tb:molecules} and         
 \ref{tb:id_qband}, calculating in particular                
 W=$\int$T$_{MB}$dv where the changing telescope beam        
 across the Q-band is taken into account,                    
 and also updates the numbers and                            
 percentages of identified lines given in the abstract and in
 the main text. This allows us to introduce criteria in our  
 analysis, and new fits or baseline corrections, that are    
 immediately checked throughout all the data in a consistent 
 consistent way, with the results updated in our manuscript  
 without having to check "by hand" again and again. This     
 also provides a systematic way of re-checking the whole     
 work if, for example, more observations are added and,      
 therefore, the noise level is improved and/or, more         
 frequencies are added.                                      
 %\clearpage
\section{Legacy and summary} \label{sect:summary}            
The main purpose of this work has been to provide a reference
 in the Q-band for further investigations on the molecular   
 content of the circumstellar envelope of CW Leo using       
 other bands. For that reason we provide the full data set   
 and line fits as on-line material available for the         
 scientific community.                                       
                                                             
 Nevertheless, we can say a few things about the results     
 published here. First of all, the large amount of lines and 
 molecules detected provides an opportunity to give a        
 very good diagnostics for the rotational temperature. To do 
 that we use those molecules in Table \ref{tb:linefits}      
 with reliable fits. We provide three results, the first two 
 are weighted averages using the number of lines detected and
 the derived column density. The third value  is just an     
 arithmetic average. The values are:                         
  16.9,  11.4, and  14.1
 K respectively.                                             
                                                             
 Concerning the isotopic ratios, the only one we can give in 
 this work is $^{12}$C/$^{13}$C using HC$_5$N due to the     
 weakness of the lines from the HC$_7$N isotopologues and the
 small number of lines (2) from HC$_3$N and its              
 isotopologues.  The result ($^{12}$C/$^{13}$C =             
 43.0$\pm$  4.7
 is in good agreement with previous works, 44\,$\pm$\,3 from 
 \cite{Kah1992} and 45\,$\pm$\,3 from \cite{Cer00}.          
                                                             
 The other important contribution of this work is that it    
 provides a large amount of data on molecular column         
 densities to be used for comparisons with                   
 results from chemical models of this source, together with  
 the full 4.3-1.0 mm data from the IRAM-30m telescope that   
 should also be published soon.                 Detected     
 molecules with very few lines in the Q-band are left for    
 analysis in that future publication, combining the data     
 from them published here (Table \ref{tb:id_qband} and       
 Figures \ref{fig01} to \ref{fig39})                         
 with those obtained with the IRAM-30m telescope.            
                                                             
%\section{Summary}                                           
In this paper we have presented the full results of Q-band   
(31.0-50.3 GHz) deep integration towards IRC+10216, one of   
the richest and more exhaustively studied molecular stellar  
 envelopes in the sky. This work represents a big improvement
 over previously available surveys in the same frequency     
 range, in terms of both noise level and line identification 
 due to the use of state of the art line catalogs. Prior to  
 this publication, a series of papers has presented the      
 identification from these data of new                       
 molecular species and/or vibrational states never before    
 detected in space of known molecular species. The list of   
 U-lines is almost entirely reduced to features at 3-4       
 $\sigma$ level with the exception of three lines that       
 still defy identification.                                  
                                                             
 In order to provide a useful legacy from this project, we   
 have proceeded to fit all the spectroscopic features in the 
 data to provide detailed figures of data versus fit; tables 
with line identifications with theoretical versus            
experimental line  parameters; and, finally, rotational      
diagrams to explore the physical conditions of IRC+10216     
 through the molecular species present in this work.         
\begin{acknowledgements}                                     
  We thank Ministerio de Ciencia e Innovaci\'on of Spain     
  for funding support through projects                       
  PID2019-106110GB-I00,                                      
 PID2019-107115GB-C21 / AEI / 10.13039/501100011033,         
 PID2019-106235GB-I00, and grant FJCI-2016-27983 for CB.     
  We also thank ERC for funding through                      
 grant ERC-2013-Syg-	610256-NANOCOSMOS. M.A. thanks MICIU    
 for grant RyC-2014-16277.                                   
                            \end{acknowledgements}

  {}                                      
\clearpage                                                   

\begin{appendix}                                             
\onecolumn                                                   
\section{Tables and Figures}  \label{sect:appendix}          
\begin{figure*}                                              
\includegraphics[width=0.93\textwidth]                       {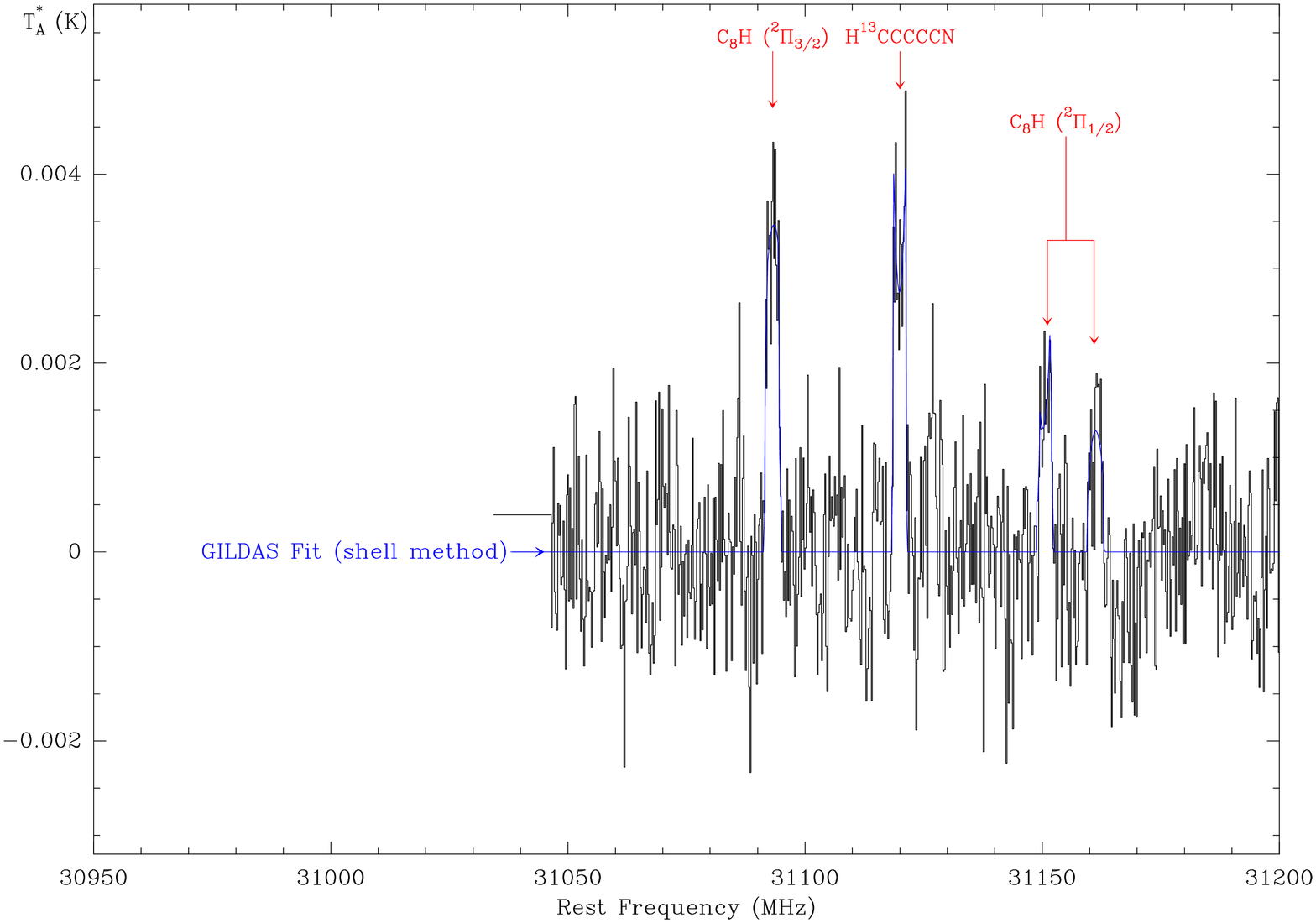}\\
\includegraphics[width=0.93\textwidth]                       {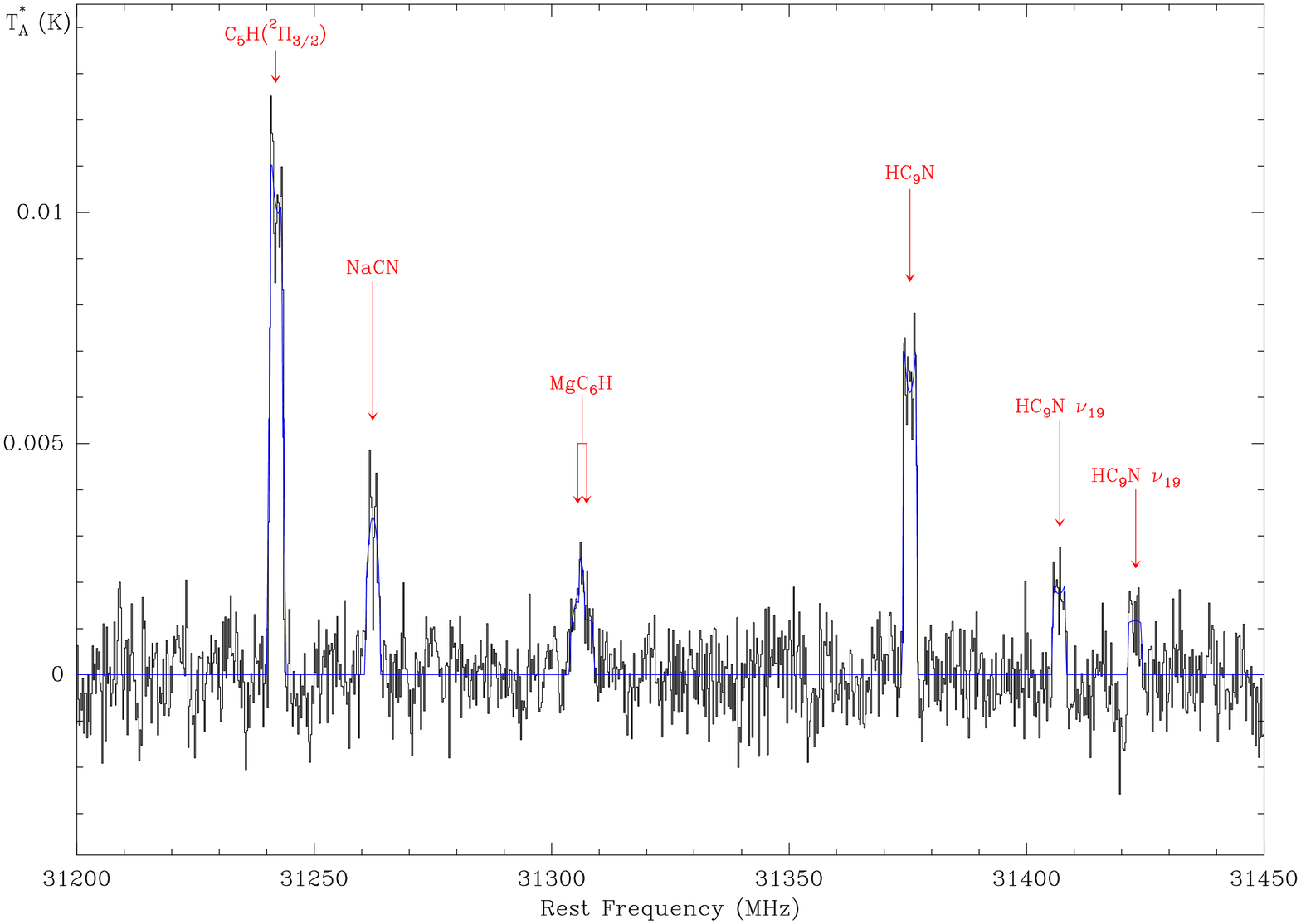}
\caption{IRC+10216 YEBES 40m data, line fits and labels from 
30950 to 31450 GHz.}
\label{fig01}
\end{figure*}

\begin{figure*}                                              
\includegraphics[width=0.93\textwidth]                       {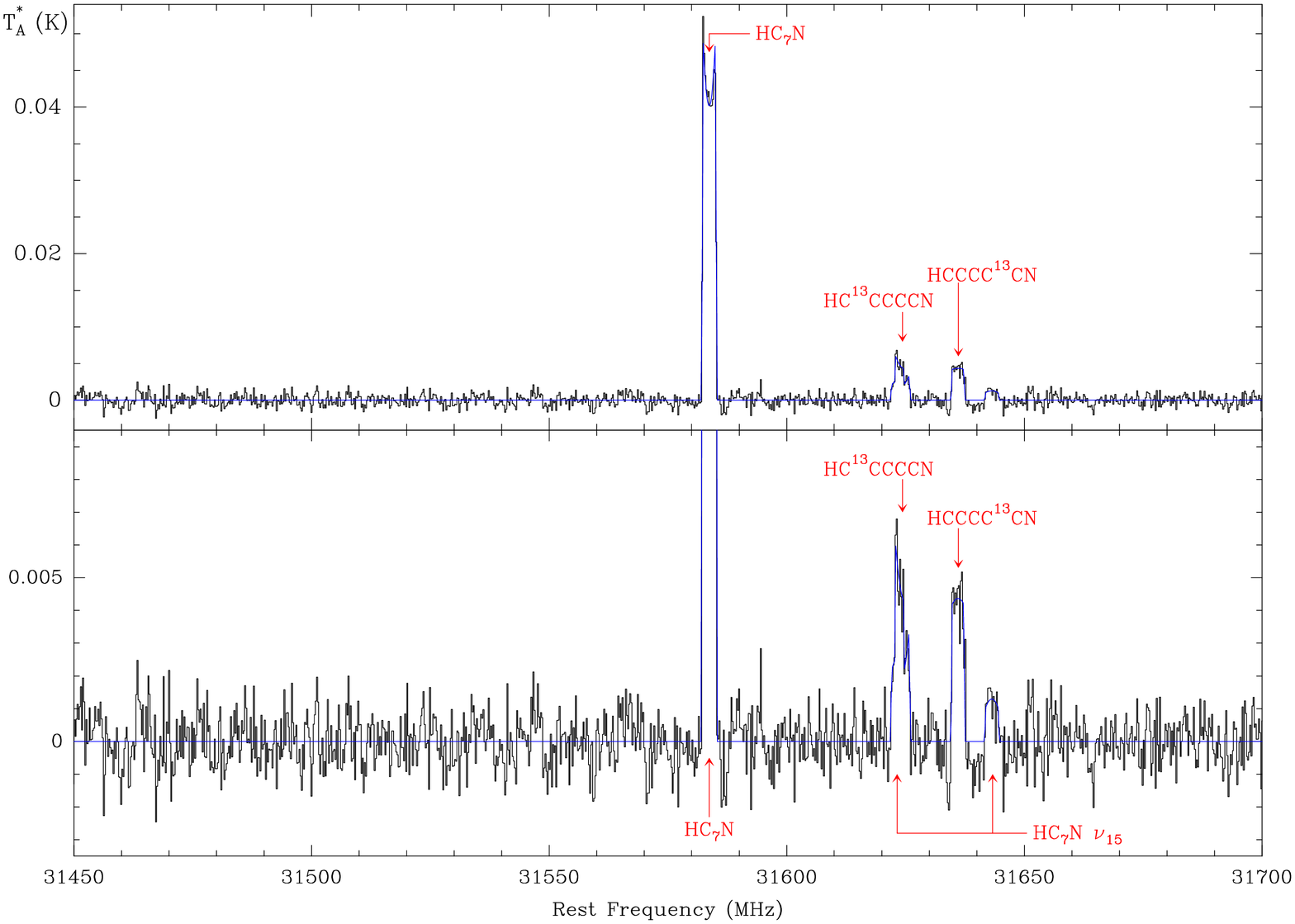}\\
\includegraphics[width=0.93\textwidth]                       {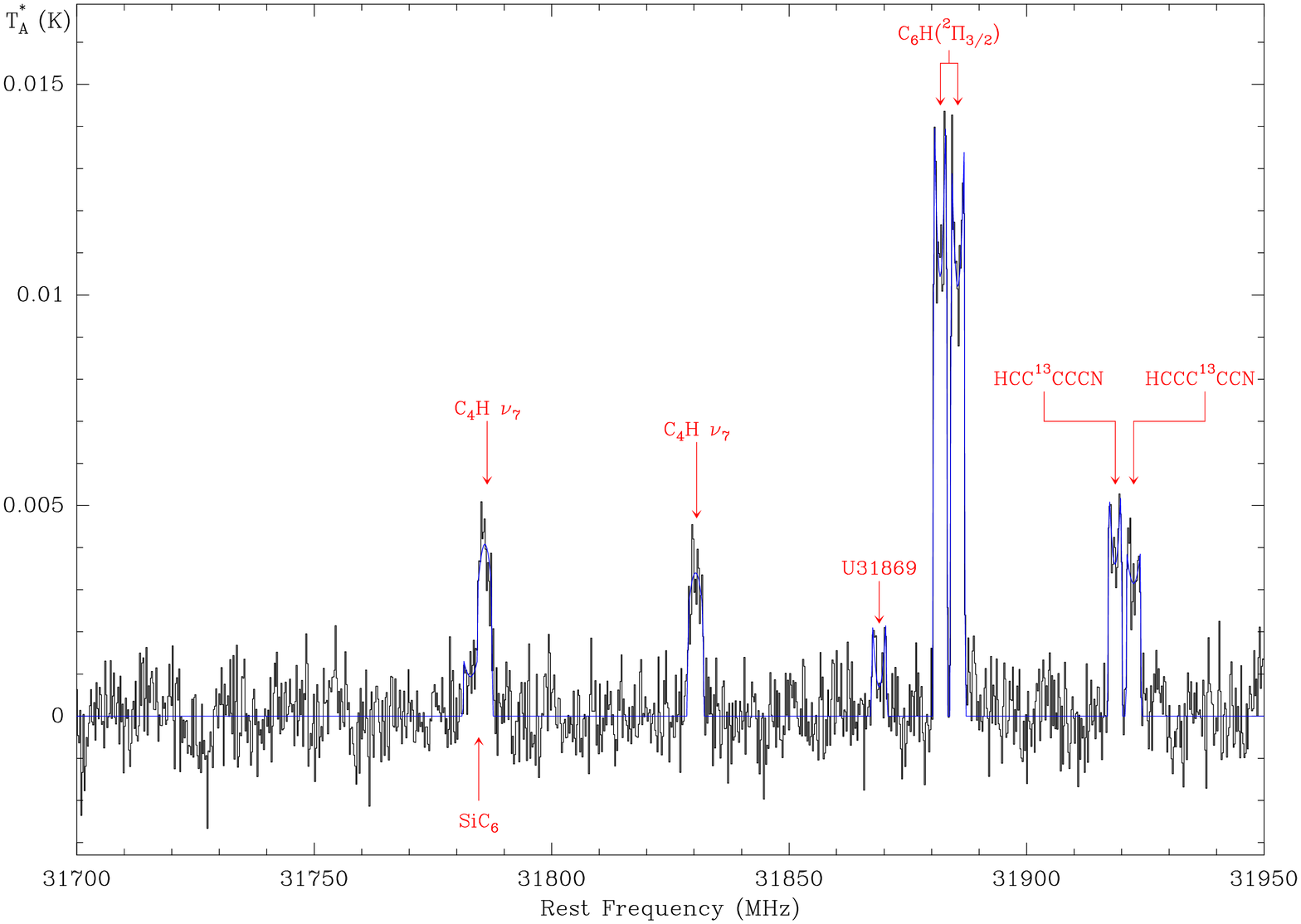}
\caption{IRC+10216 YEBES 40m data, line fits and labels from 
31450 to 31950 GHz.}
\label{fig02}
\end{figure*}

\begin{figure*}                                              
\includegraphics[width=0.93\textwidth]                       {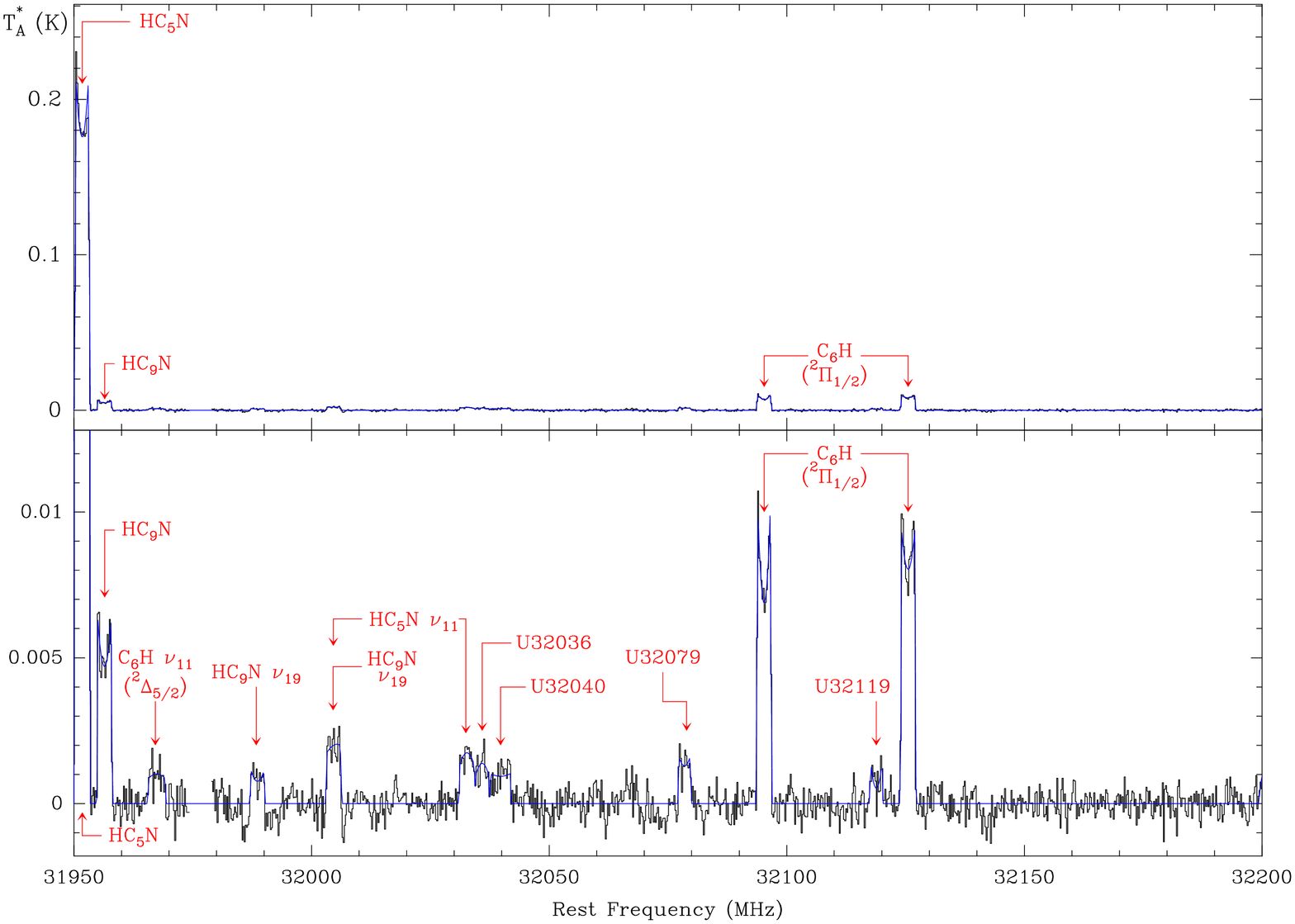}\\
\includegraphics[width=0.93\textwidth]                       {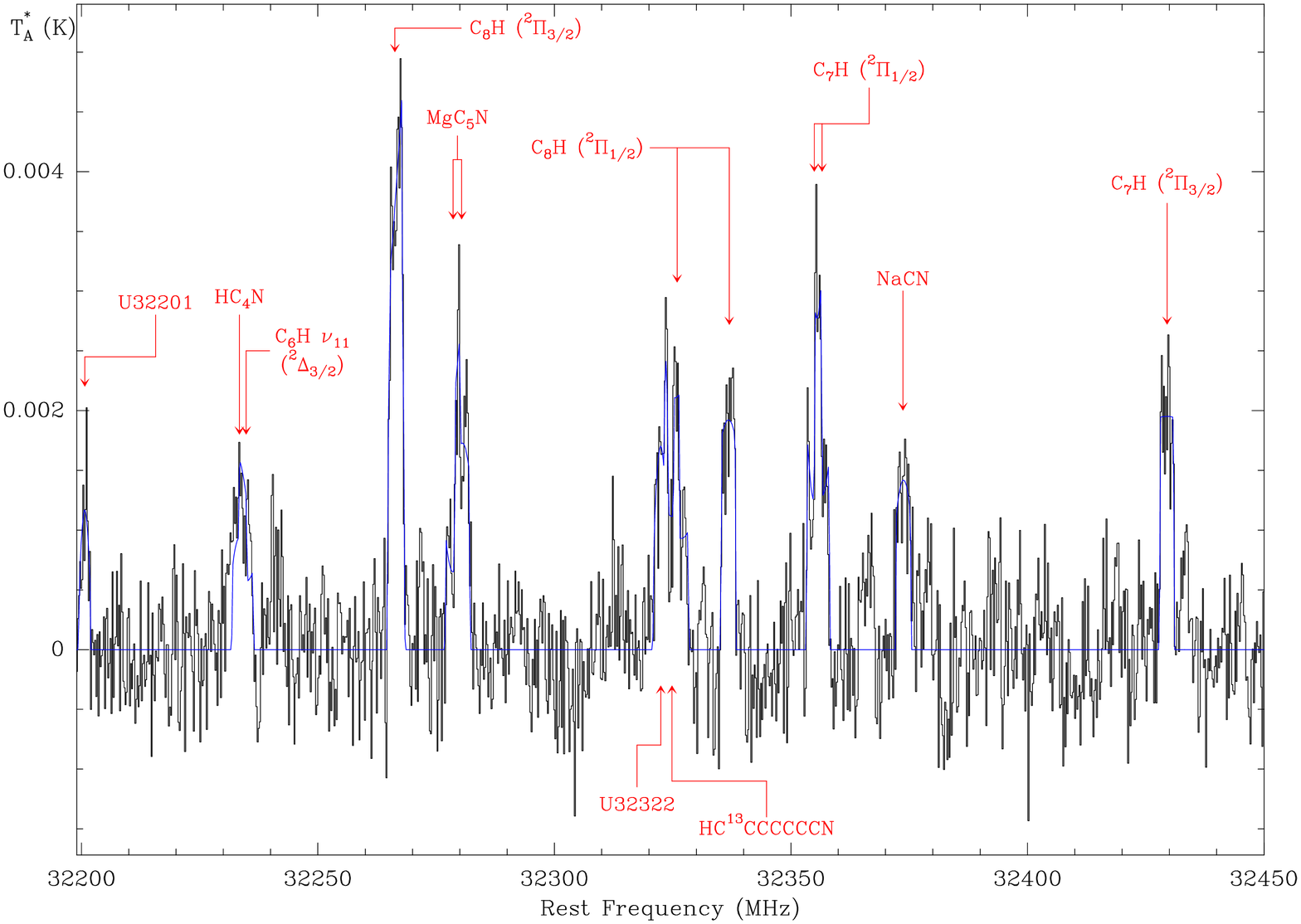}
\caption{IRC+10216 YEBES 40m data, line fits and labels from 
31950 to 32450 GHz.}
\label{fig03}
\end{figure*}

\begin{figure*}                                              
\includegraphics[width=0.93\textwidth]                       {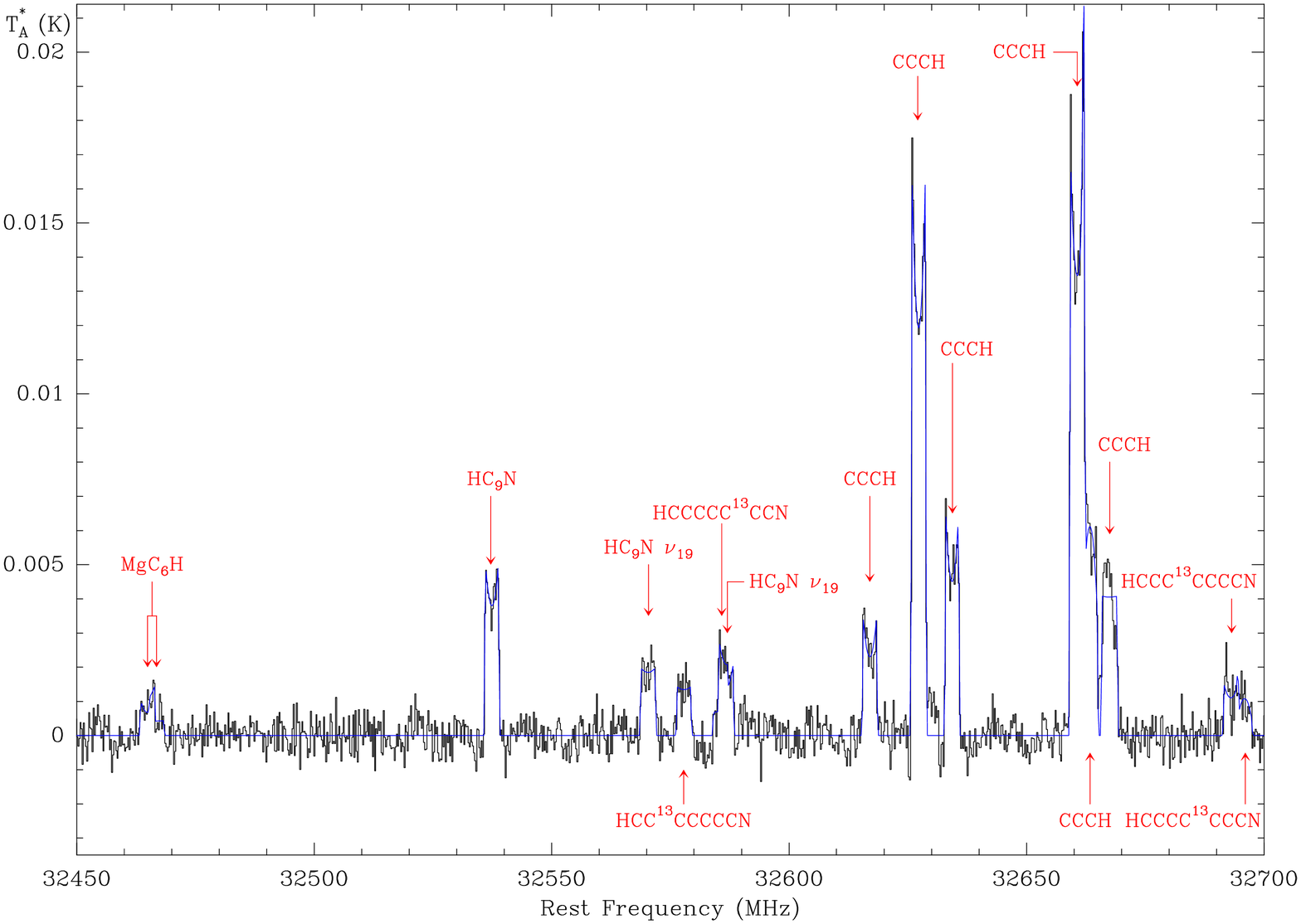}\\
\includegraphics[width=0.93\textwidth]                       {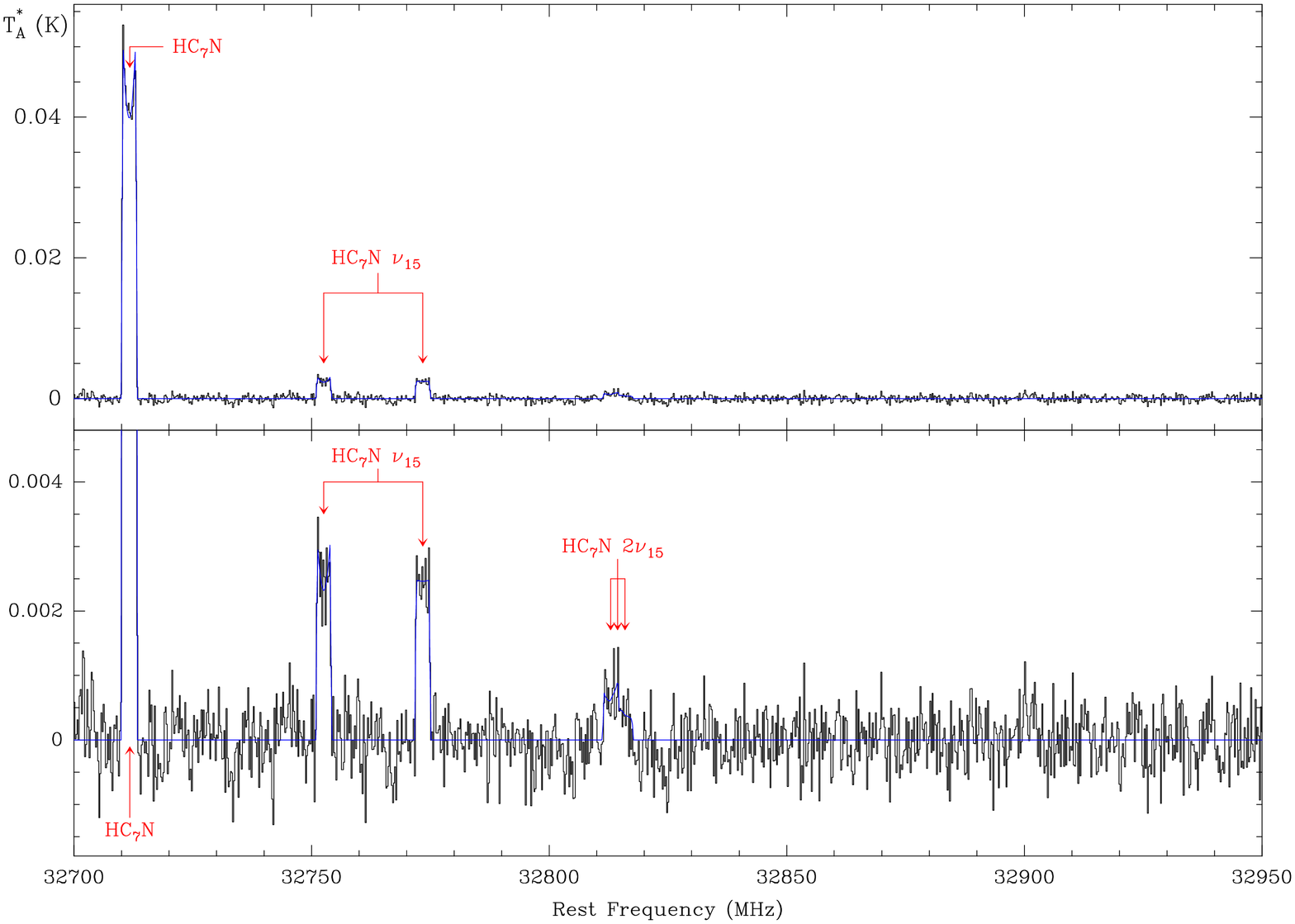}
\caption{IRC+10216 YEBES 40m data, line fits and labels from 
32450 to 32950 GHz.}
\label{fig04}
\end{figure*}

\begin{figure*}                                              
\includegraphics[width=0.93\textwidth]                       {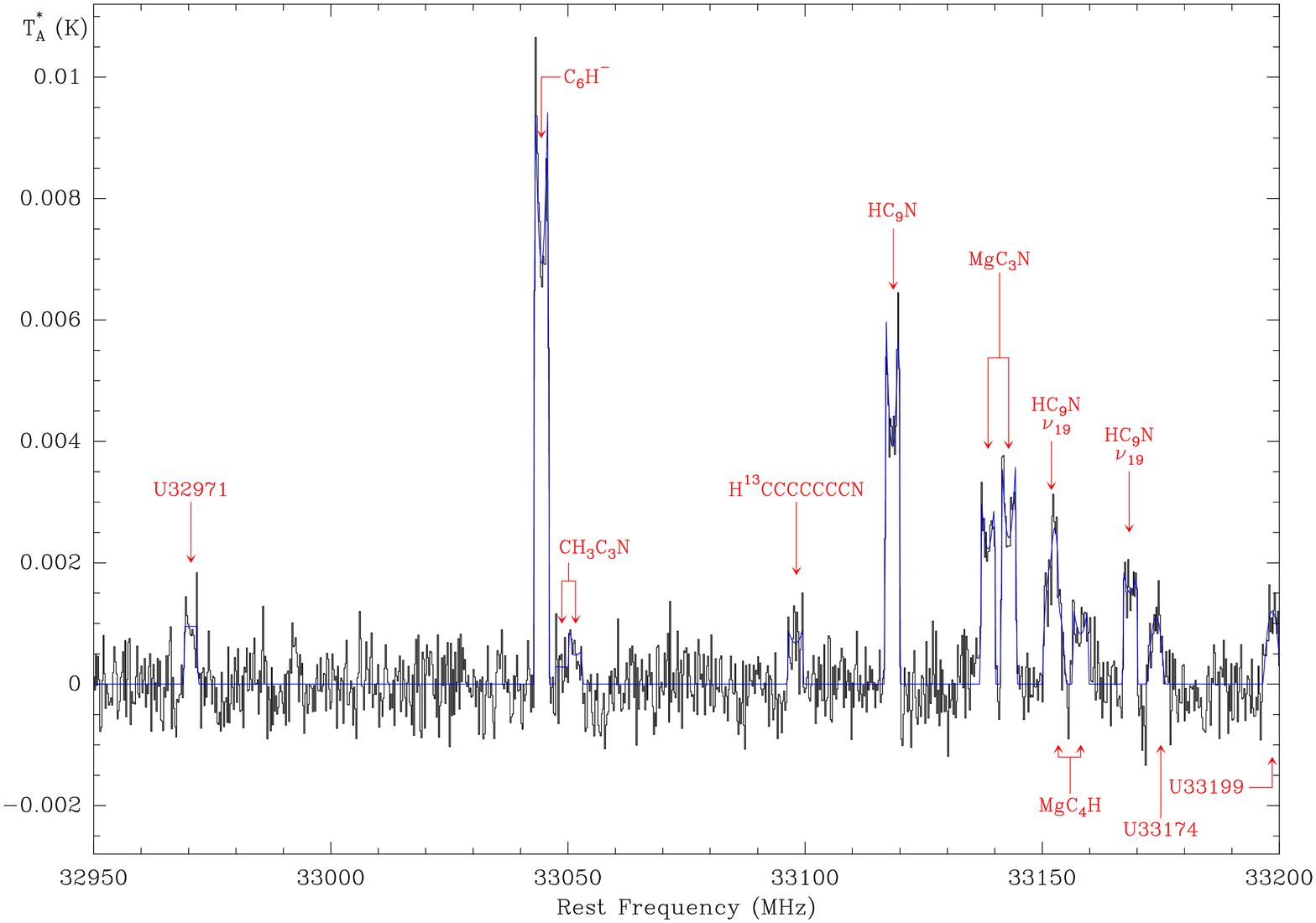}\\
\includegraphics[width=0.93\textwidth]                       {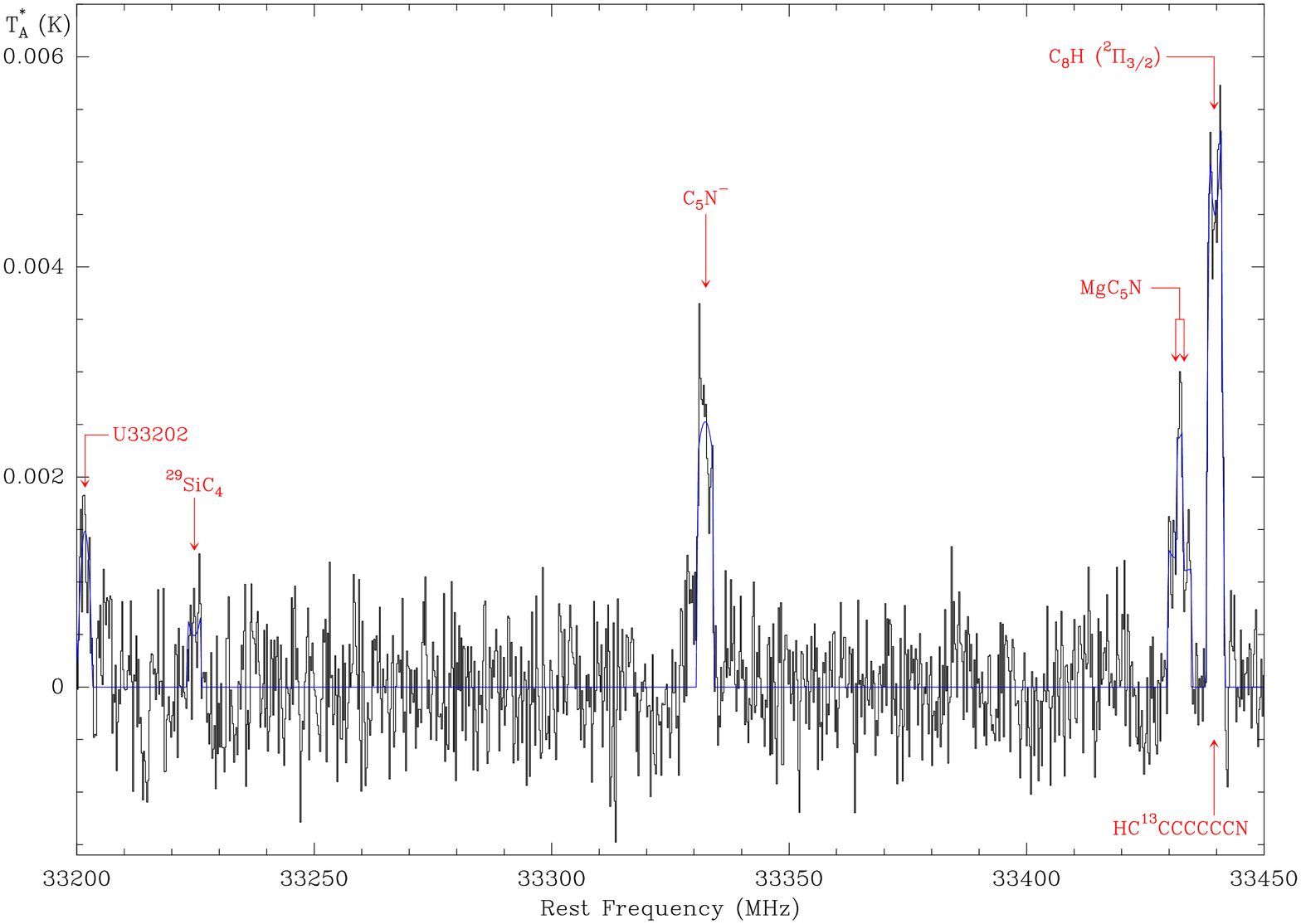}
\caption{IRC+10216 YEBES 40m data, line fits and labels from 
32950 to 33450 GHz.}
\label{fig05}
\end{figure*}

\begin{figure*}                                              
\includegraphics[width=0.93\textwidth]                       {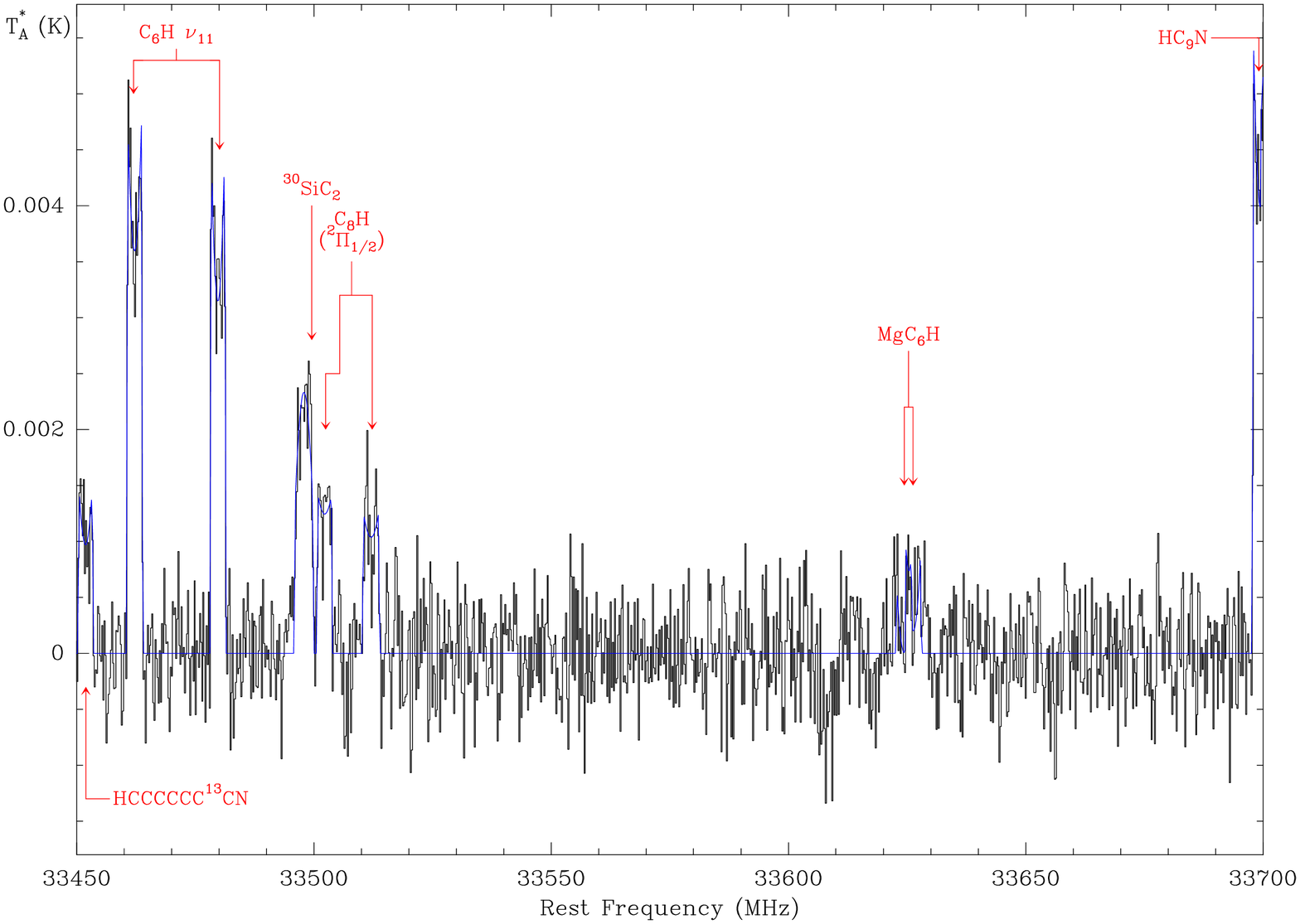}\\
\includegraphics[width=0.93\textwidth]                       {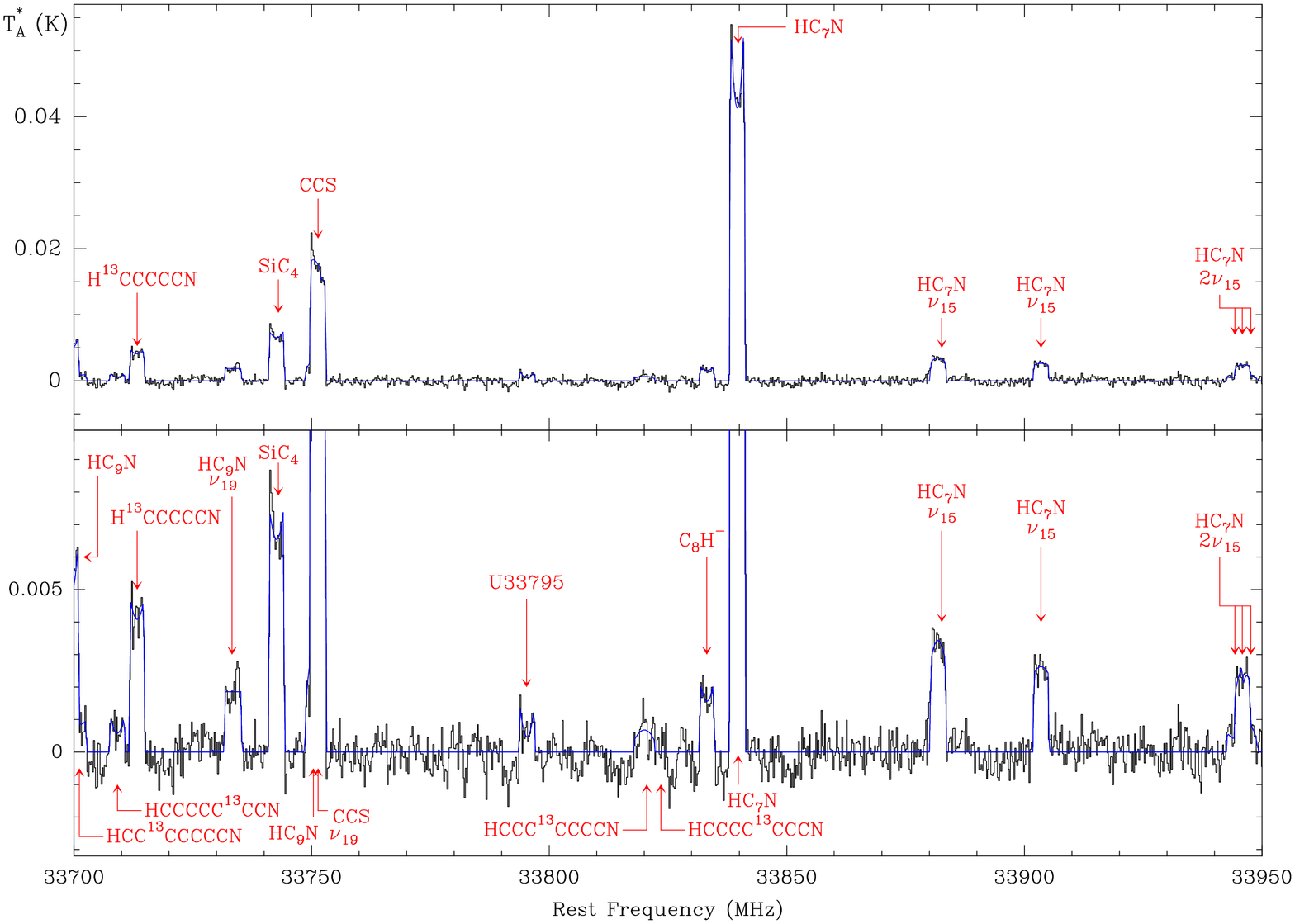}
\caption{IRC+10216 YEBES 40m data, line fits and labels from 
33450 to 33950 GHz.}
\label{fig06}
\end{figure*}

\begin{figure*}                                              
\includegraphics[width=0.93\textwidth]                       {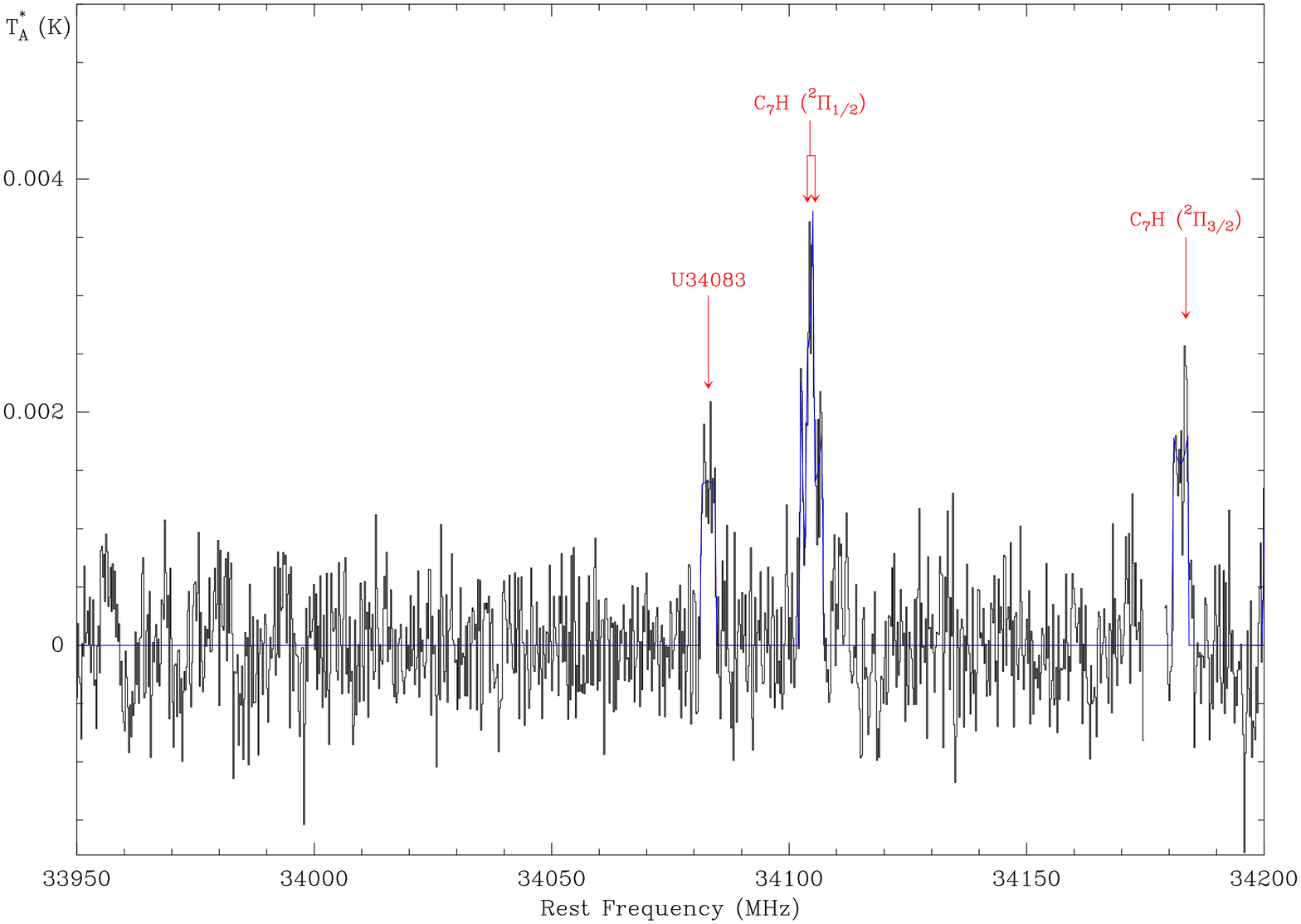}\\
\includegraphics[width=0.93\textwidth]                       {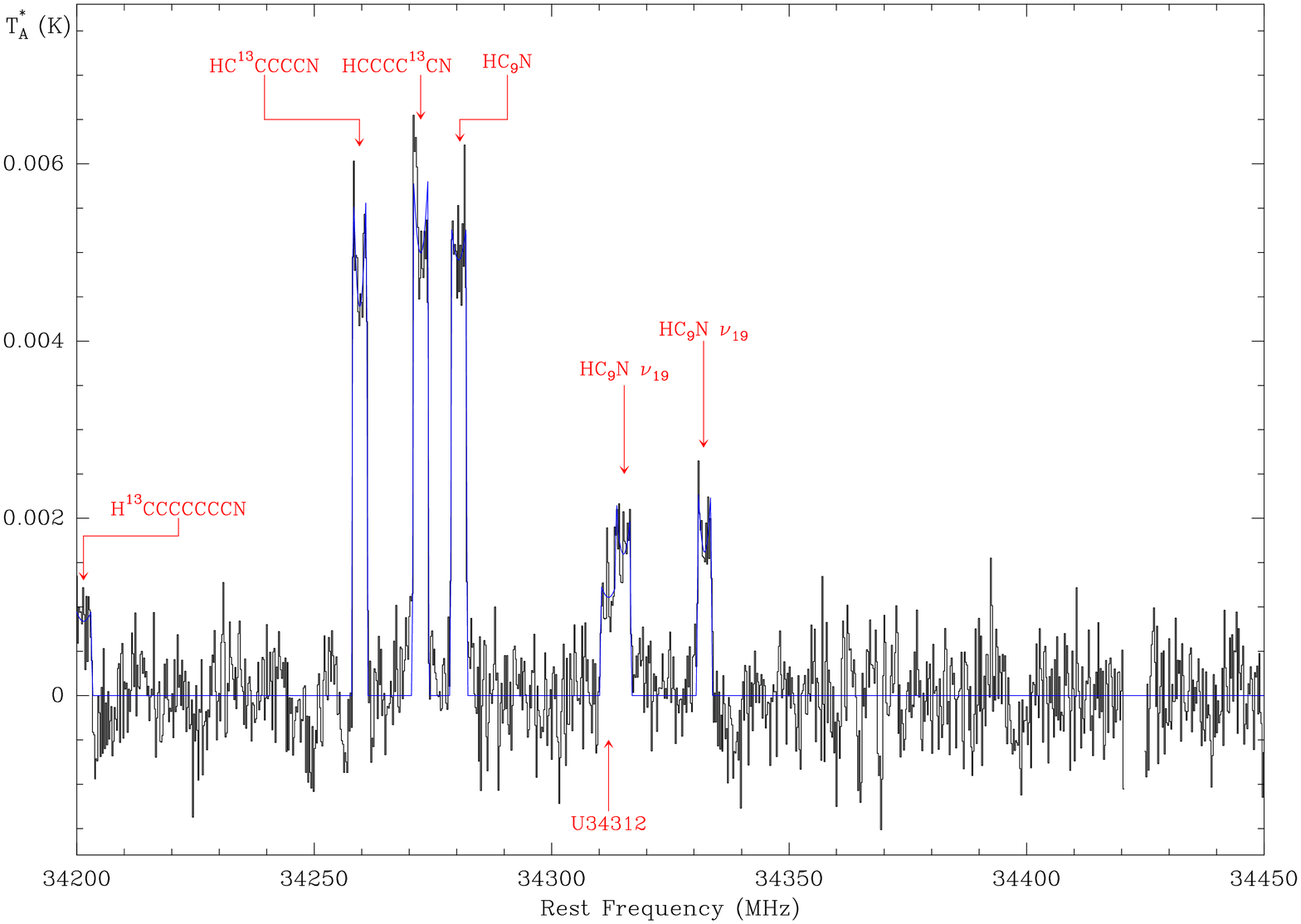}
\caption{IRC+10216 YEBES 40m data, line fits and labels from 
33950 to 34450 GHz.}
\label{fig07}
\end{figure*}

\begin{figure*}                                              
\includegraphics[width=0.93\textwidth]                       {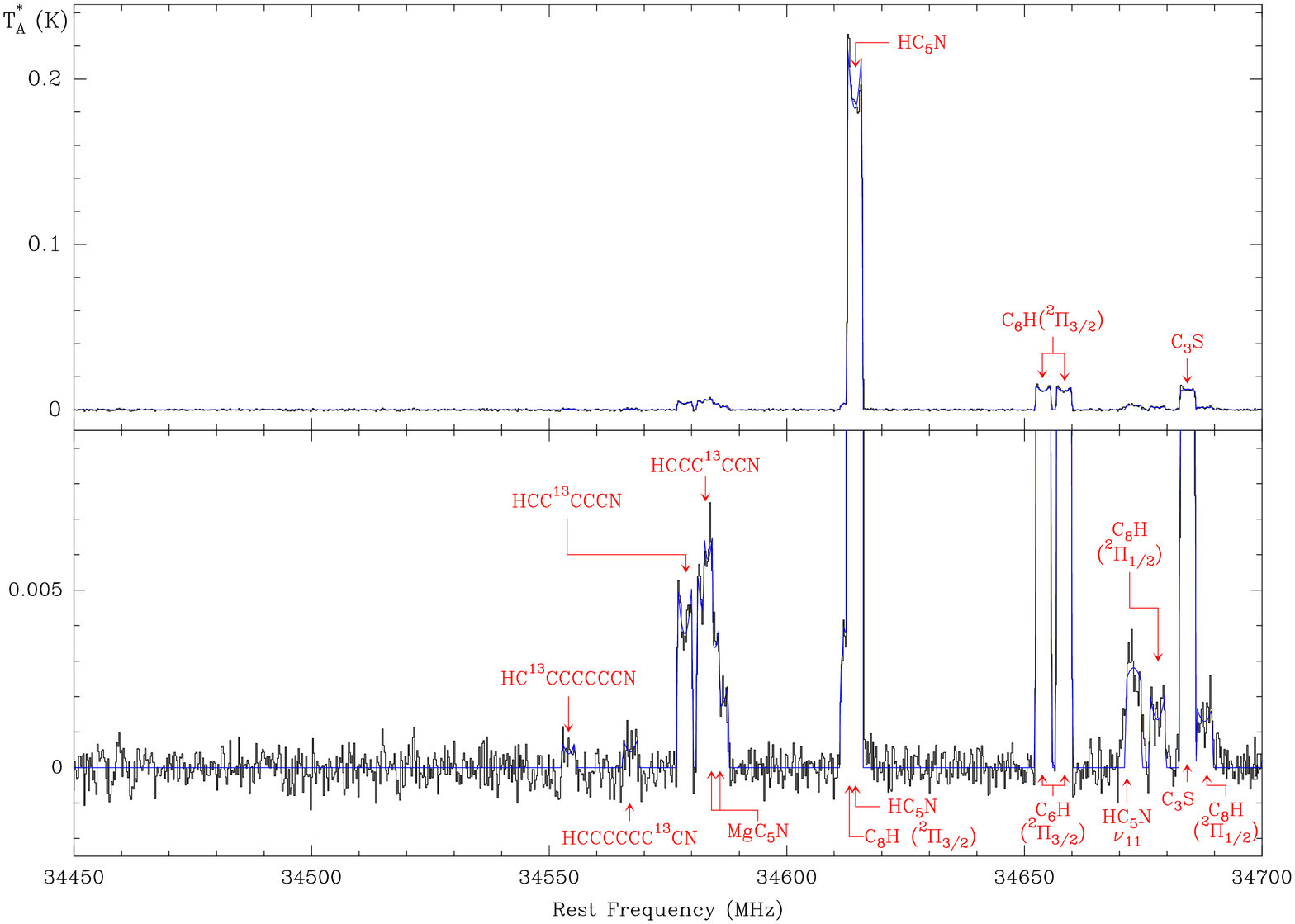}\\
\includegraphics[width=0.93\textwidth]                       {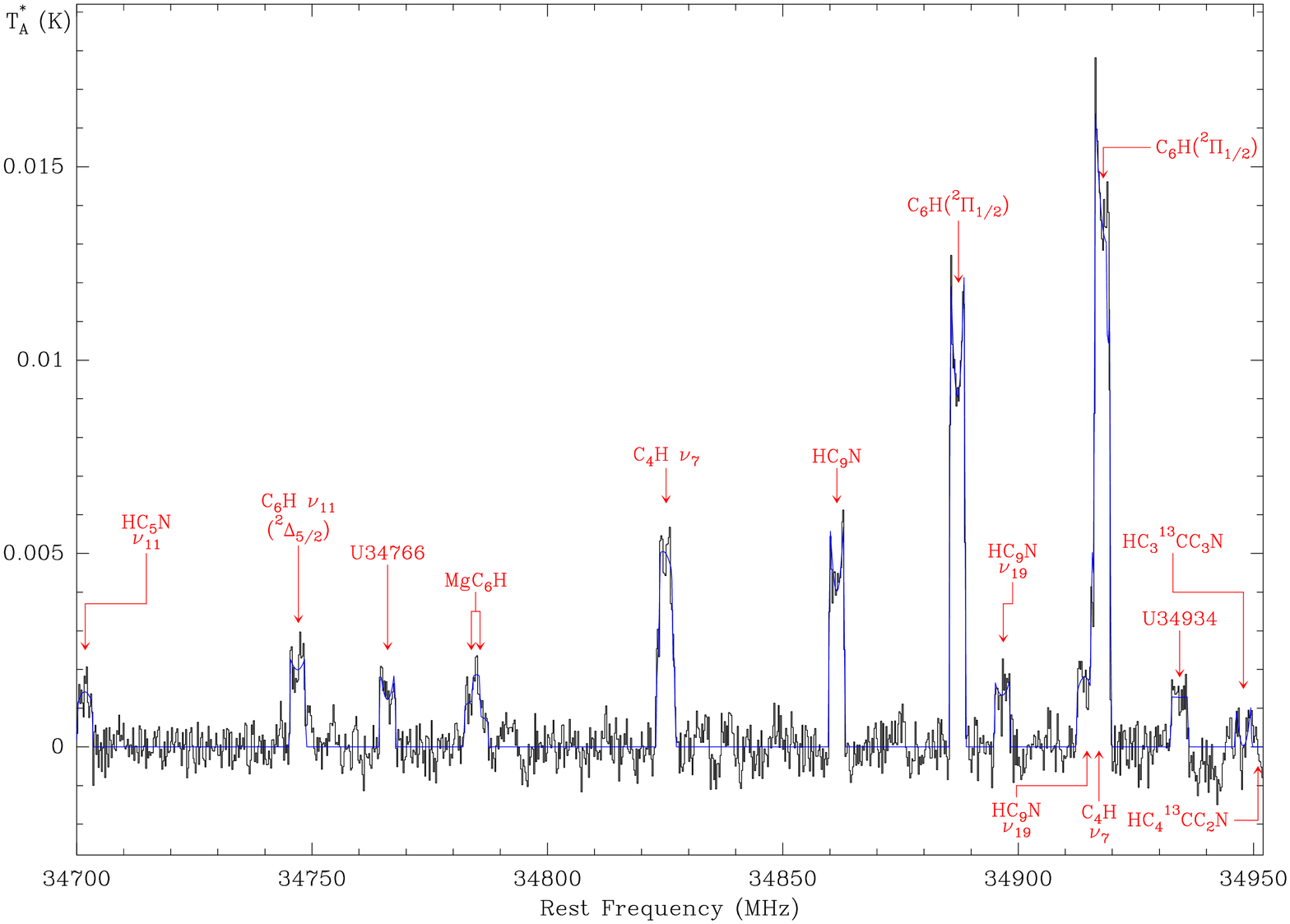}
\caption{IRC+10216 YEBES 40m data, line fits and labels from 
34450 to 34950 GHz.}
\label{fig08}
\end{figure*}

\begin{figure*}                                              
\includegraphics[width=0.93\textwidth]                       {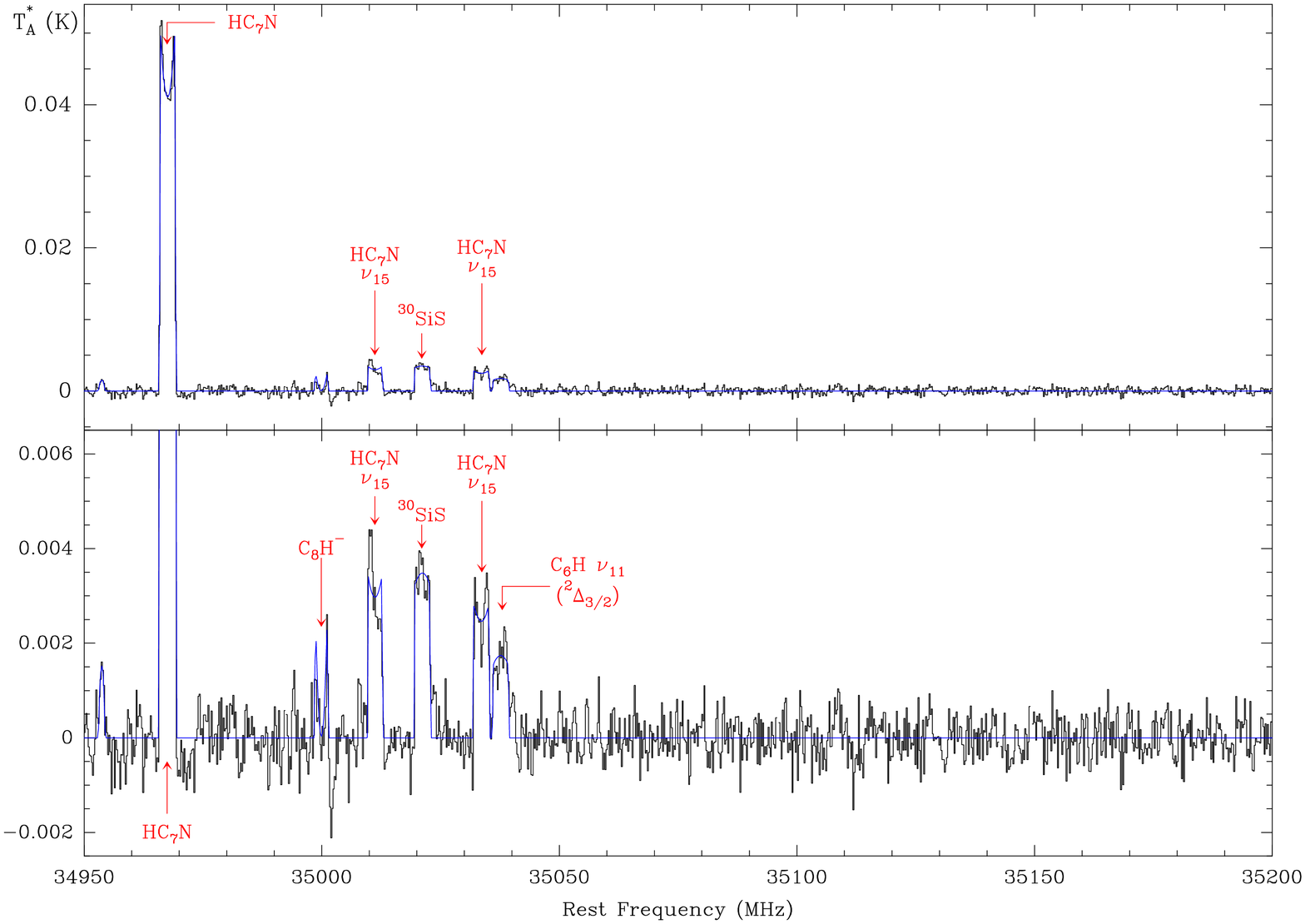}\\
\includegraphics[width=0.93\textwidth]                       {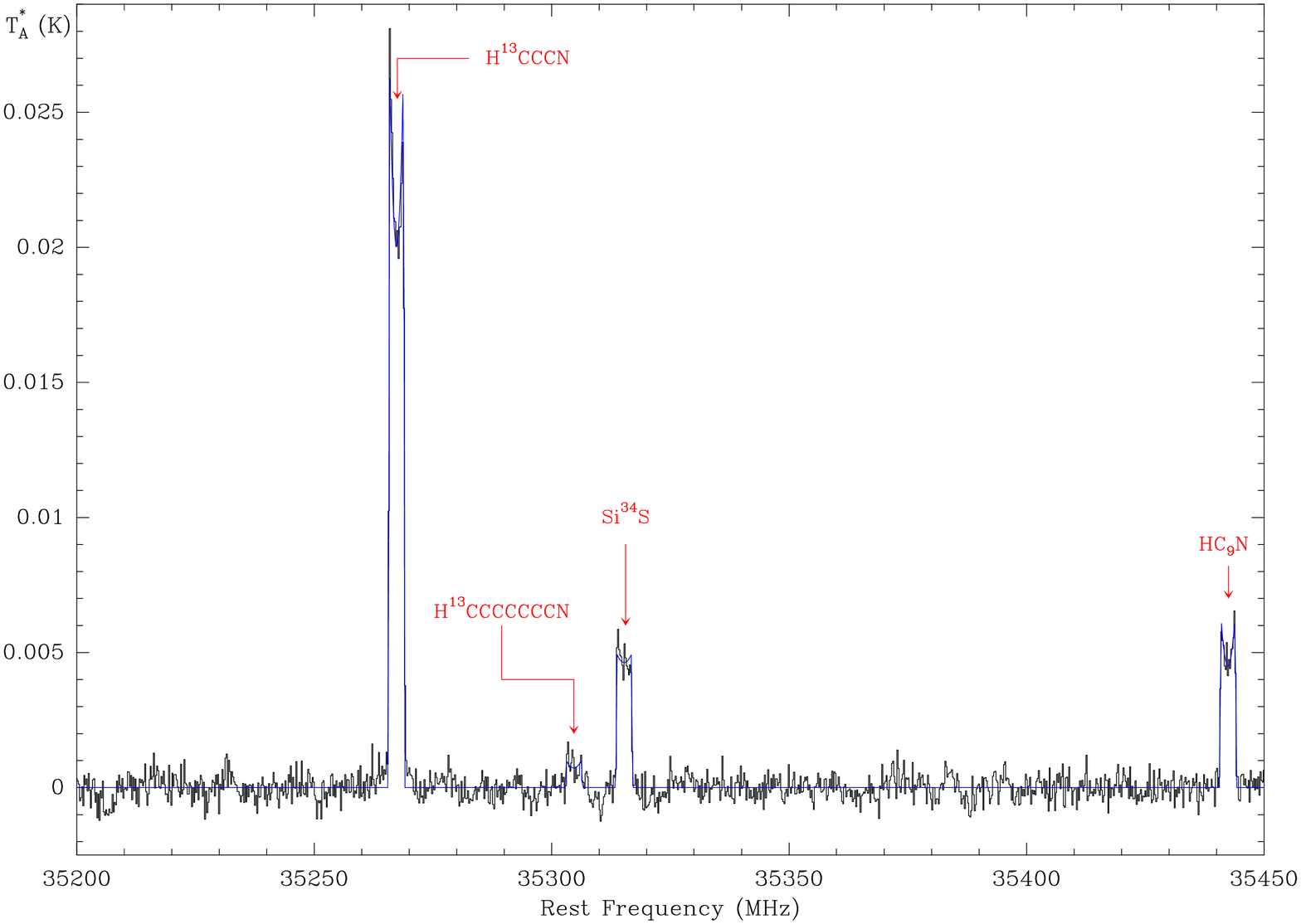}
\caption{IRC+10216 YEBES 40m data, line fits and labels from 
34950 to 35450 GHz.}
\label{fig09}
\end{figure*}

\begin{figure*}                                              
\includegraphics[width=0.93\textwidth]                       {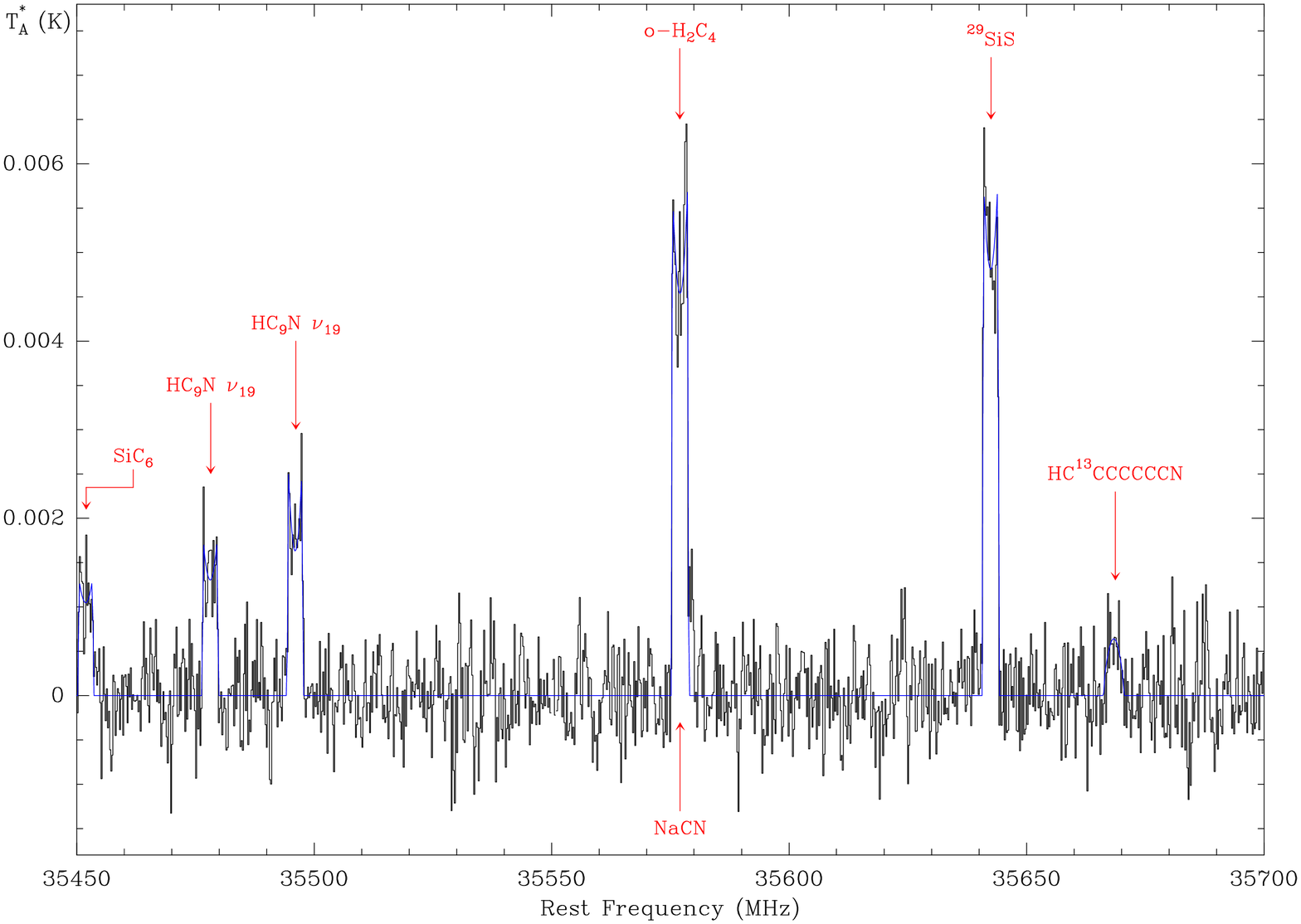}\\
\includegraphics[width=0.93\textwidth]                       {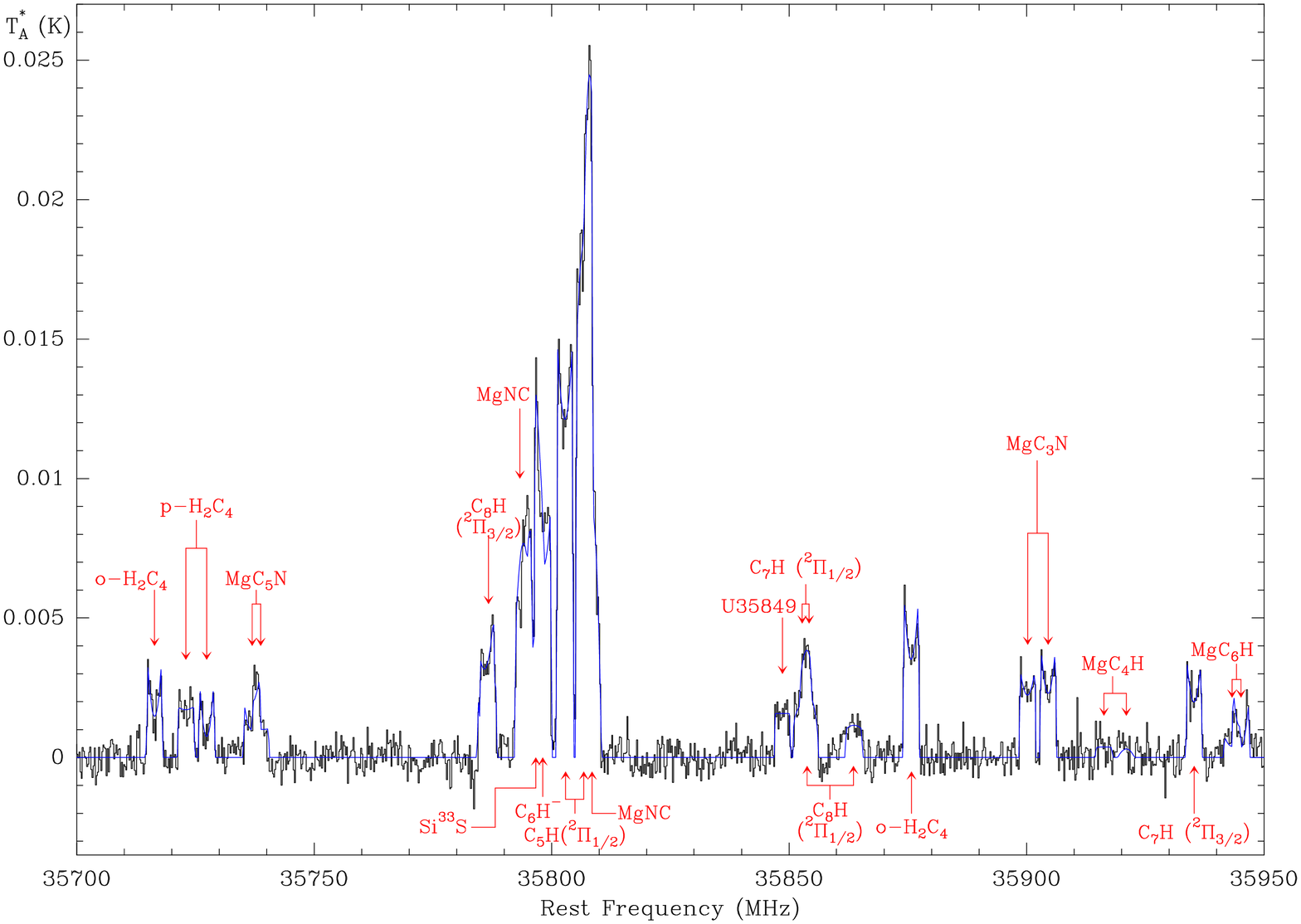}
\caption{IRC+10216 YEBES 40m data, line fits and labels from 
35450 to 35950 GHz.}
\label{fig10}
\end{figure*}

\begin{figure*}                                              
\includegraphics[width=0.93\textwidth]                       {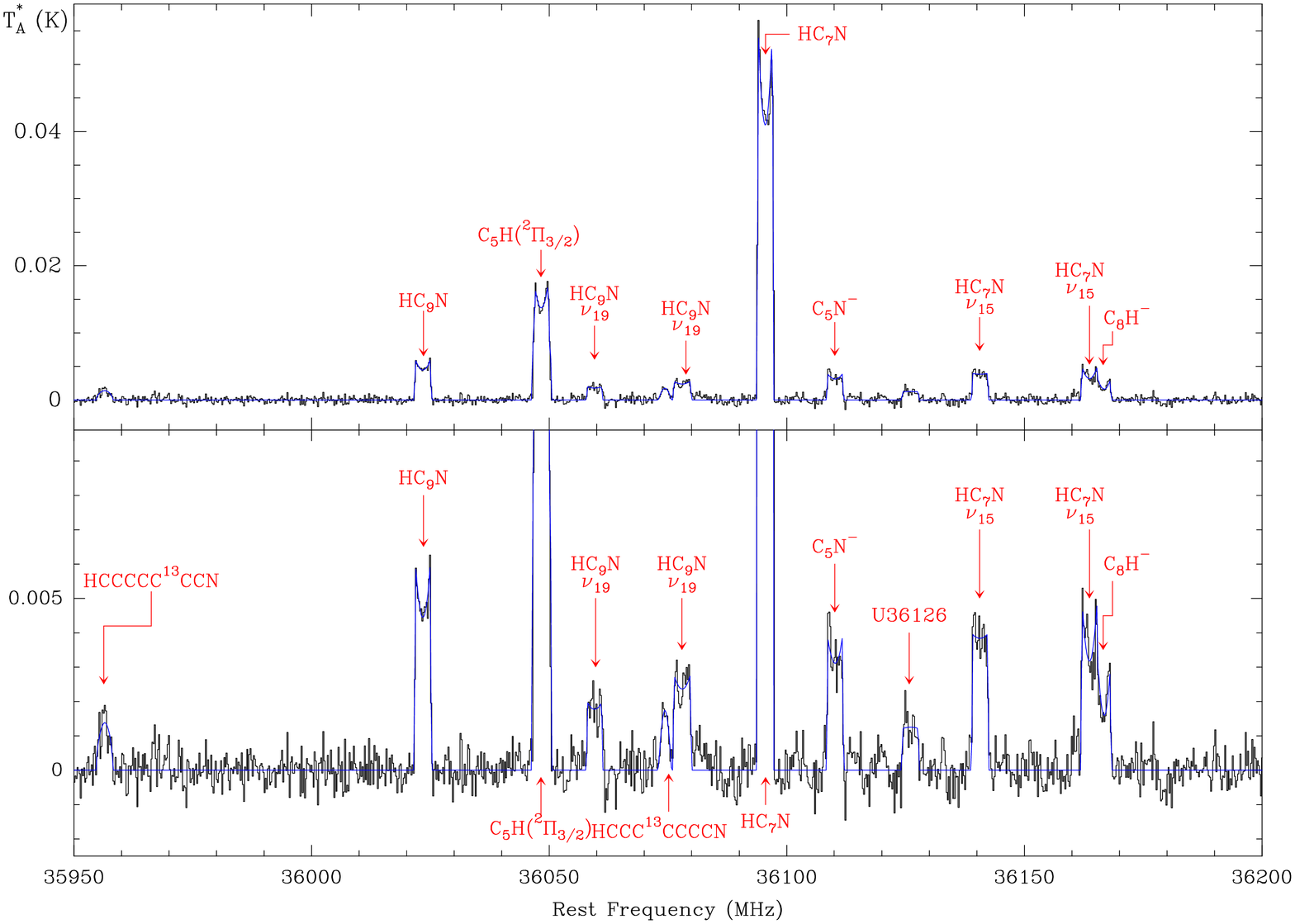}\\
\includegraphics[width=0.93\textwidth]                       {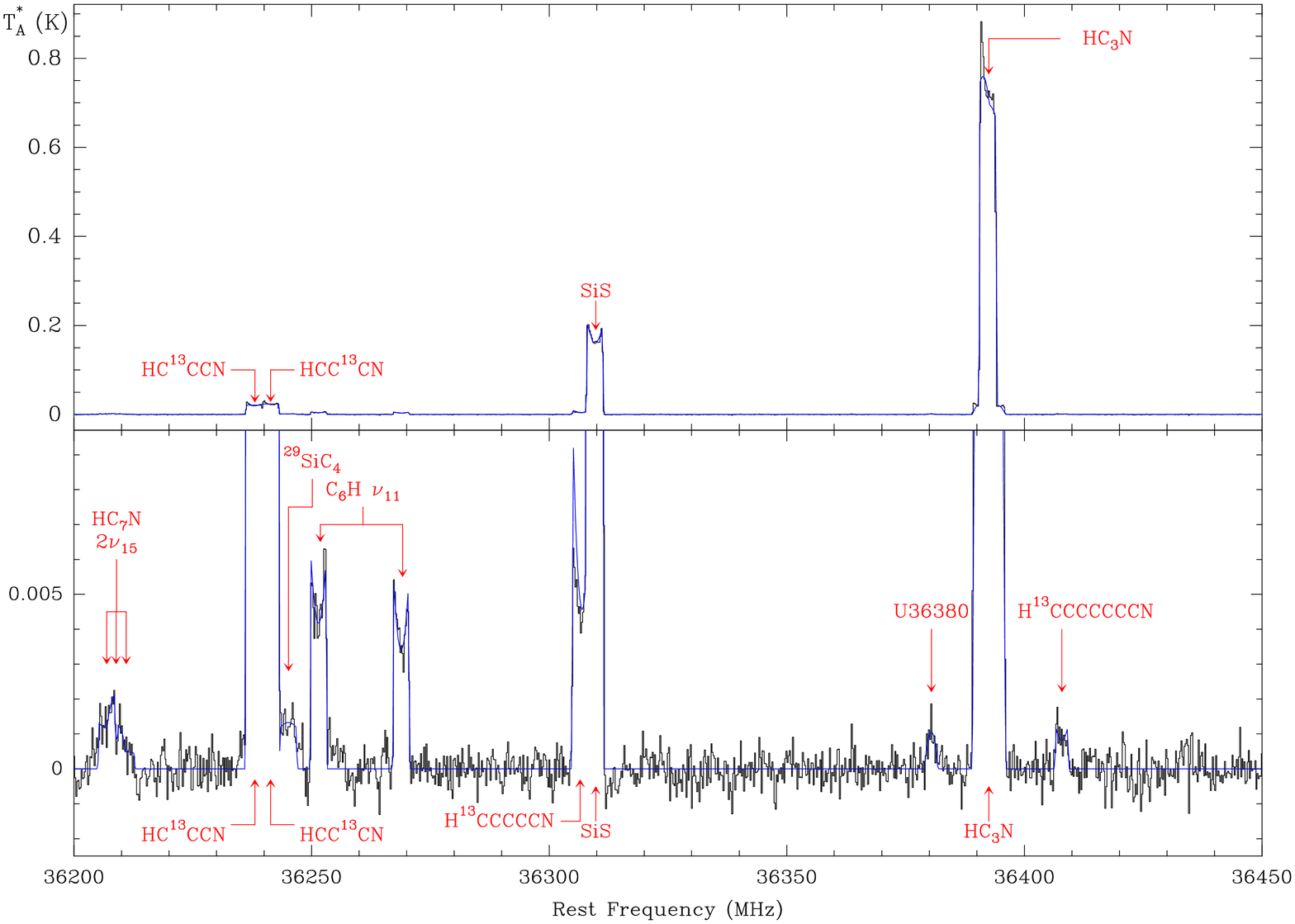}
\caption{IRC+10216 YEBES 40m data, line fits and labels from 
35950 to 36450 GHz.}
\label{fig11}
\end{figure*}

\begin{figure*}                                              
\includegraphics[width=0.93\textwidth]                       {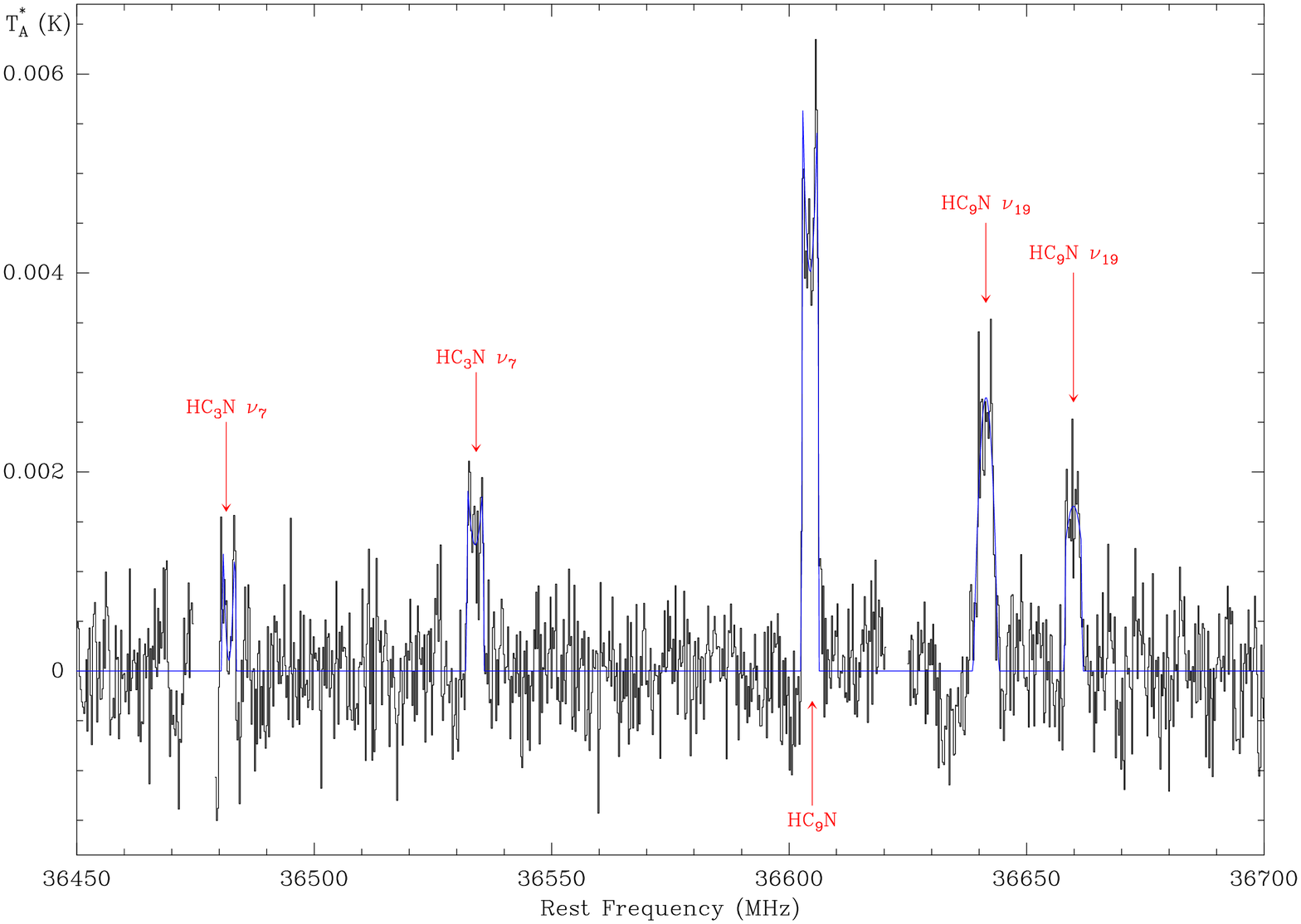}\\
\includegraphics[width=0.93\textwidth]                       {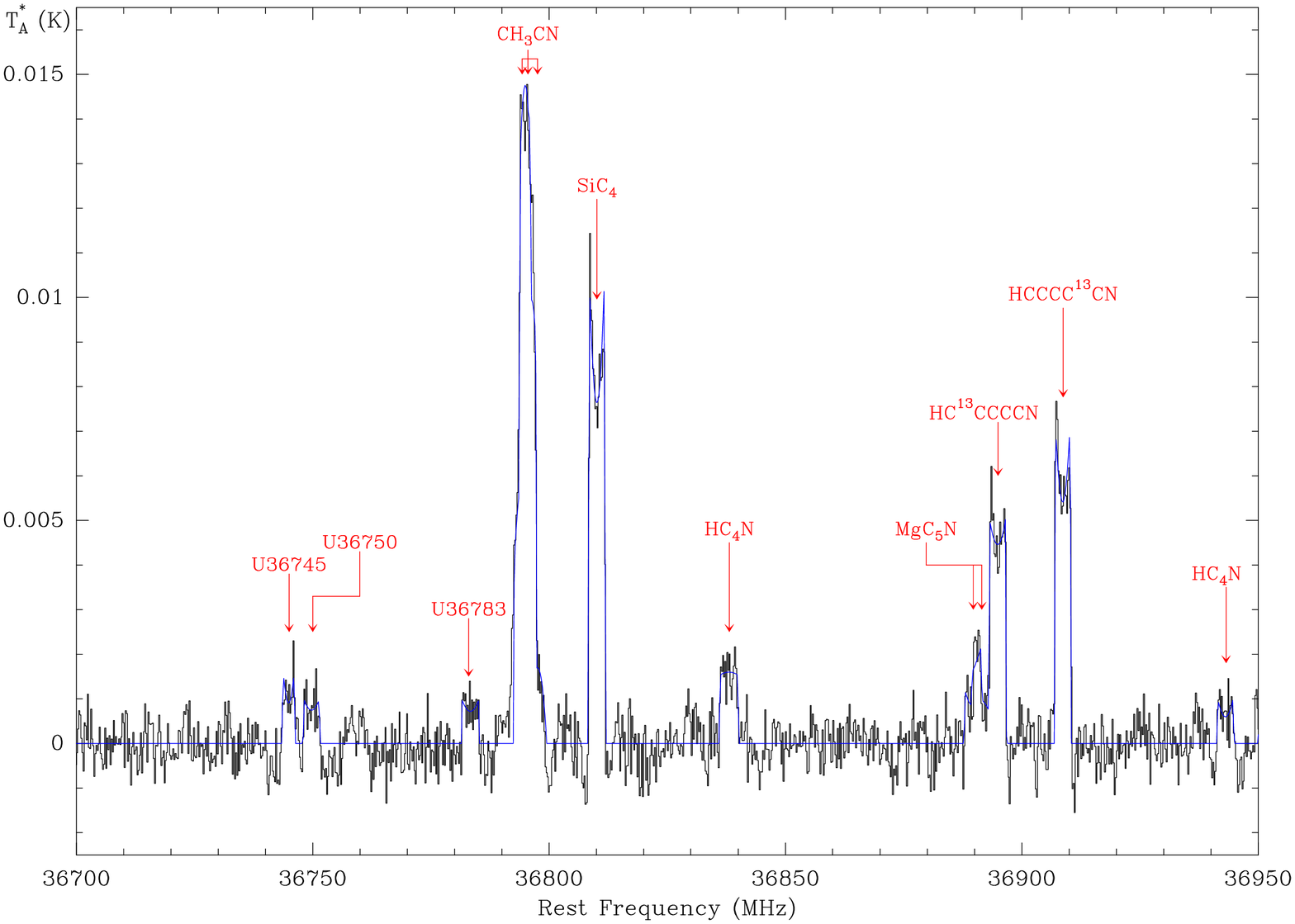}
\caption{IRC+10216 YEBES 40m data, line fits and labels from 
36450 to 36950 GHz.}
\label{fig12}
\end{figure*}

\begin{figure*}                                              
\includegraphics[width=0.93\textwidth]                       {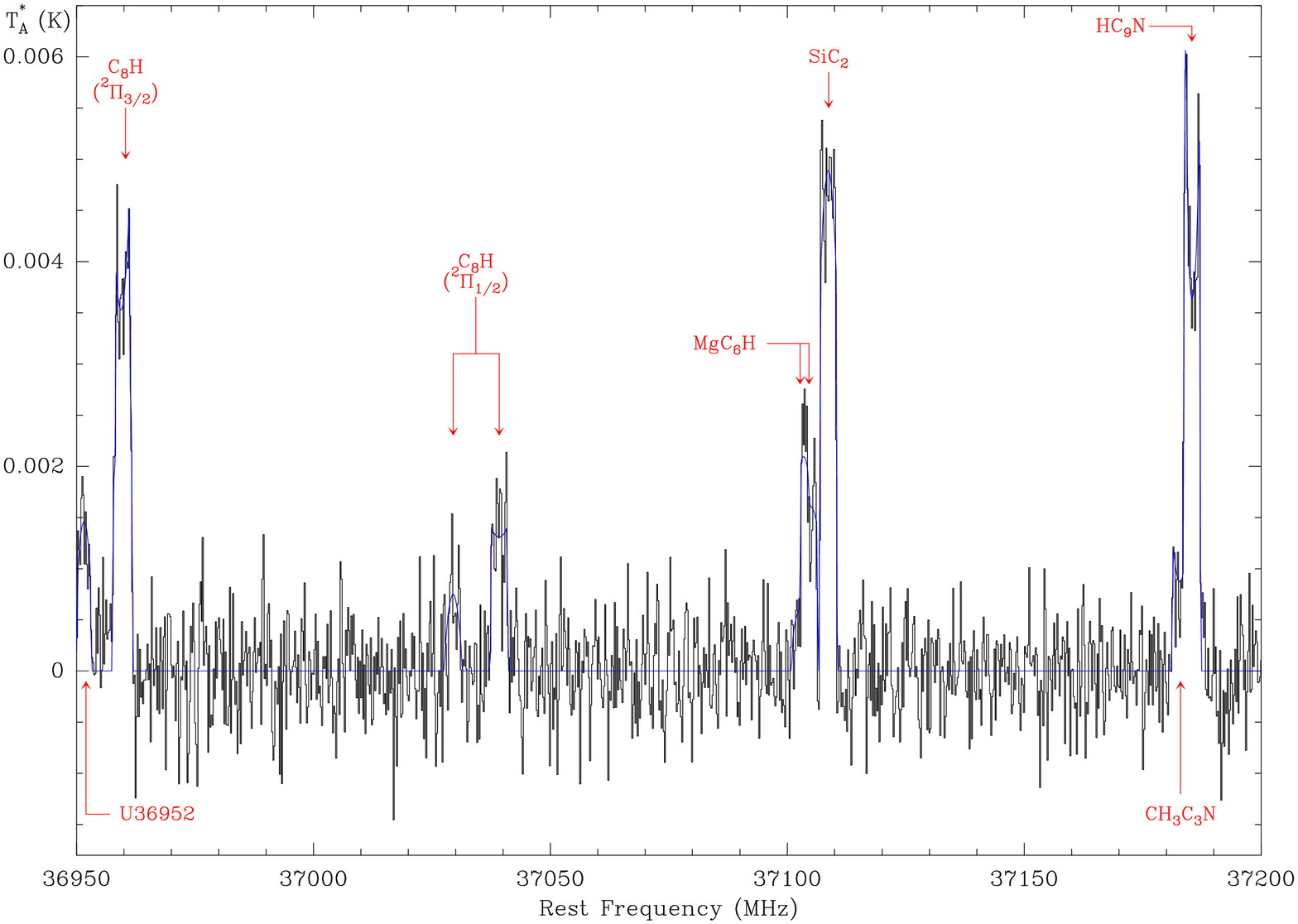}\\
\includegraphics[width=0.93\textwidth]                       {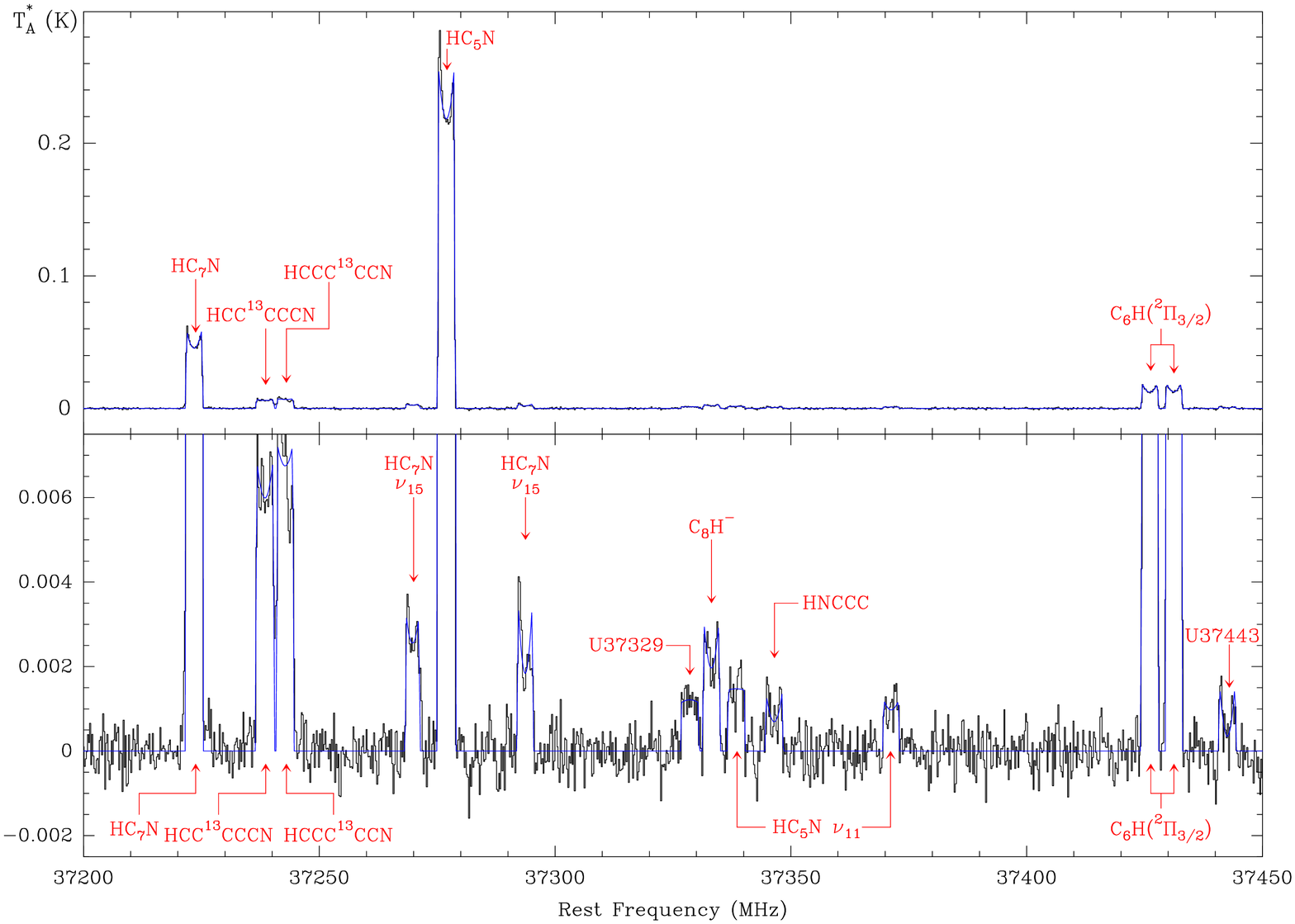}
\caption{IRC+10216 YEBES 40m data, line fits and labels from 
36950 to 37450 GHz.}
\label{fig13}
\end{figure*}

\begin{figure*}                                              
\includegraphics[width=0.93\textwidth]                       {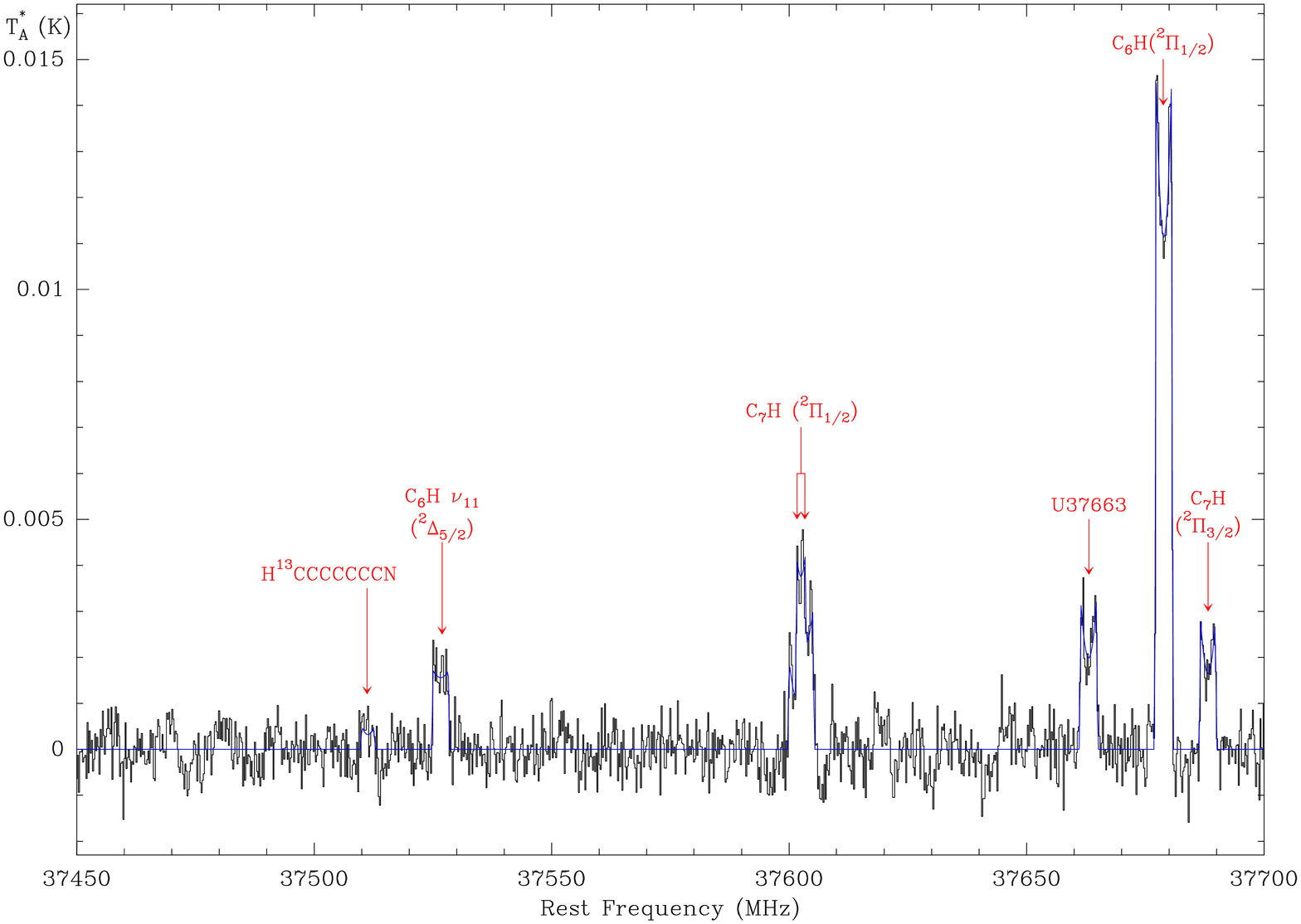}\\
\includegraphics[width=0.93\textwidth]                       {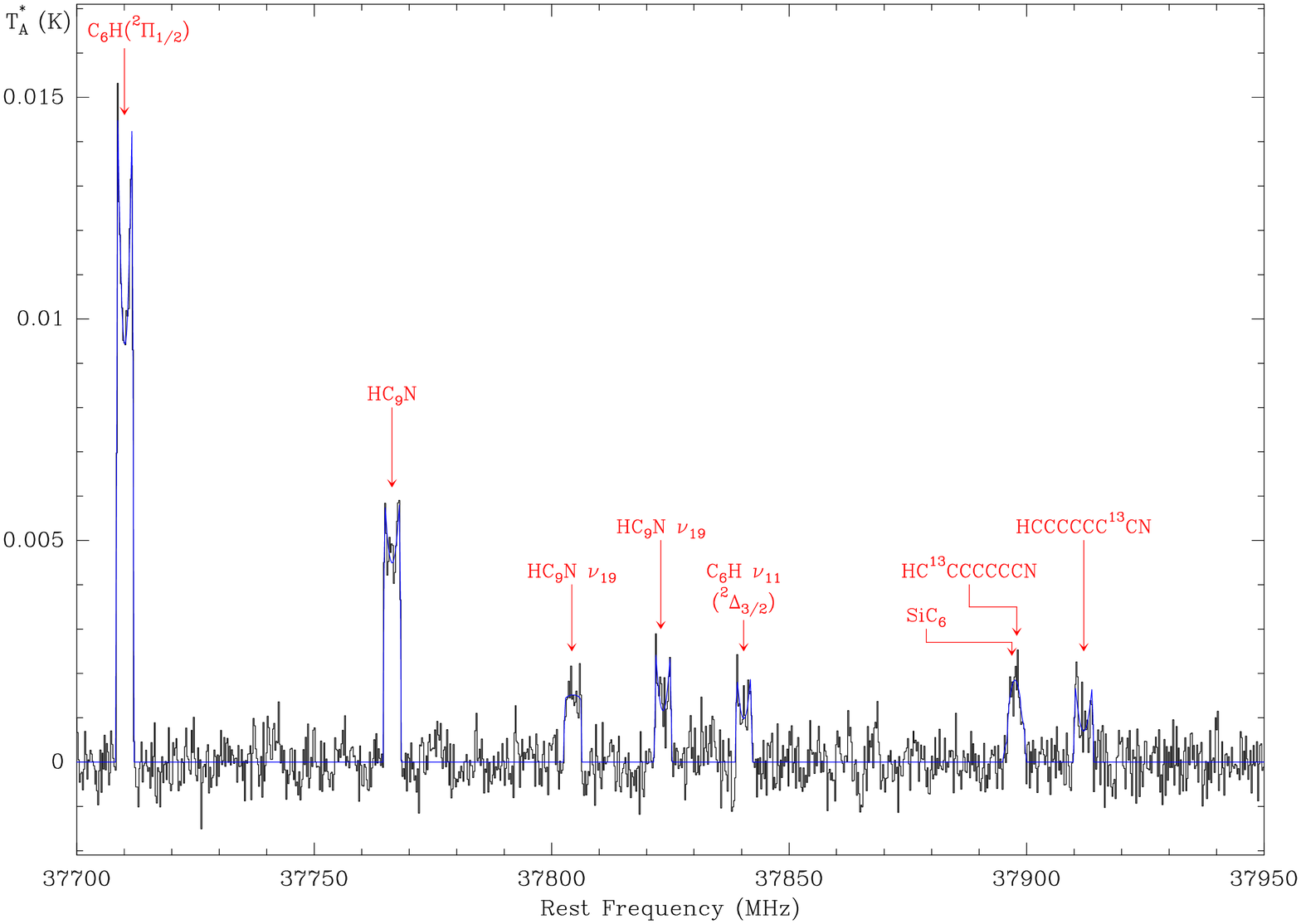}
\caption{IRC+10216 YEBES 40m data, line fits and labels from 
37450 to 37950 GHz.}
\label{fig14}
\end{figure*}

\begin{figure*}                                              
\includegraphics[width=0.93\textwidth]                       {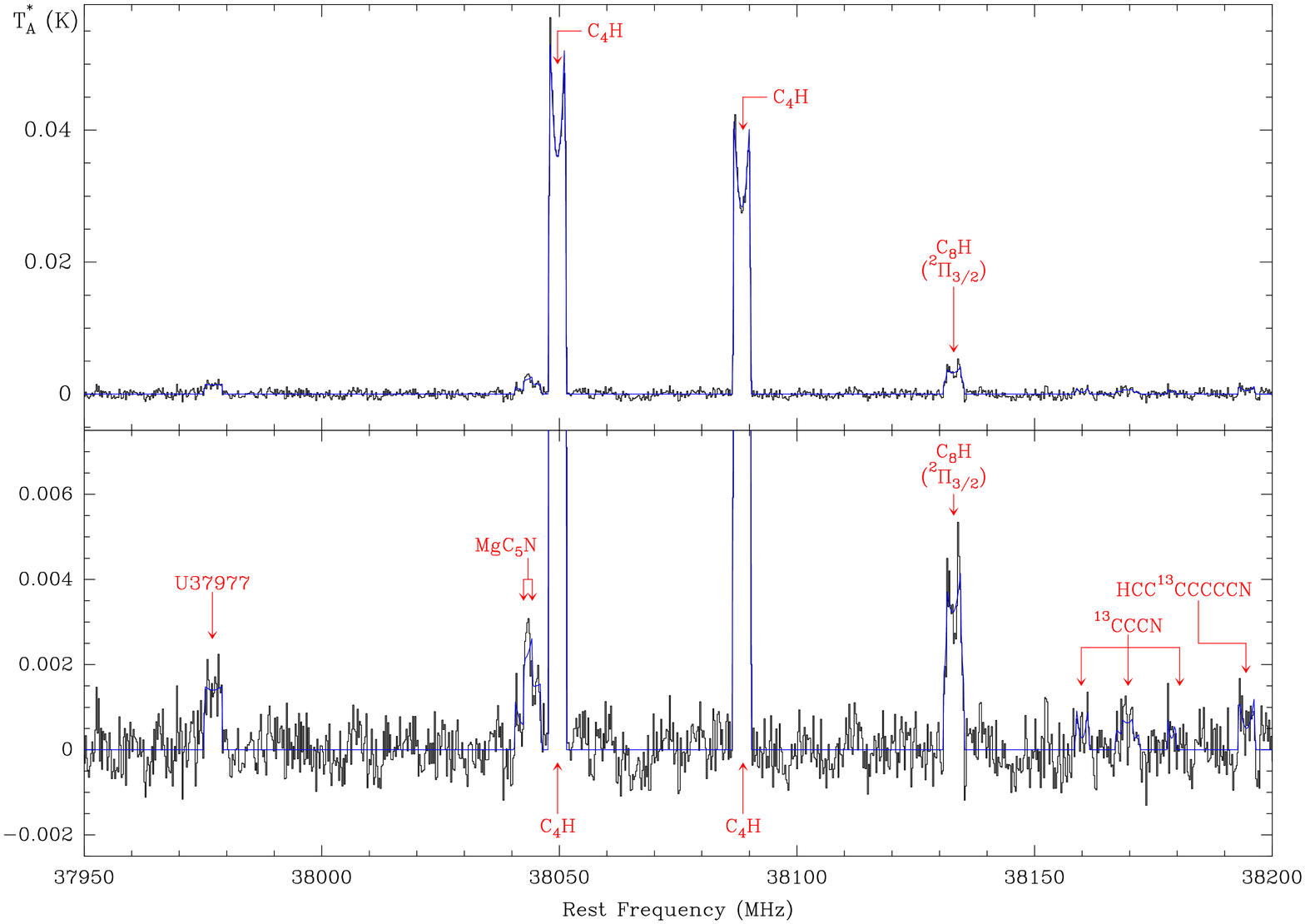}\\
\includegraphics[width=0.93\textwidth]                       {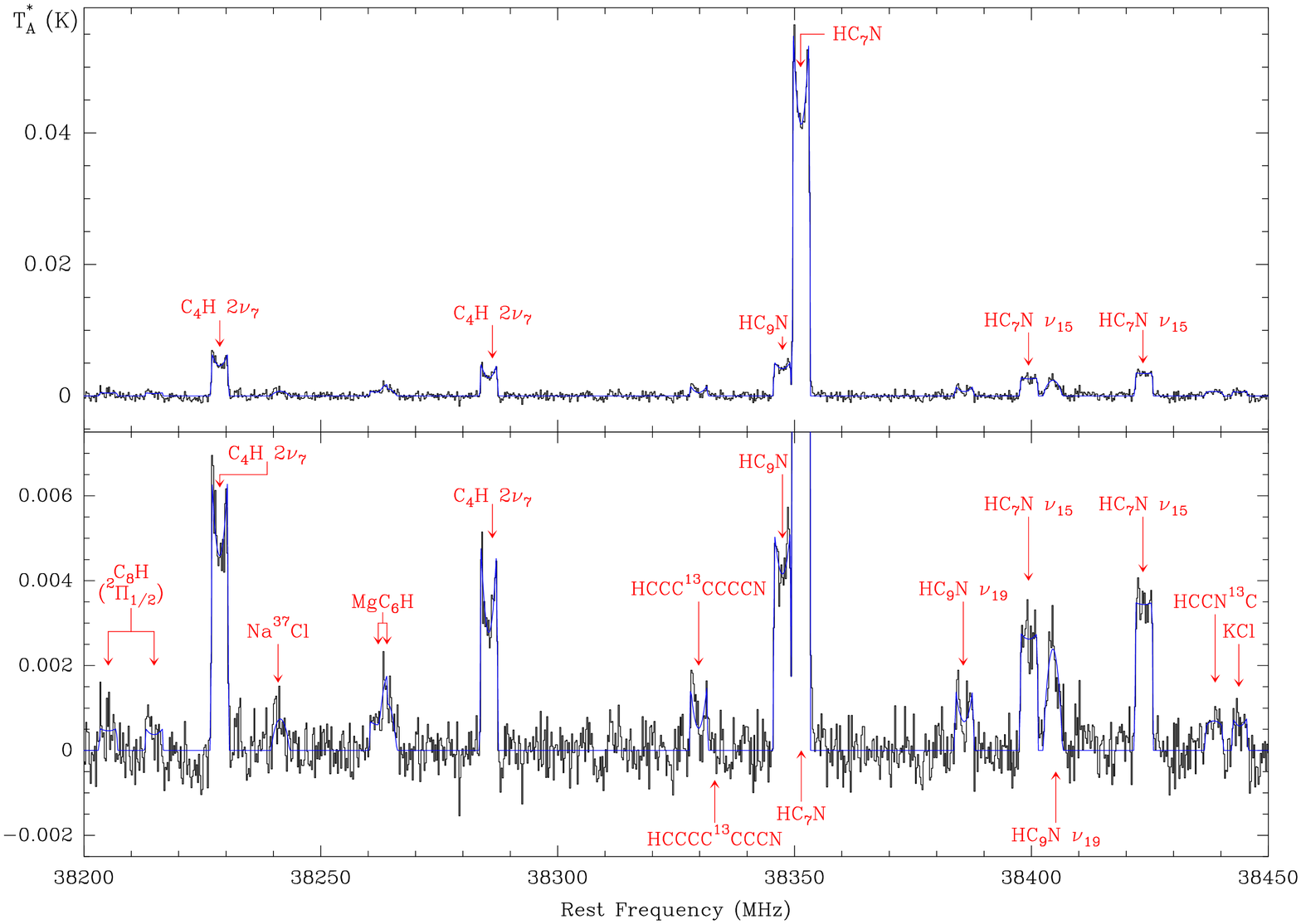}
\caption{IRC+10216 YEBES 40m data, line fits and labels from 
37950 to 38450 GHz.}
\label{fig15}
\end{figure*}

\begin{figure*}                                              
\includegraphics[width=0.93\textwidth]                       {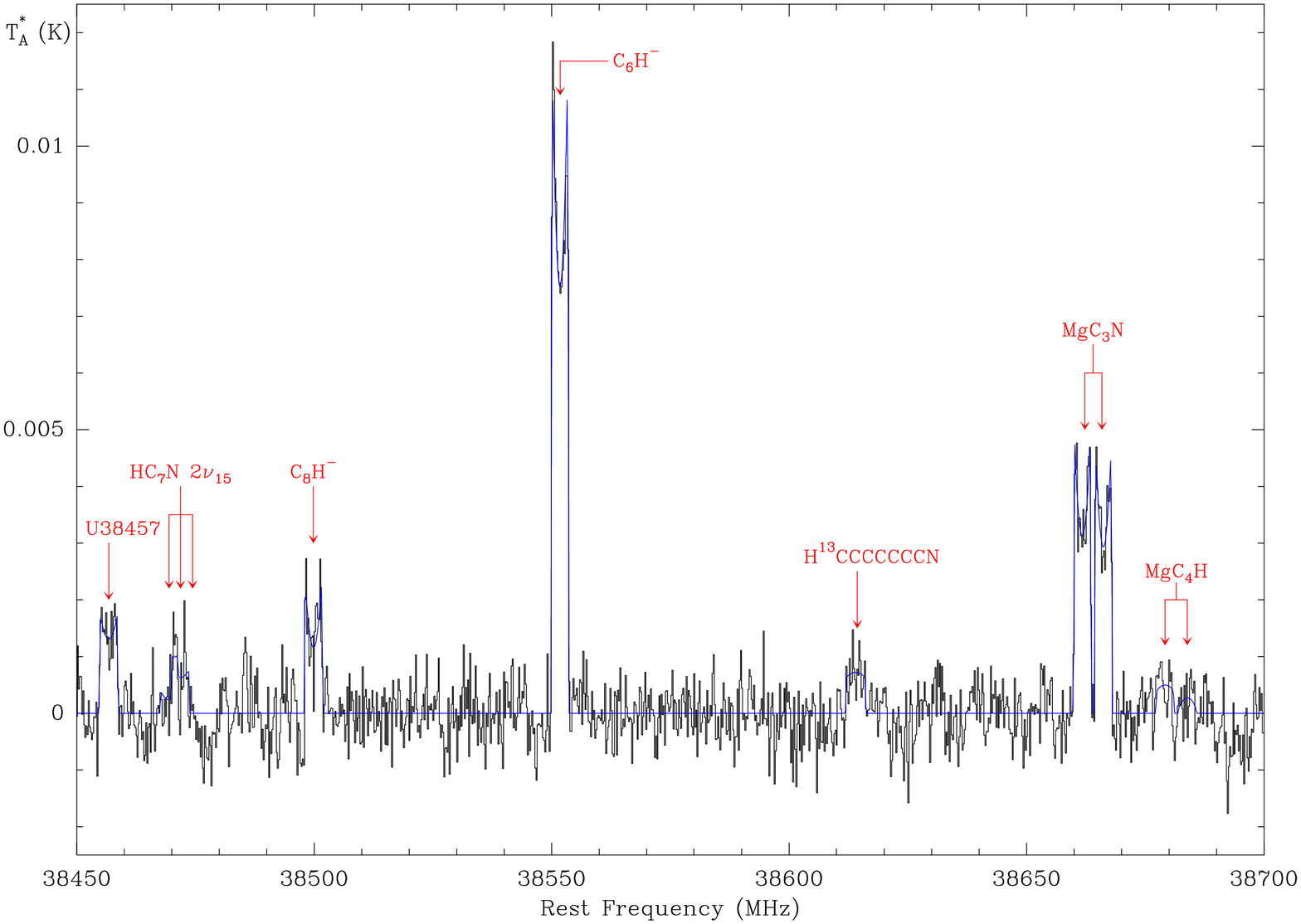}\\
\includegraphics[width=0.93\textwidth]                       {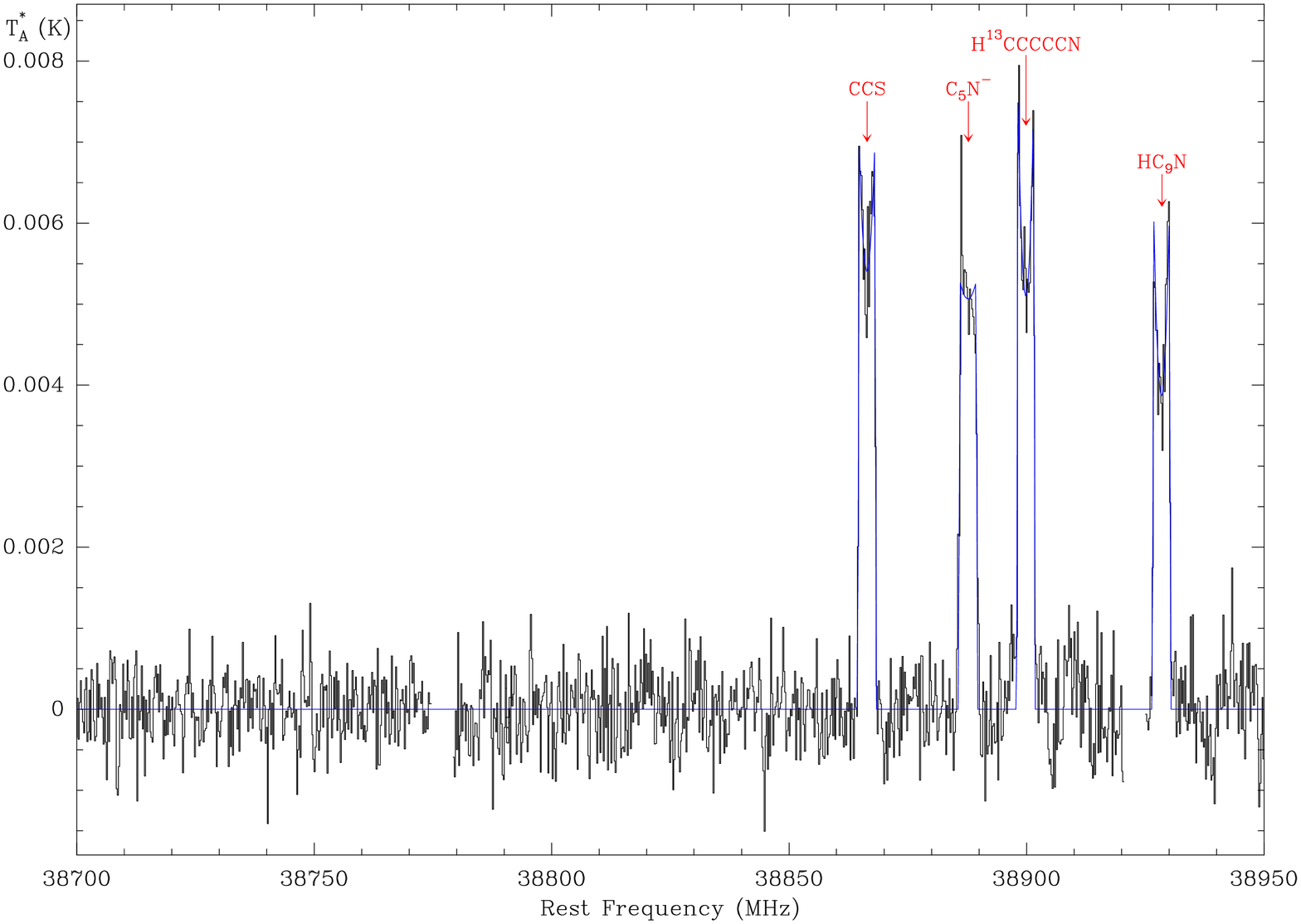}
\caption{IRC+10216 YEBES 40m data, line fits and labels from 
38450 to 38950 GHz.}
\label{fig16}
\end{figure*}

\begin{figure*}                                              
\includegraphics[width=0.93\textwidth]                       {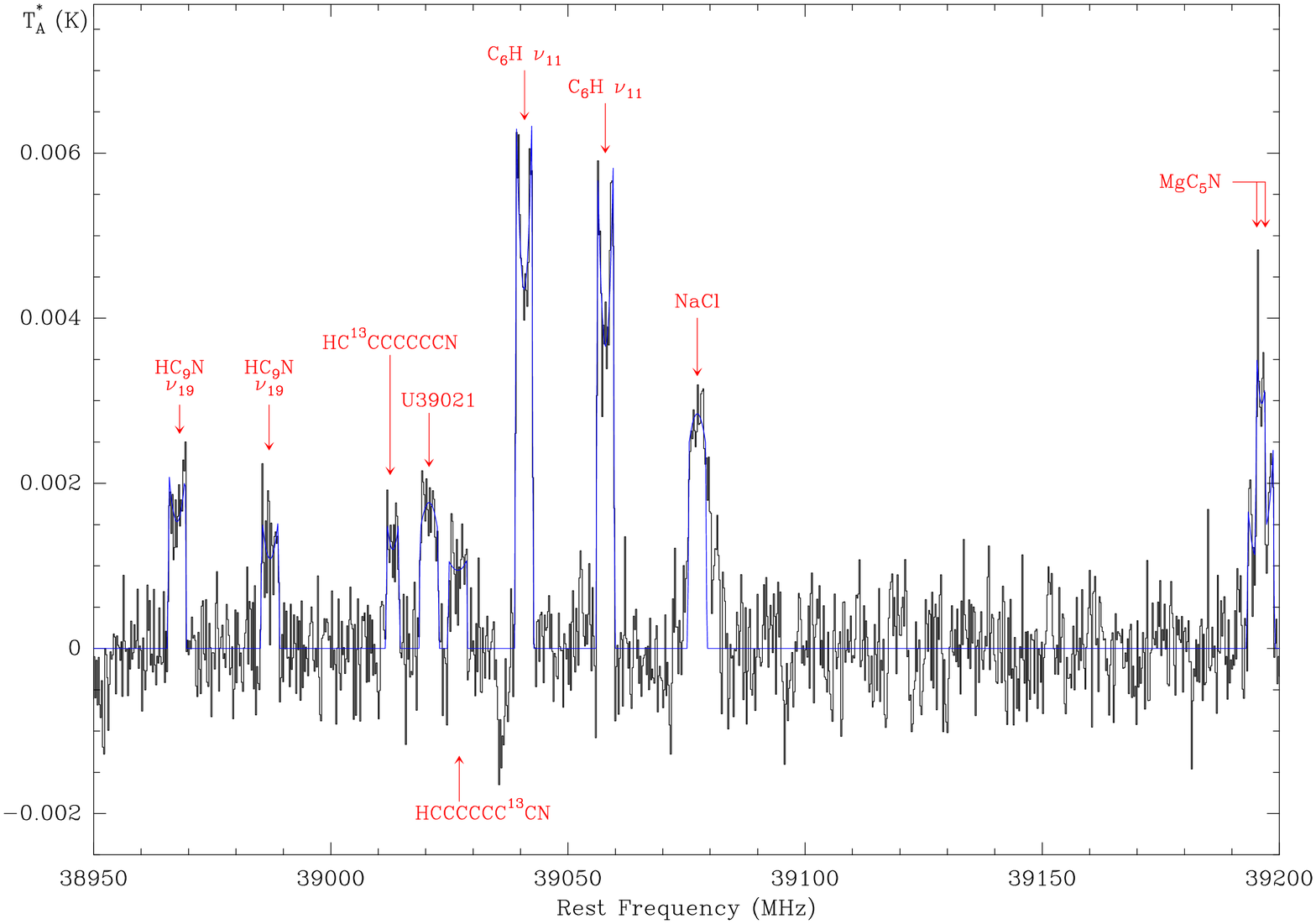}\\
\includegraphics[width=0.93\textwidth]                       {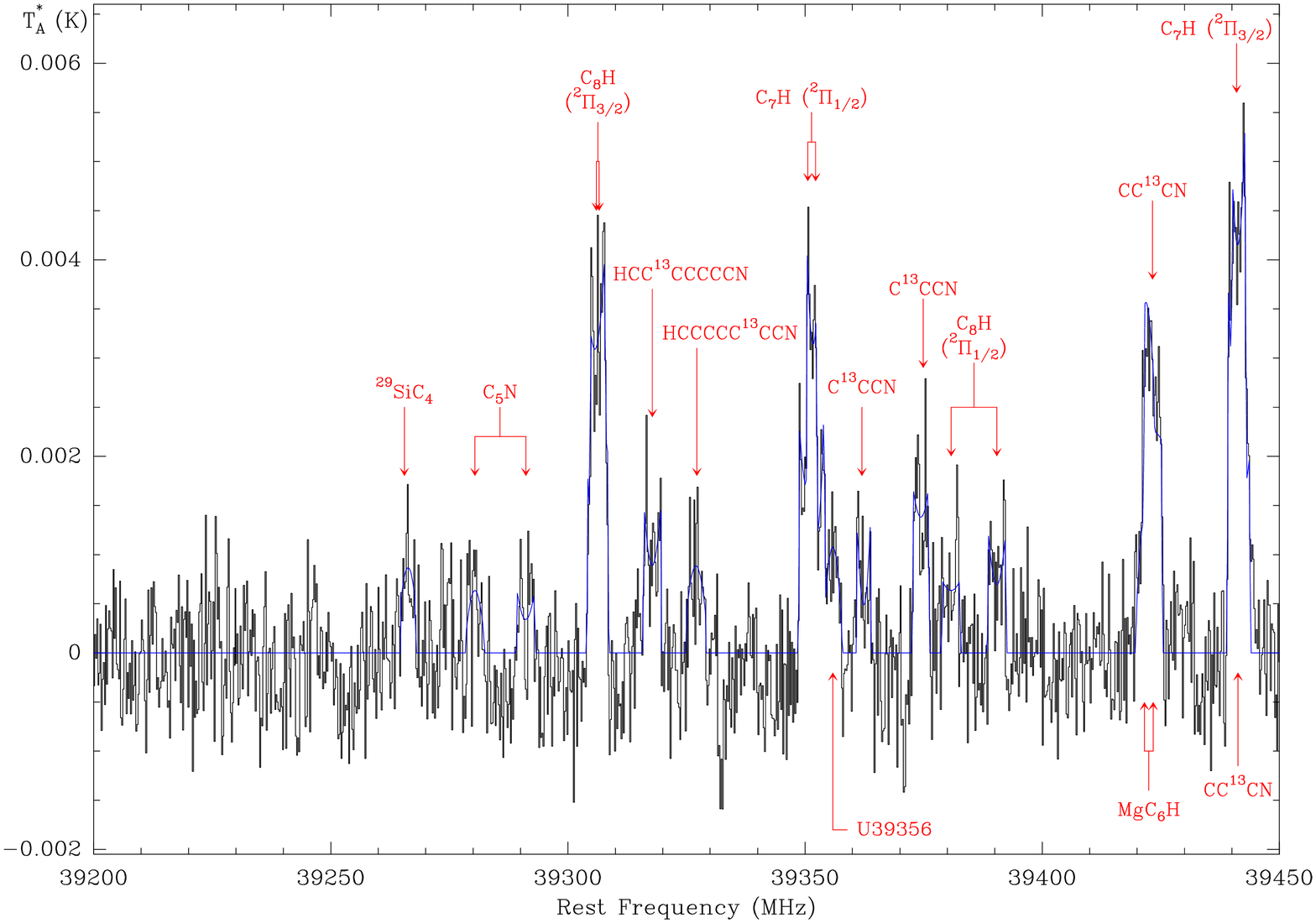}
\caption{IRC+10216 YEBES 40m data, line fits and labels from 
38950 to 39450 GHz.}
\label{fig17}
\end{figure*}

\begin{figure*}                                              
\includegraphics[width=0.93\textwidth]                       {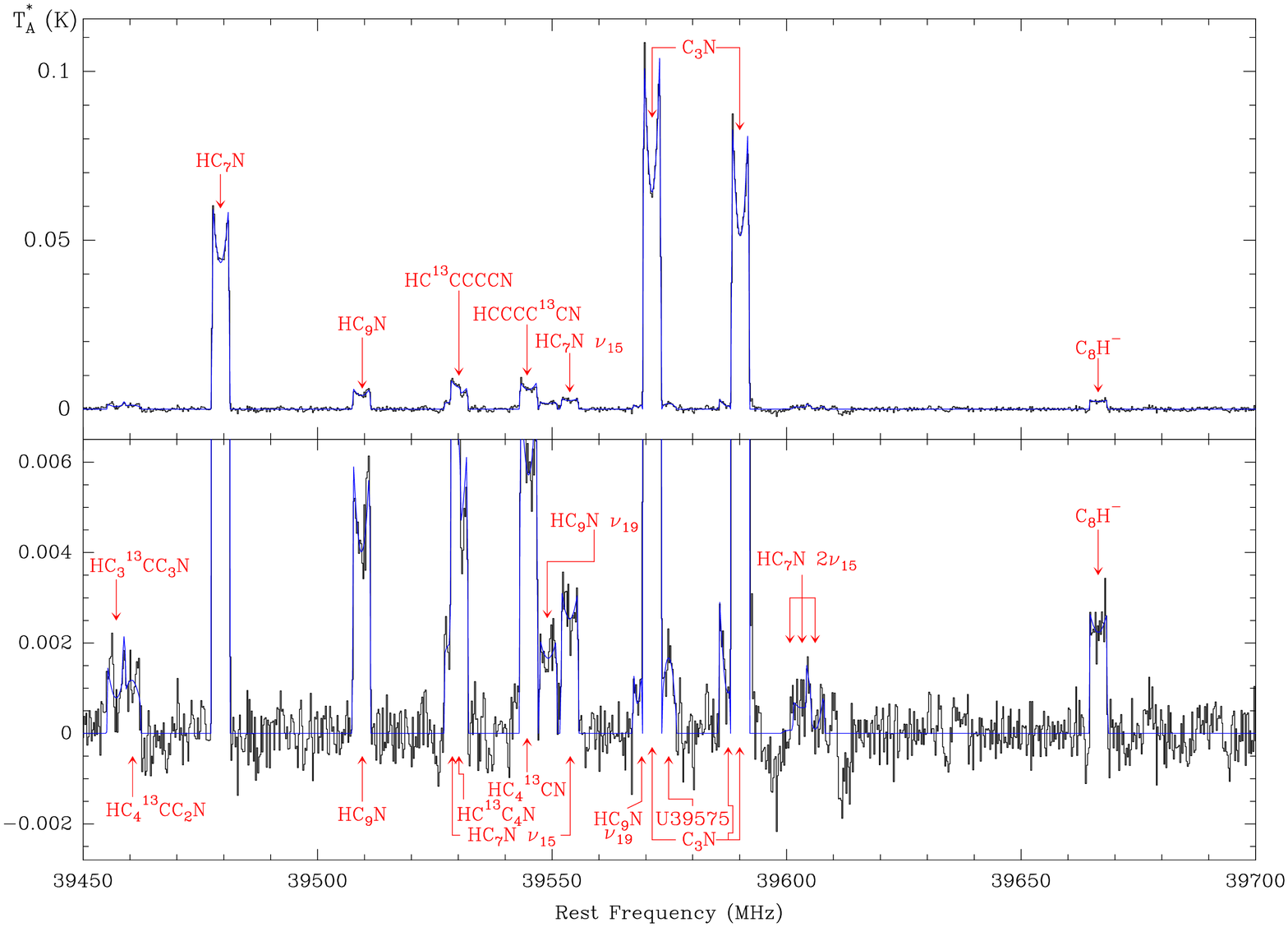}\\
\includegraphics[width=0.93\textwidth]                       {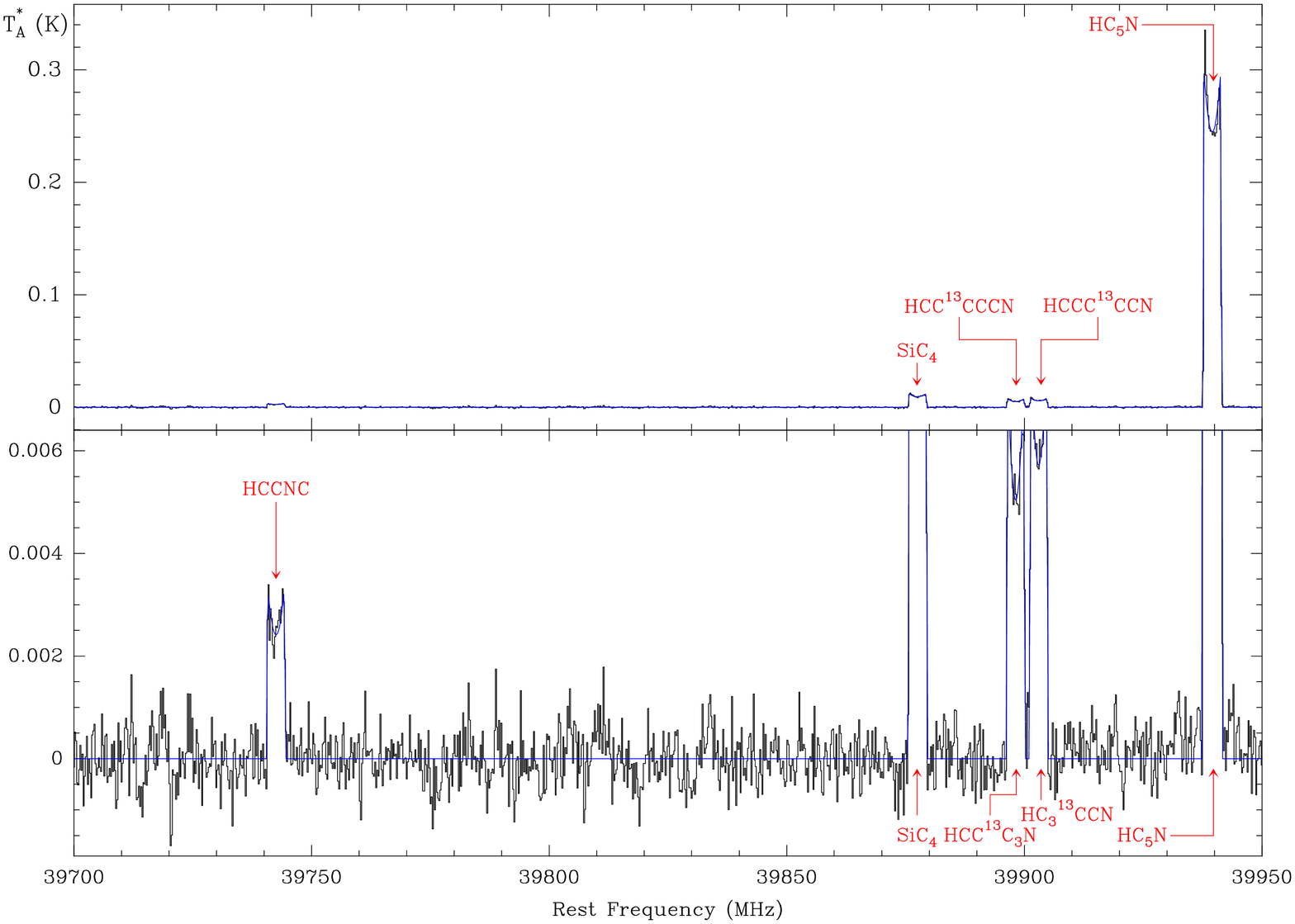}
\caption{IRC+10216 YEBES 40m data, line fits and labels from 
39450 to 39950 GHz.}
\label{fig18}
\end{figure*}

\begin{figure*}                                              
\includegraphics[width=0.93\textwidth]                       {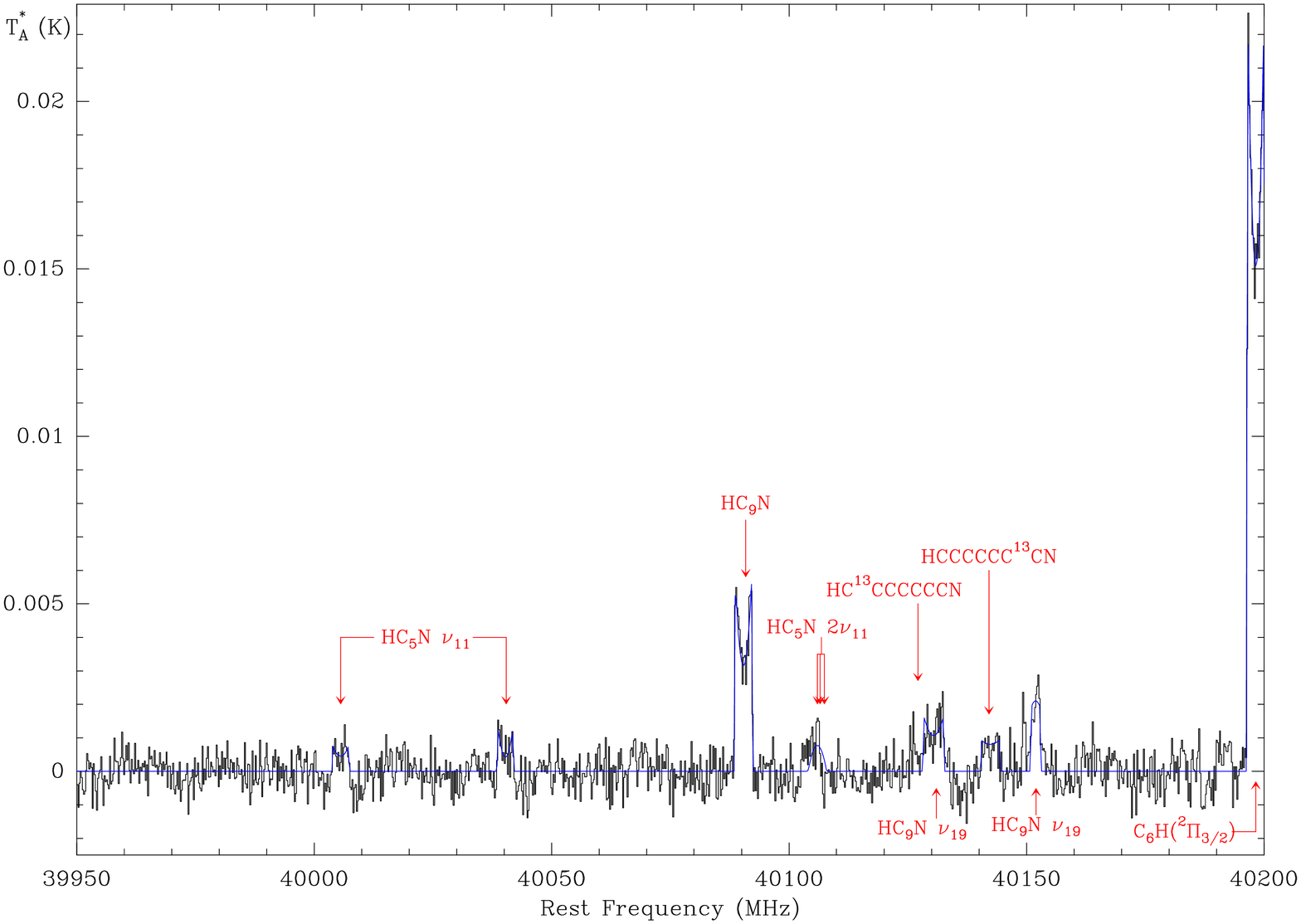}\\
\includegraphics[width=0.93\textwidth]                       {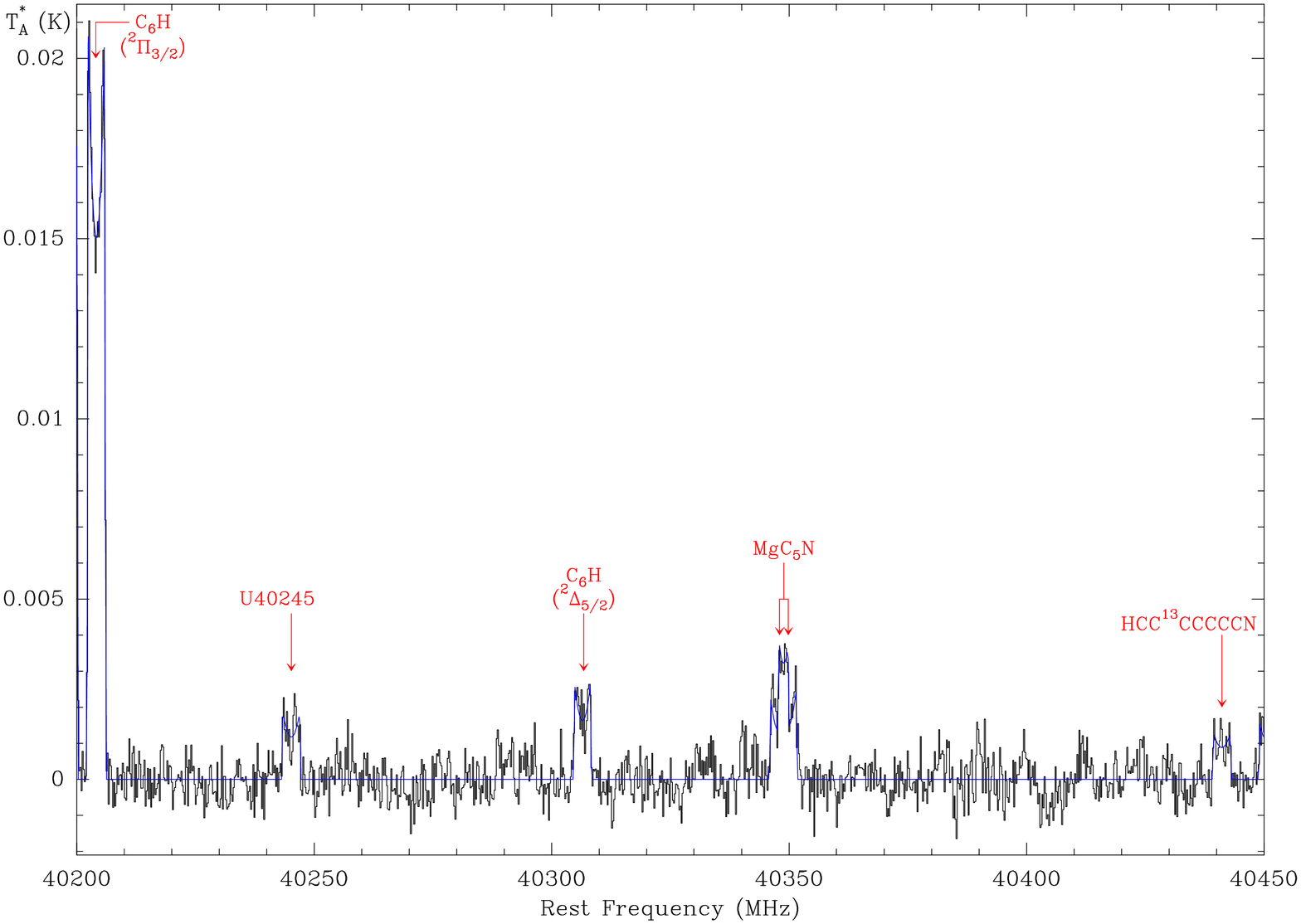}
\caption{IRC+10216 YEBES 40m data, line fits and labels from 
39950 to 40450 GHz.}
\label{fig19}
\end{figure*}

\begin{figure*}                                              
\includegraphics[width=0.93\textwidth]                       {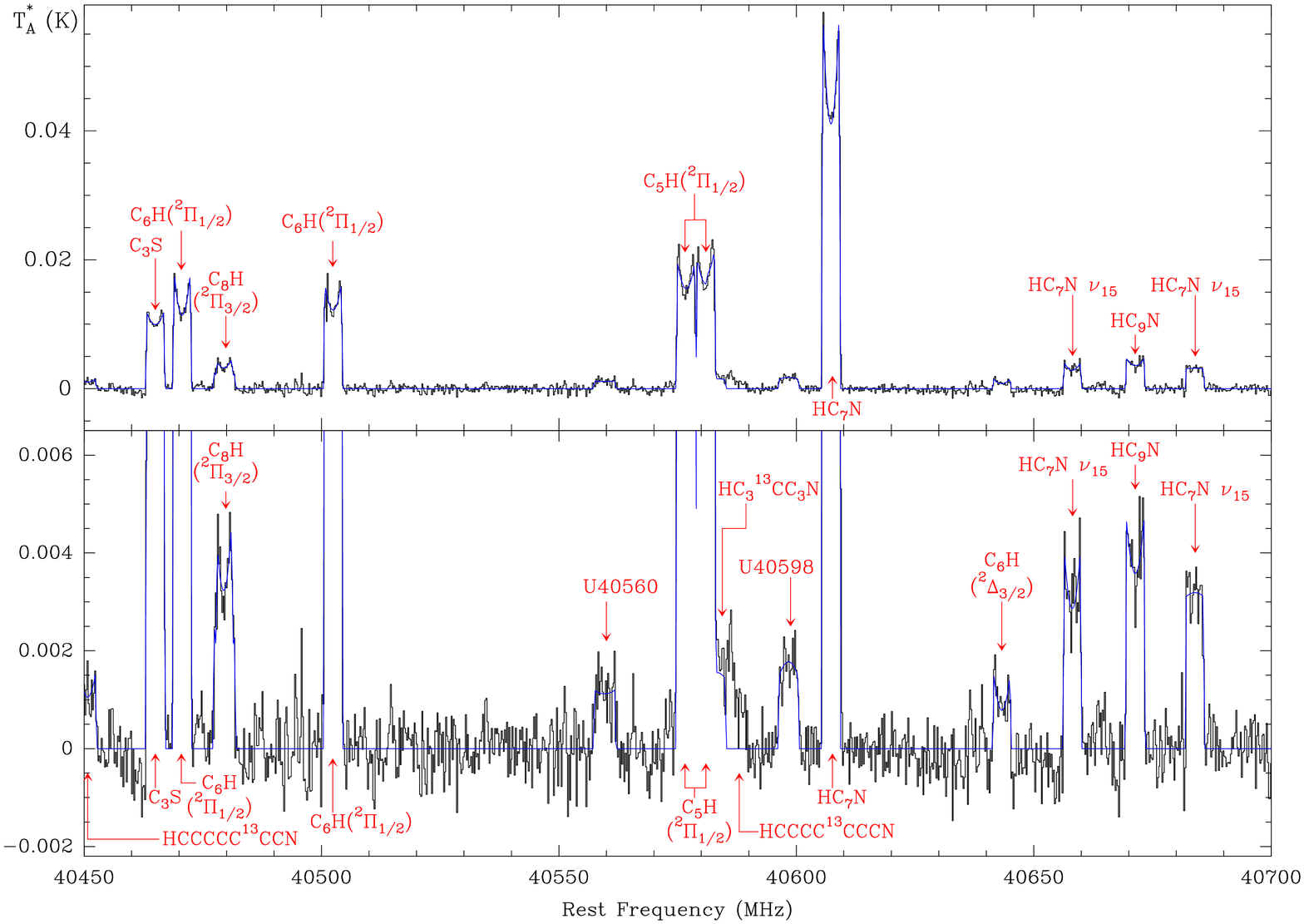}\\
\includegraphics[width=0.93\textwidth]                       {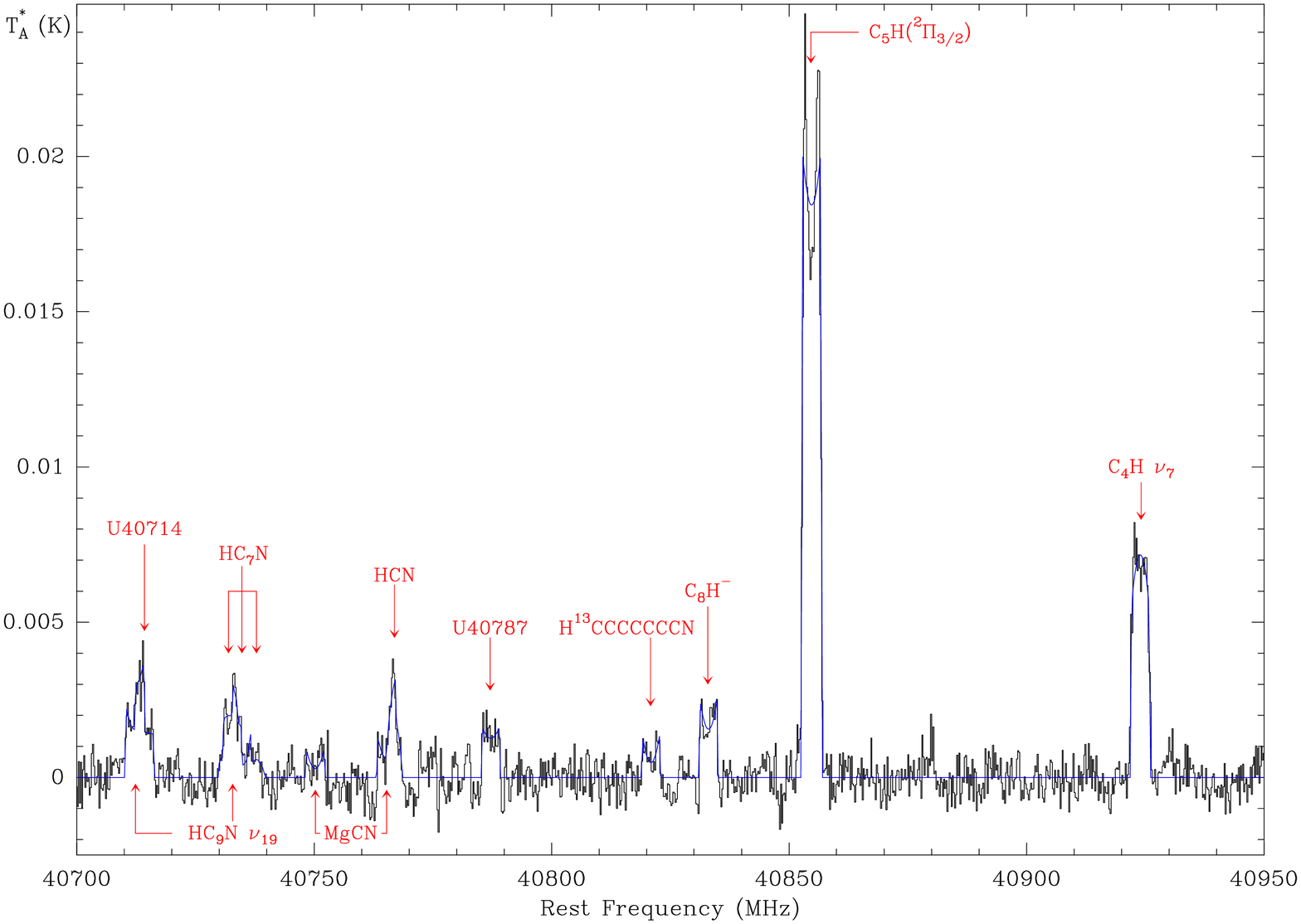}
\caption{IRC+10216 YEBES 40m data, line fits and labels from 
40450 to 40950 GHz.}
\label{fig20}
\end{figure*}

\begin{figure*}                                              
\includegraphics[width=0.93\textwidth]                       {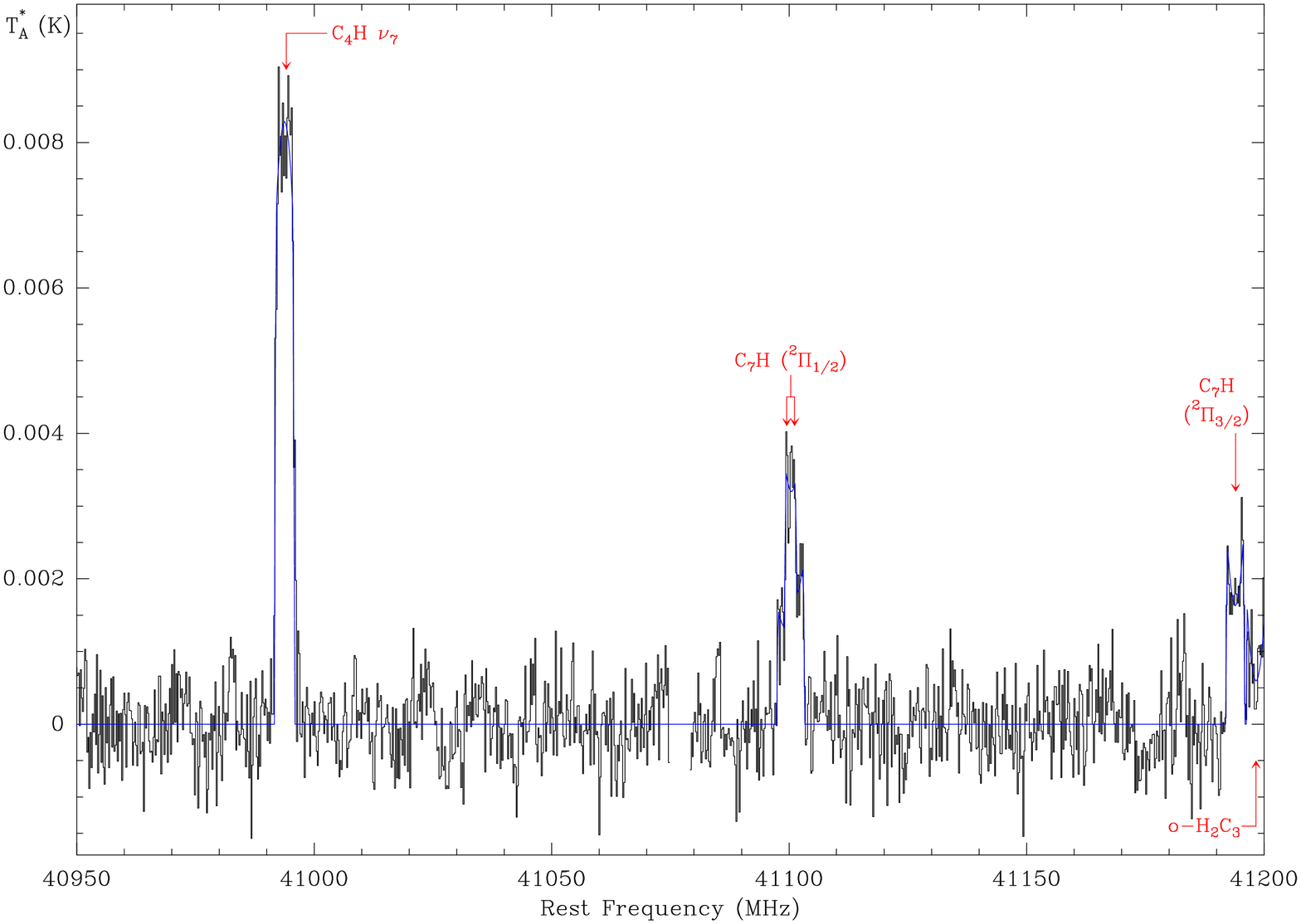}\\
\includegraphics[width=0.93\textwidth]                       {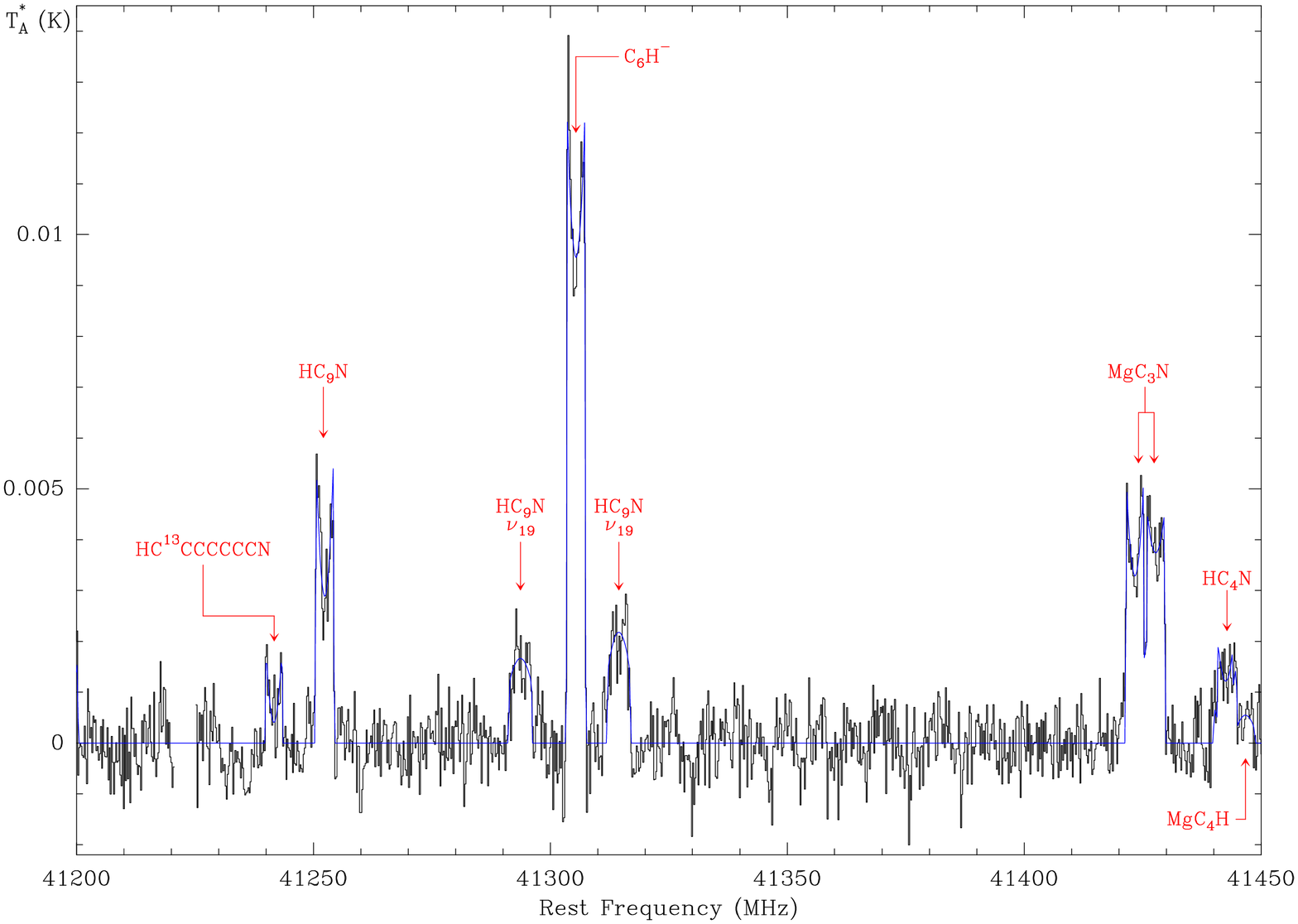}
\caption{IRC+10216 YEBES 40m data, line fits and labels from 
40950 to 41450 GHz.}
\label{fig21}
\end{figure*}

\begin{figure*}                                              
\includegraphics[width=0.93\textwidth]                       {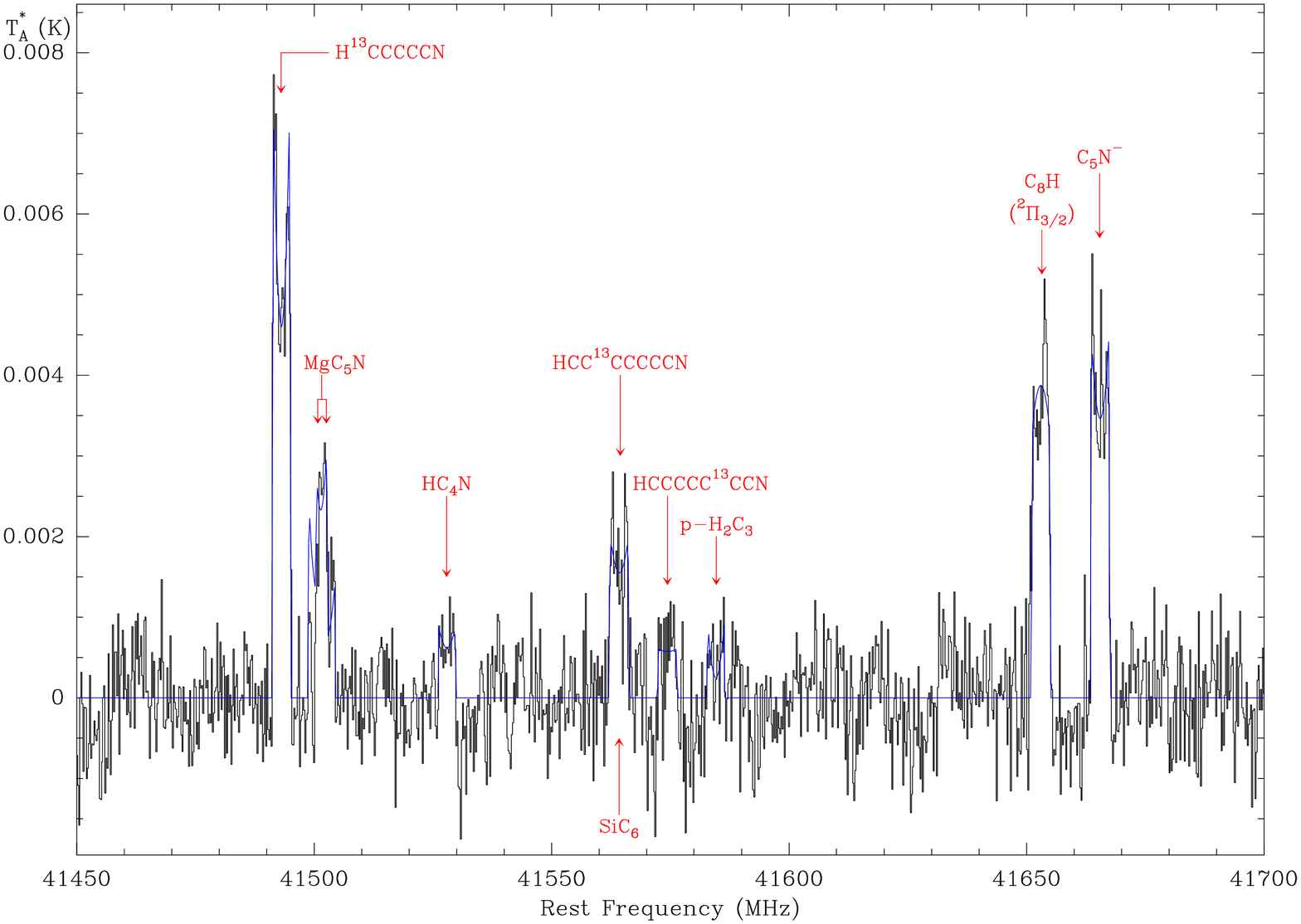}\\
\includegraphics[width=0.93\textwidth]                       {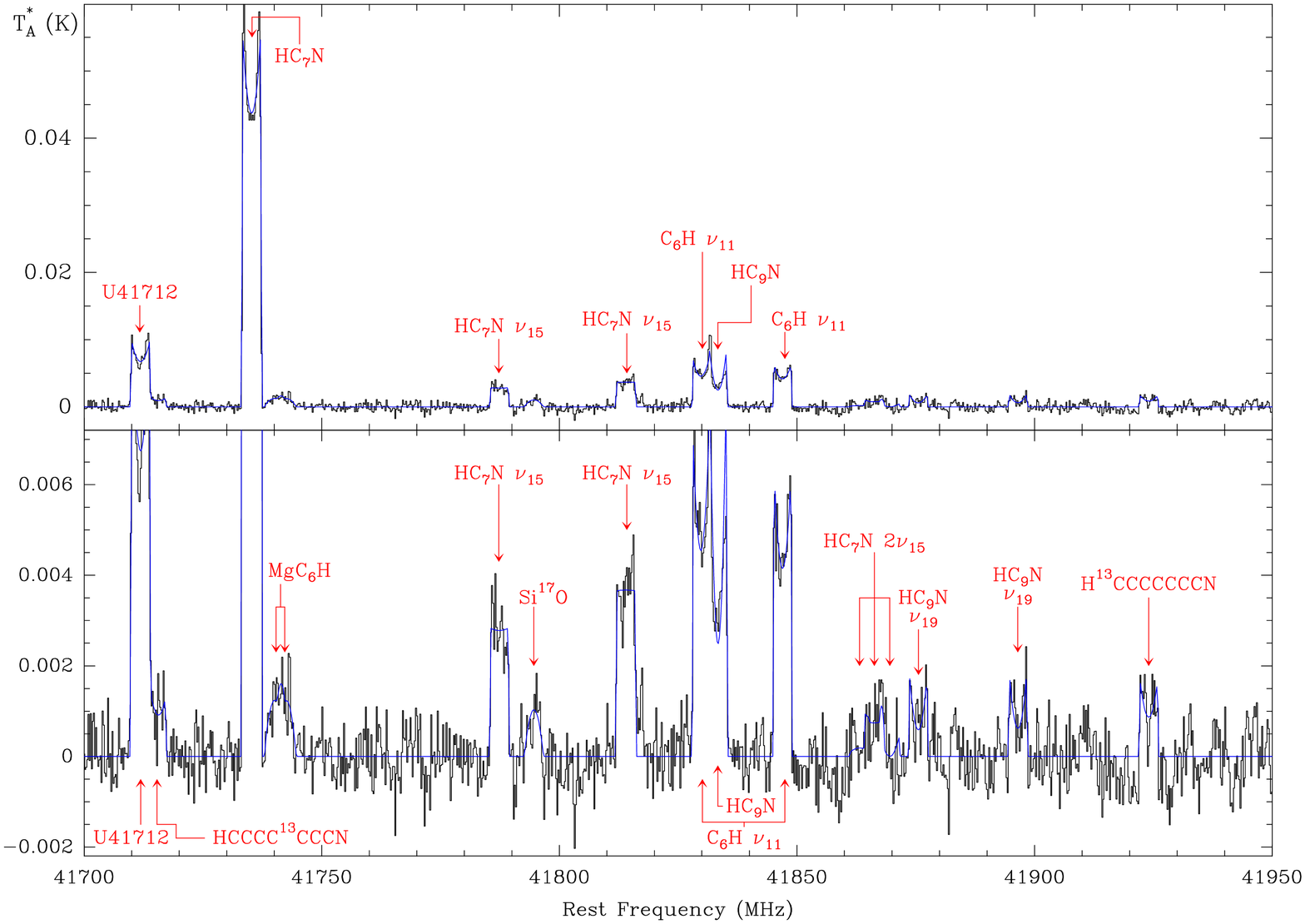}
\caption{IRC+10216 YEBES 40m data, line fits and labels from 
41450 to 41950 GHz.}
\label{fig22}
\end{figure*}

\begin{figure*}                                              
\includegraphics[width=0.93\textwidth]                       {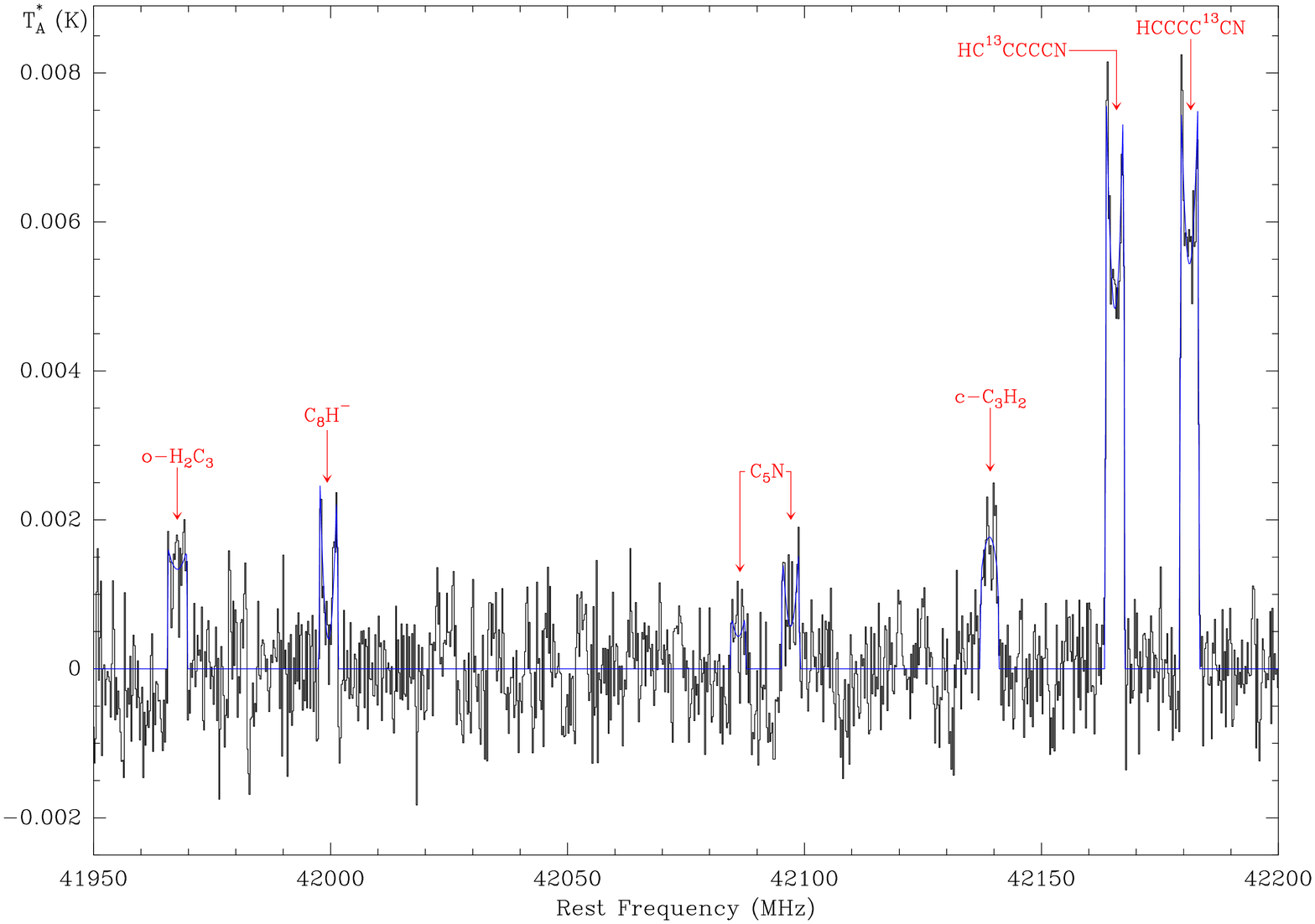}\\
\includegraphics[width=0.93\textwidth]                       {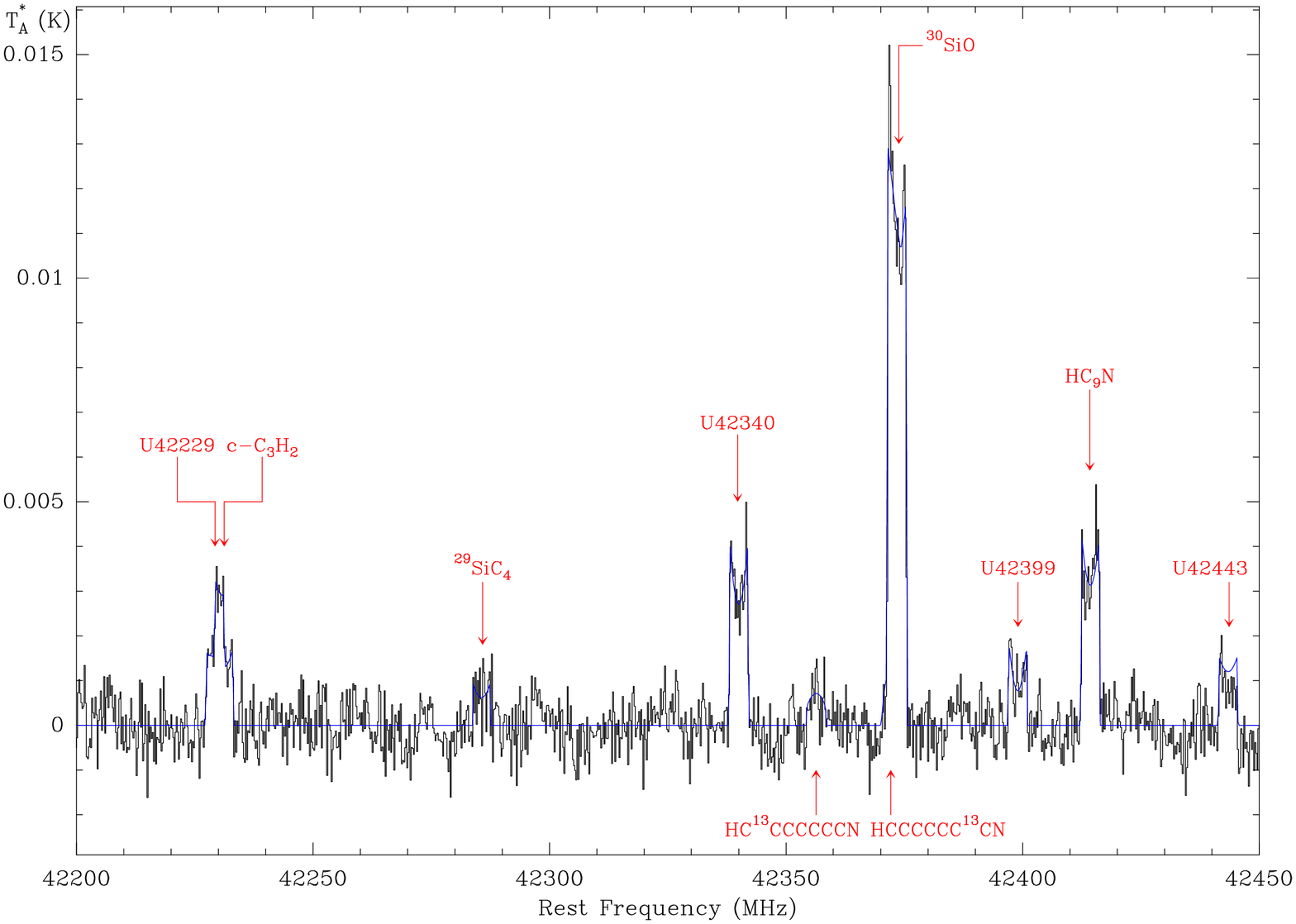}
\caption{IRC+10216 YEBES 40m data, line fits and labels from 
41950 to 42450 GHz.}
\label{fig23}
\end{figure*}

\begin{figure*}                                              
\includegraphics[width=0.93\textwidth]                       {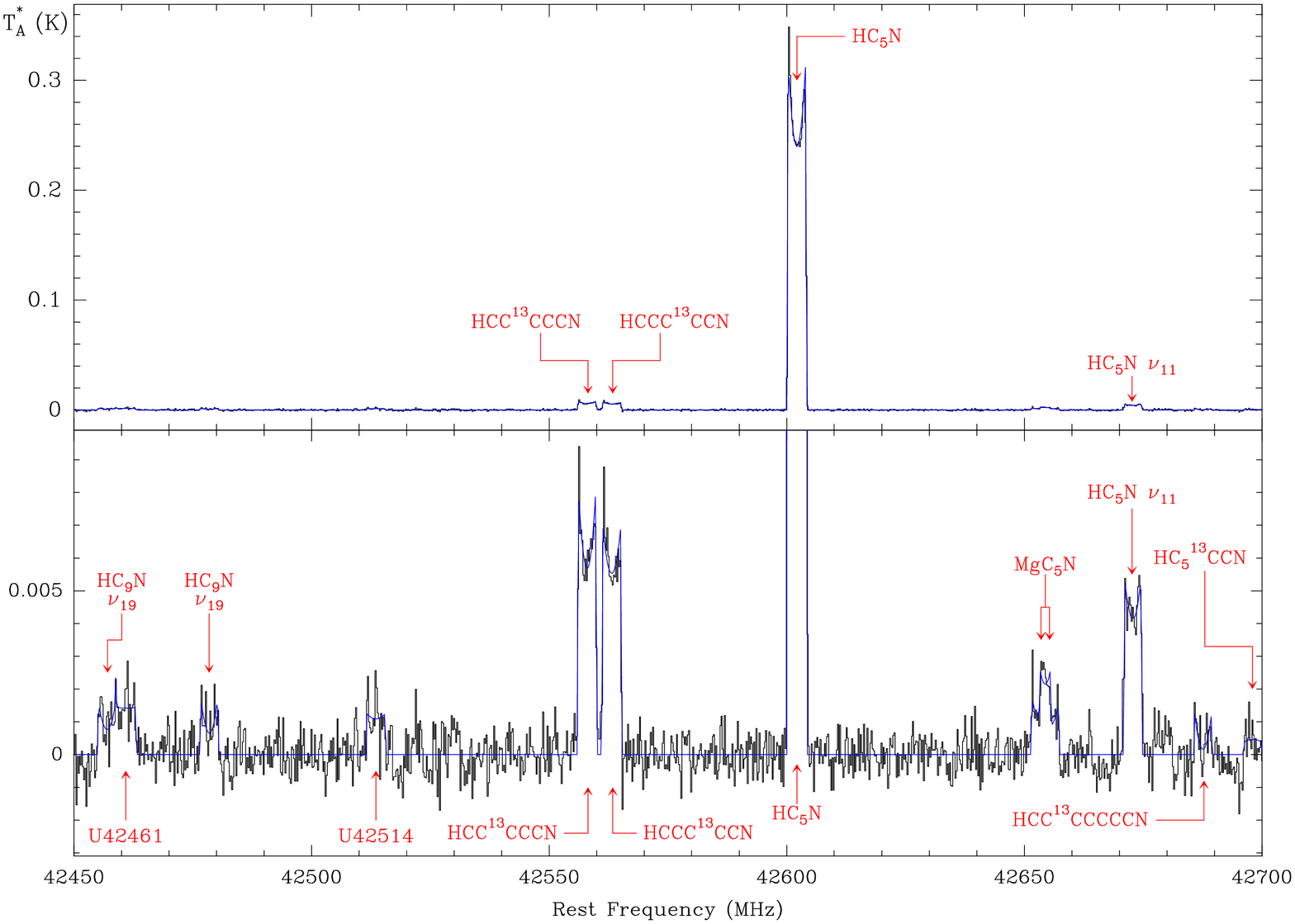}\\
\includegraphics[width=0.93\textwidth]                       {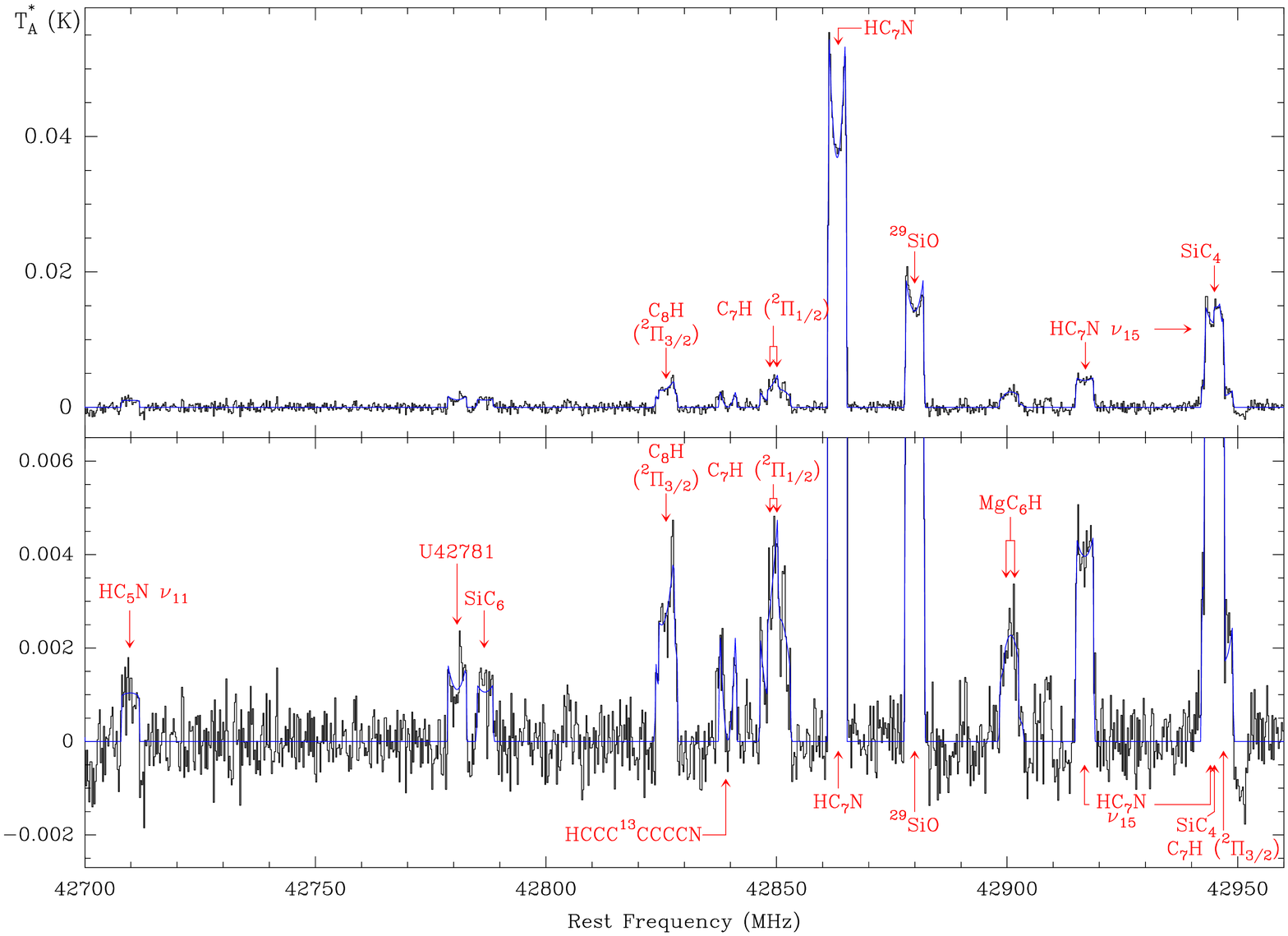}
\caption{IRC+10216 YEBES 40m data, line fits and labels from 
42450 to 42950 GHz.}
\label{fig24}
\end{figure*}

\begin{figure*}                                              
\includegraphics[width=0.93\textwidth]                       {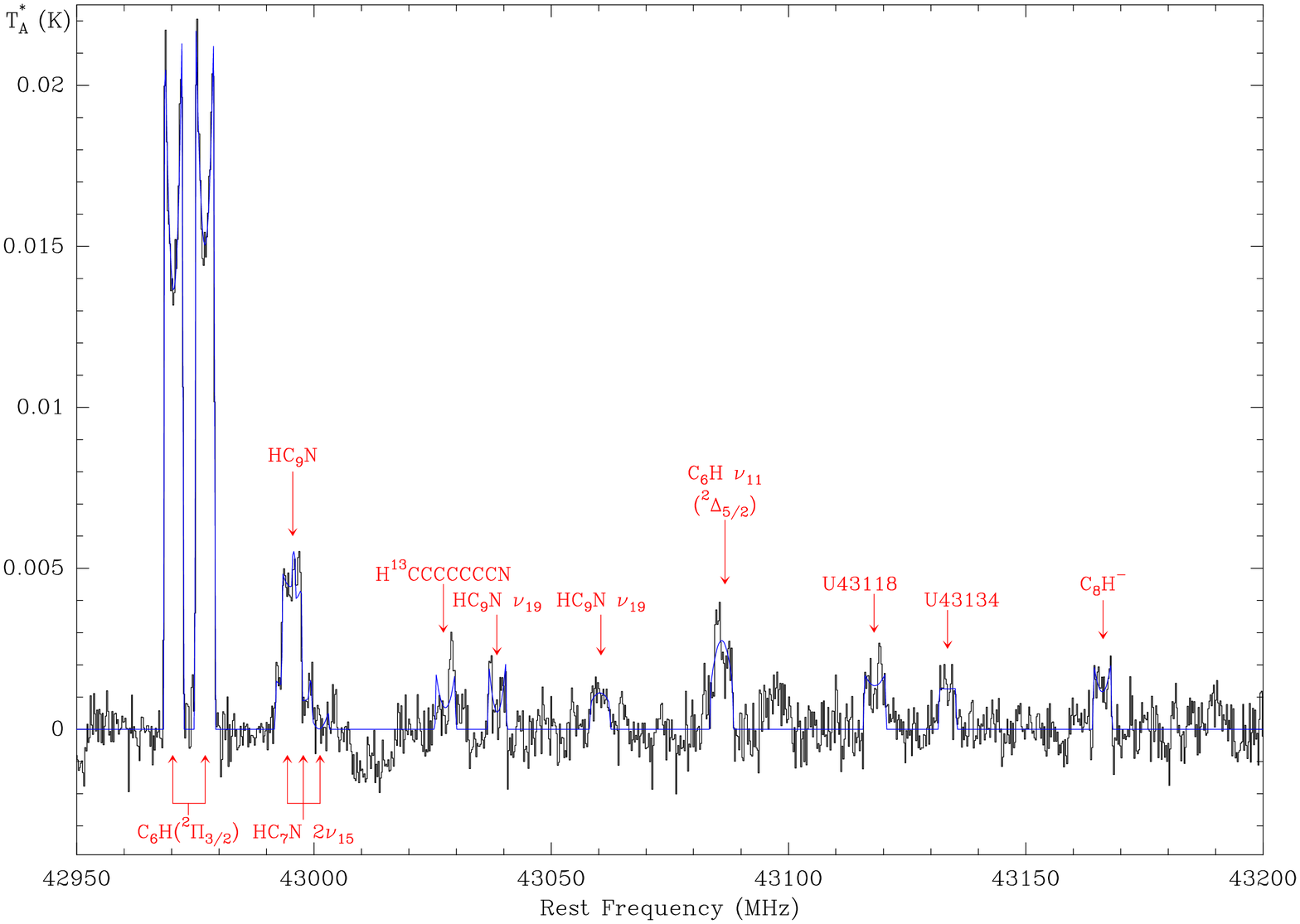}\\
\includegraphics[width=0.93\textwidth]                       {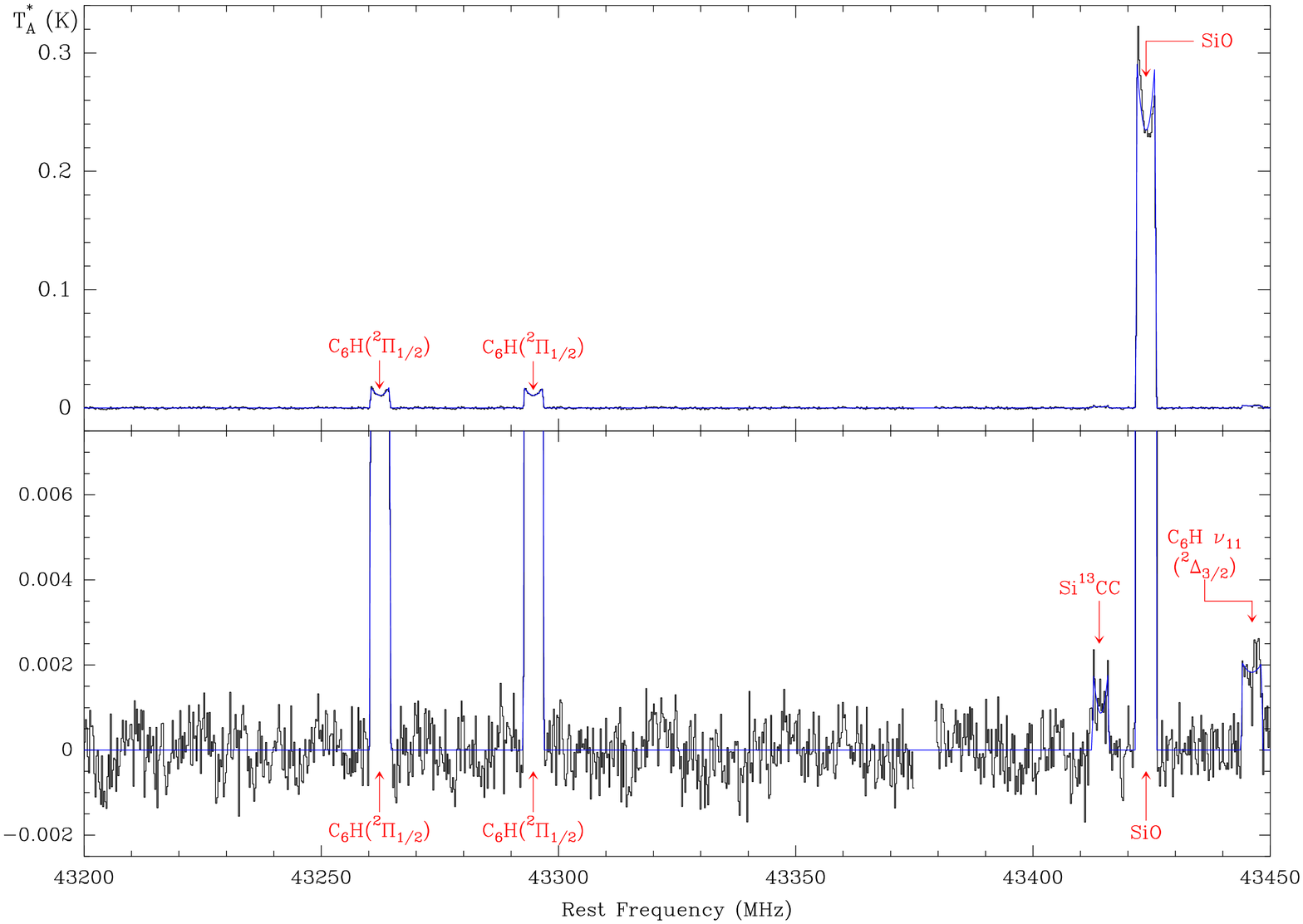}
\caption{IRC+10216 YEBES 40m data, line fits and labels from 
42950 to 43450 GHz.}
\label{fig25}
\end{figure*}

\begin{figure*}                                              
\includegraphics[width=0.93\textwidth]                       {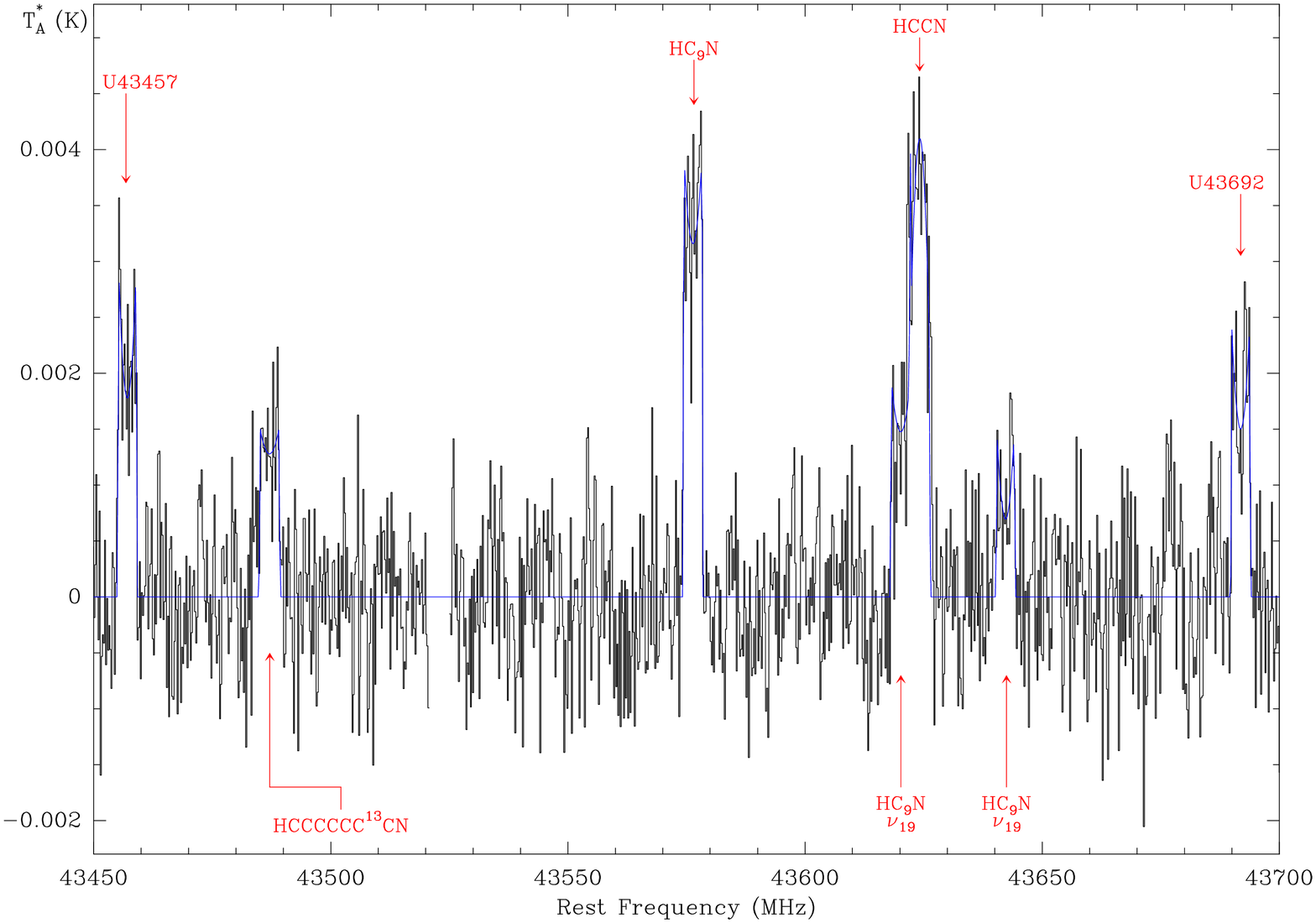}\\
\includegraphics[width=0.93\textwidth]                       {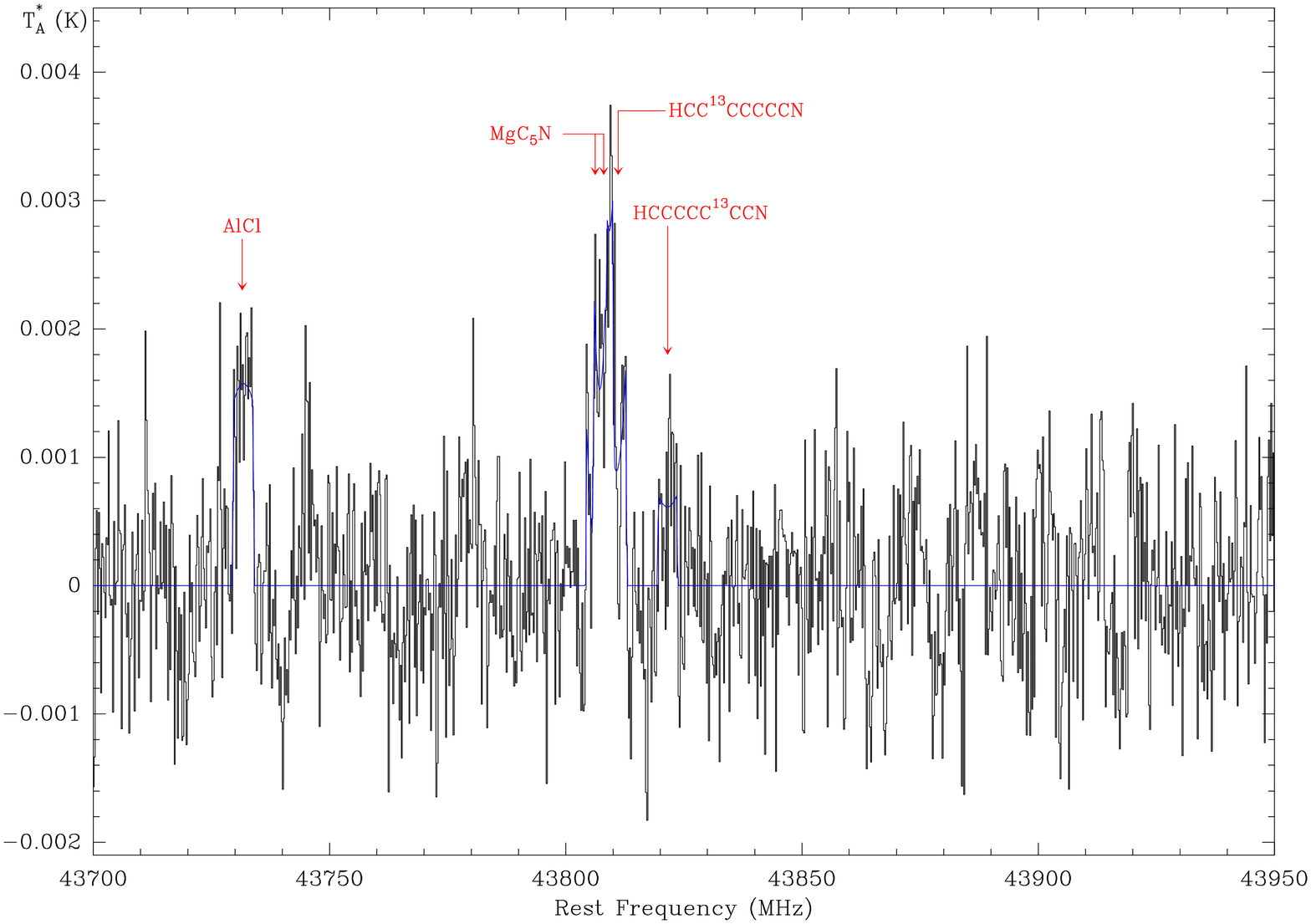}
\caption{IRC+10216 YEBES 40m data, line fits and labels from 
43450 to 43950 GHz.}
\label{fig26}
\end{figure*}

\begin{figure*}                                              
\includegraphics[width=0.93\textwidth]                       {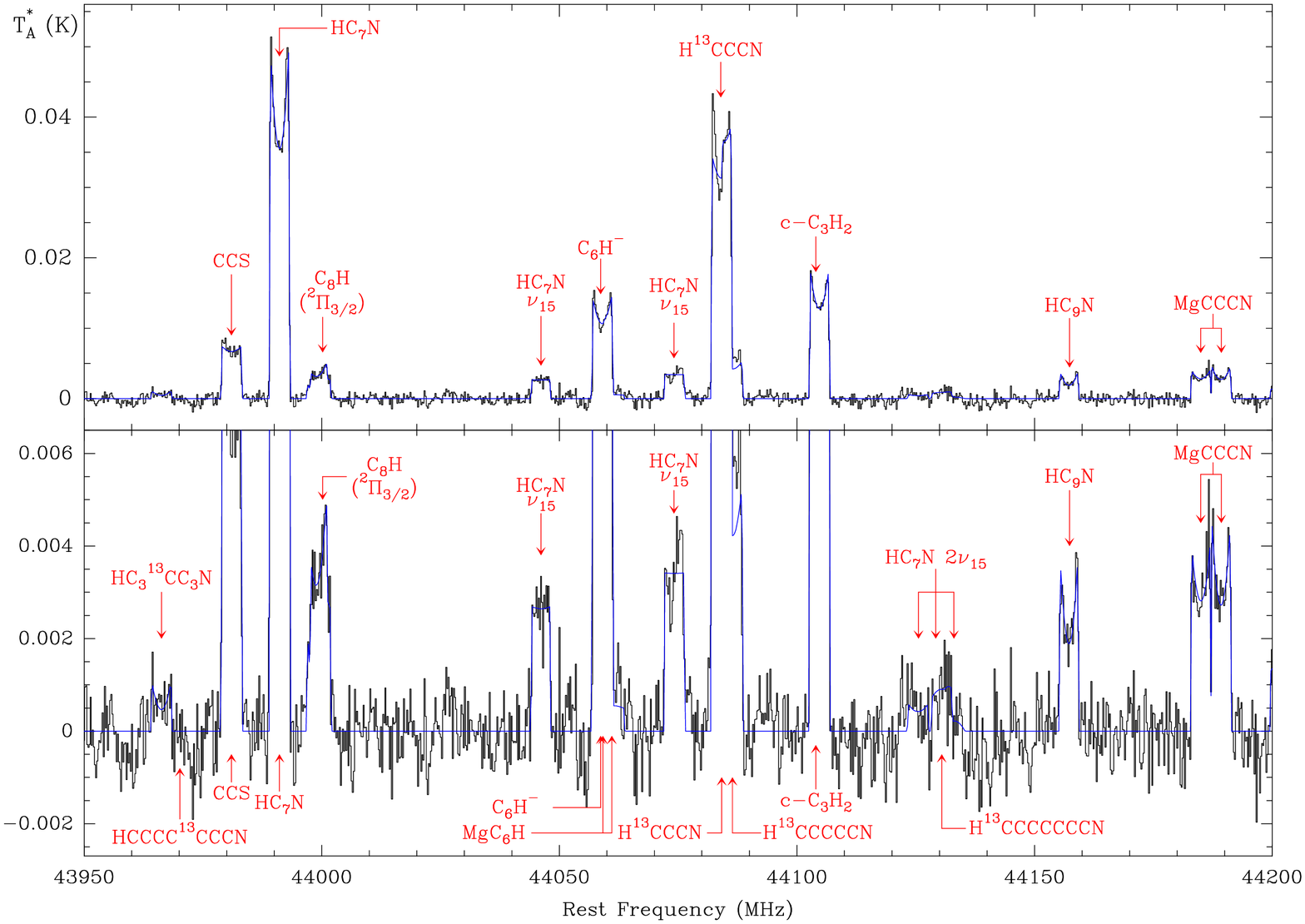}\\
\includegraphics[width=0.93\textwidth]                       {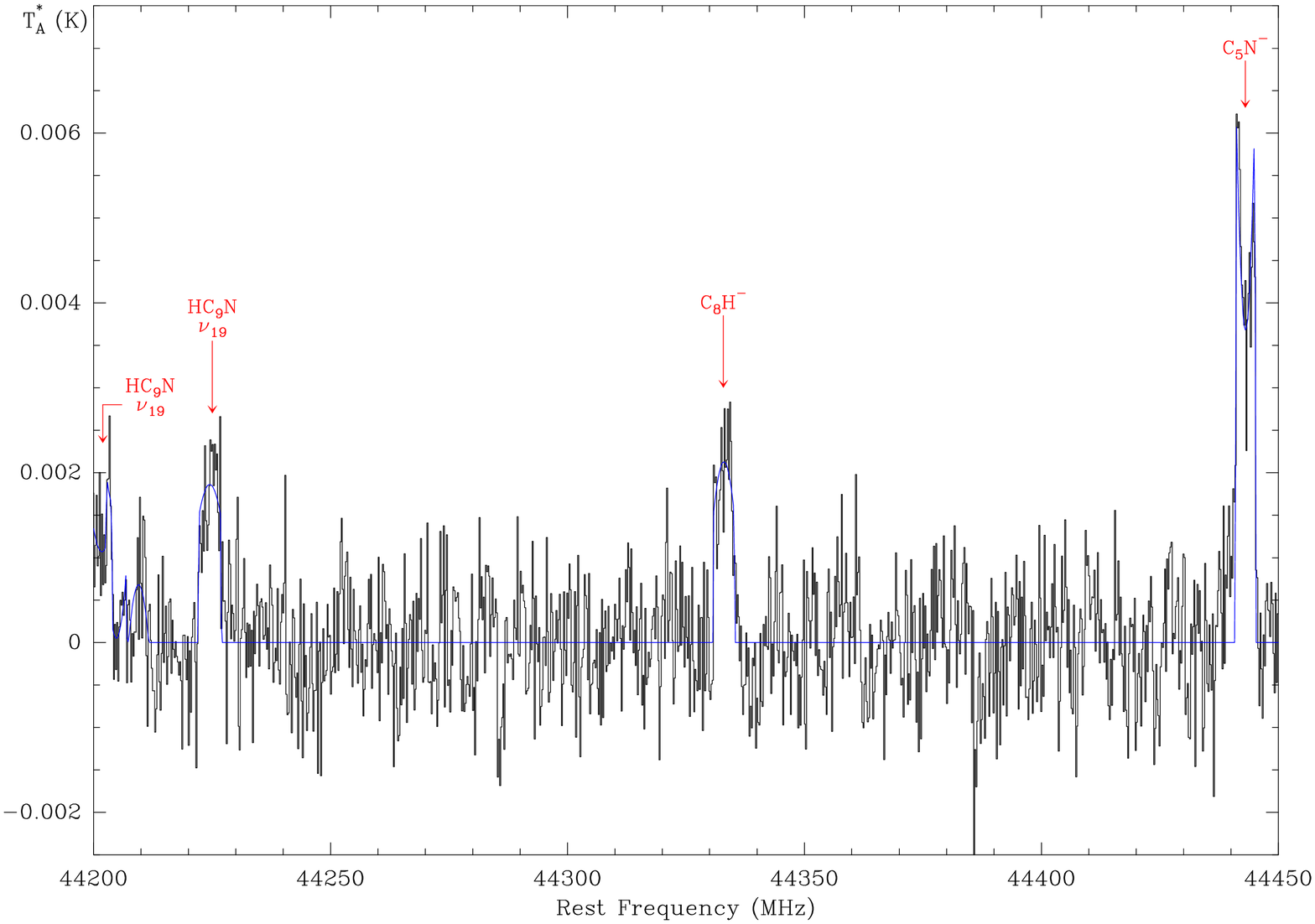}
\caption{IRC+10216 YEBES 40m data, line fits and labels from 
43950 to 44450 GHz.}
\label{fig27}
\end{figure*}

\begin{figure*}                                              
\includegraphics[width=0.93\textwidth]                       {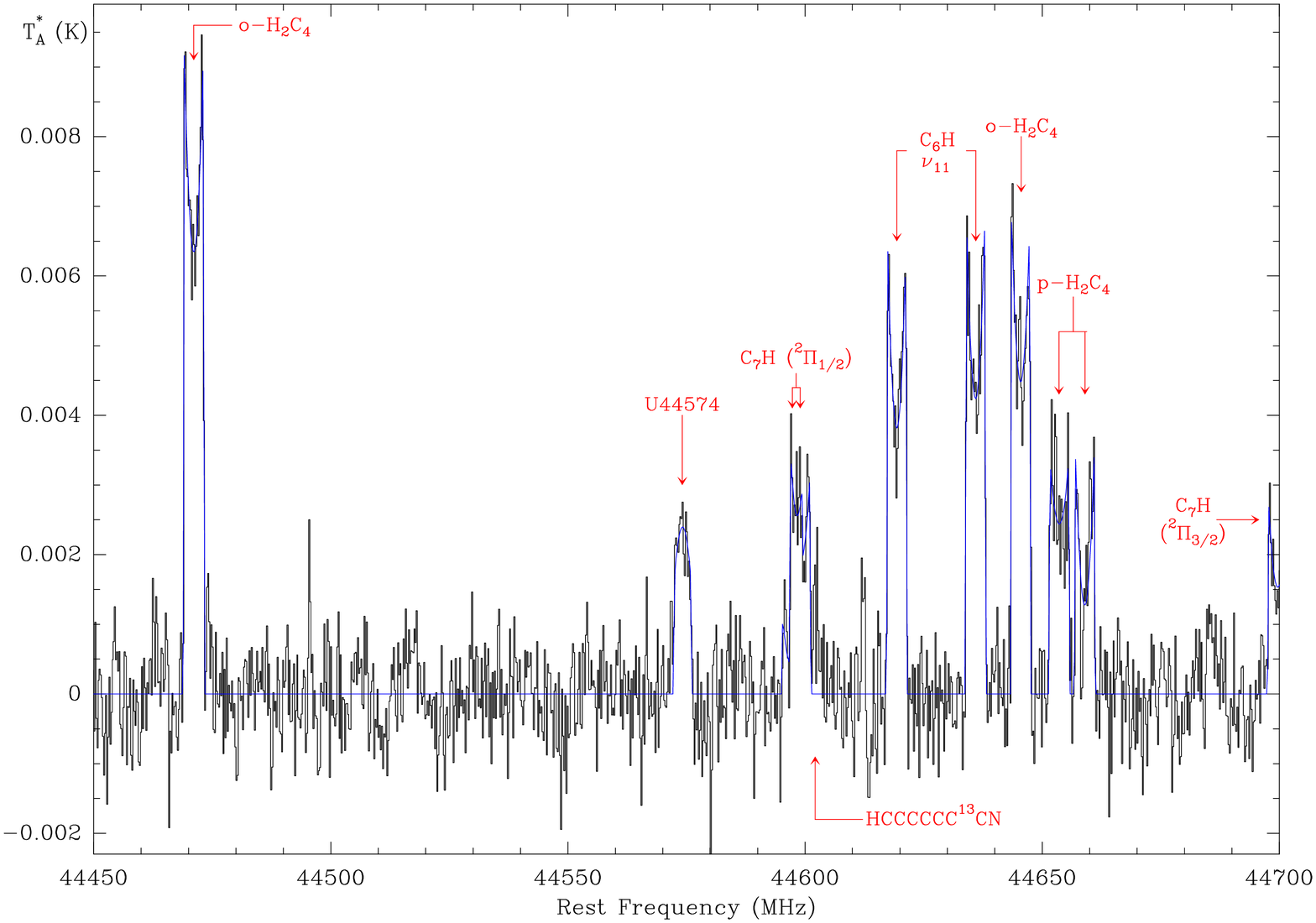}\\
\includegraphics[width=0.93\textwidth]                       {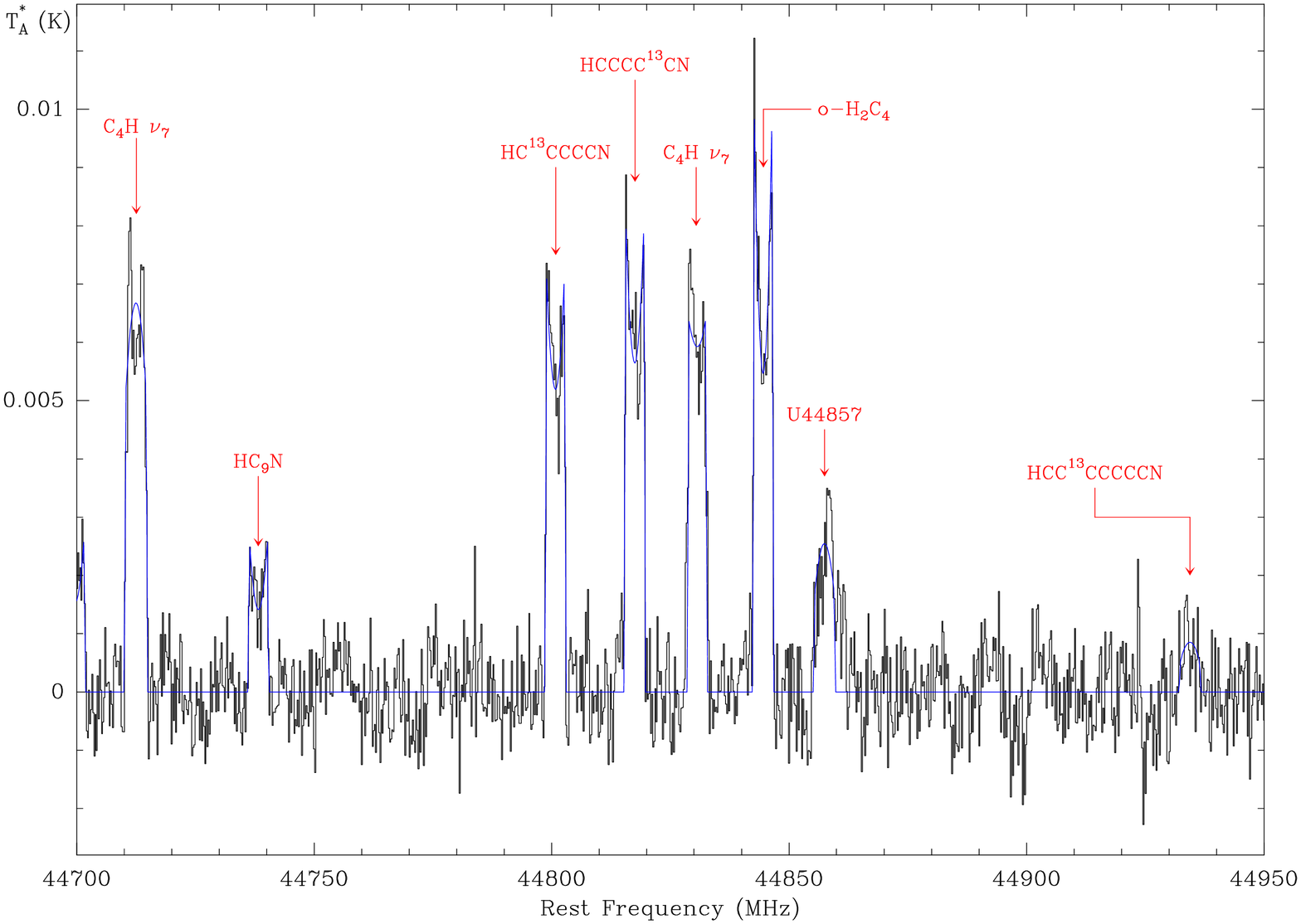}
\caption{IRC+10216 YEBES 40m data, line fits and labels from 
44450 to 44950 GHz.}
\label{fig28}
\end{figure*}

\begin{figure*}                                              
\includegraphics[width=0.93\textwidth]                       {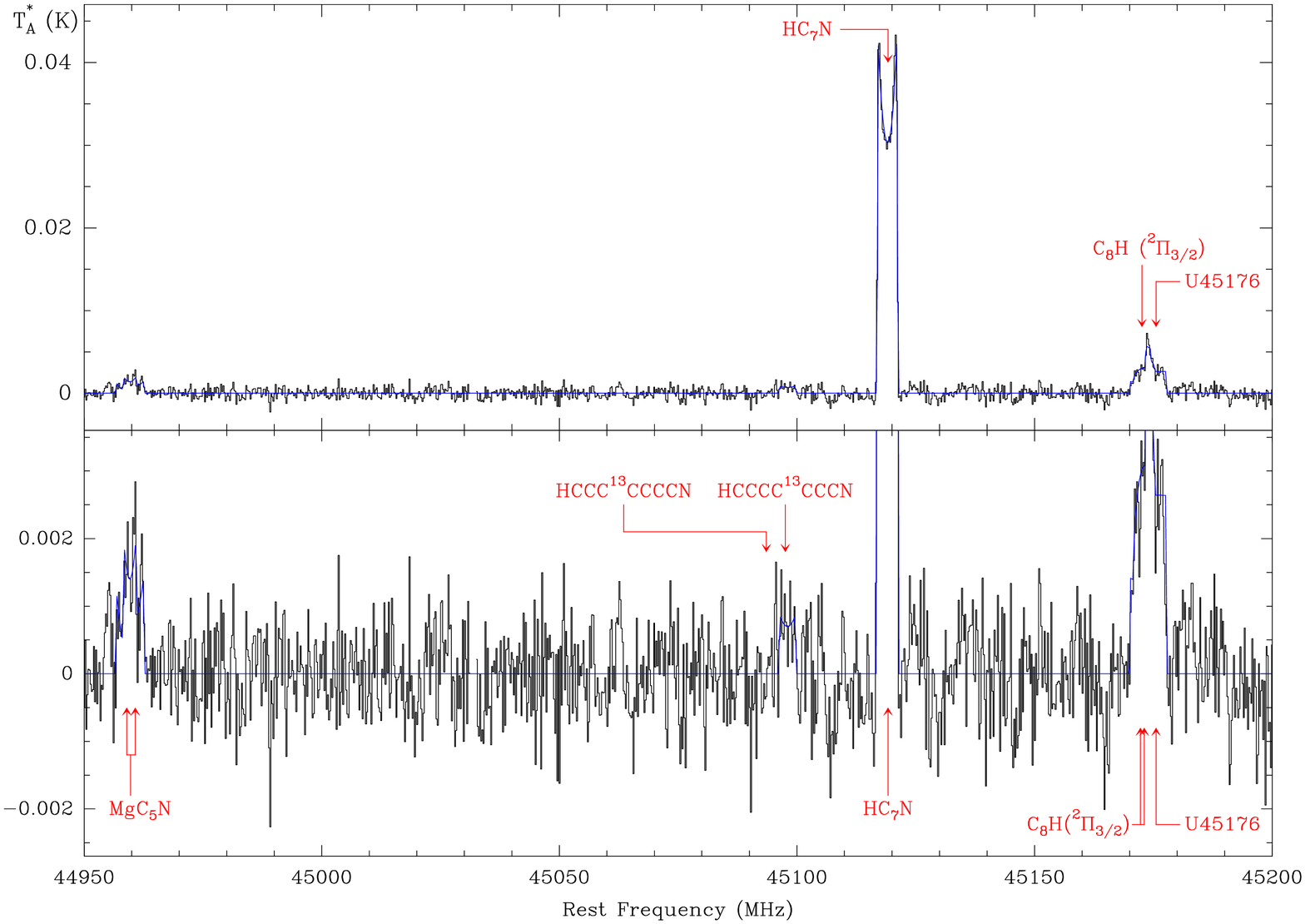}\\
\includegraphics[width=0.93\textwidth]                       {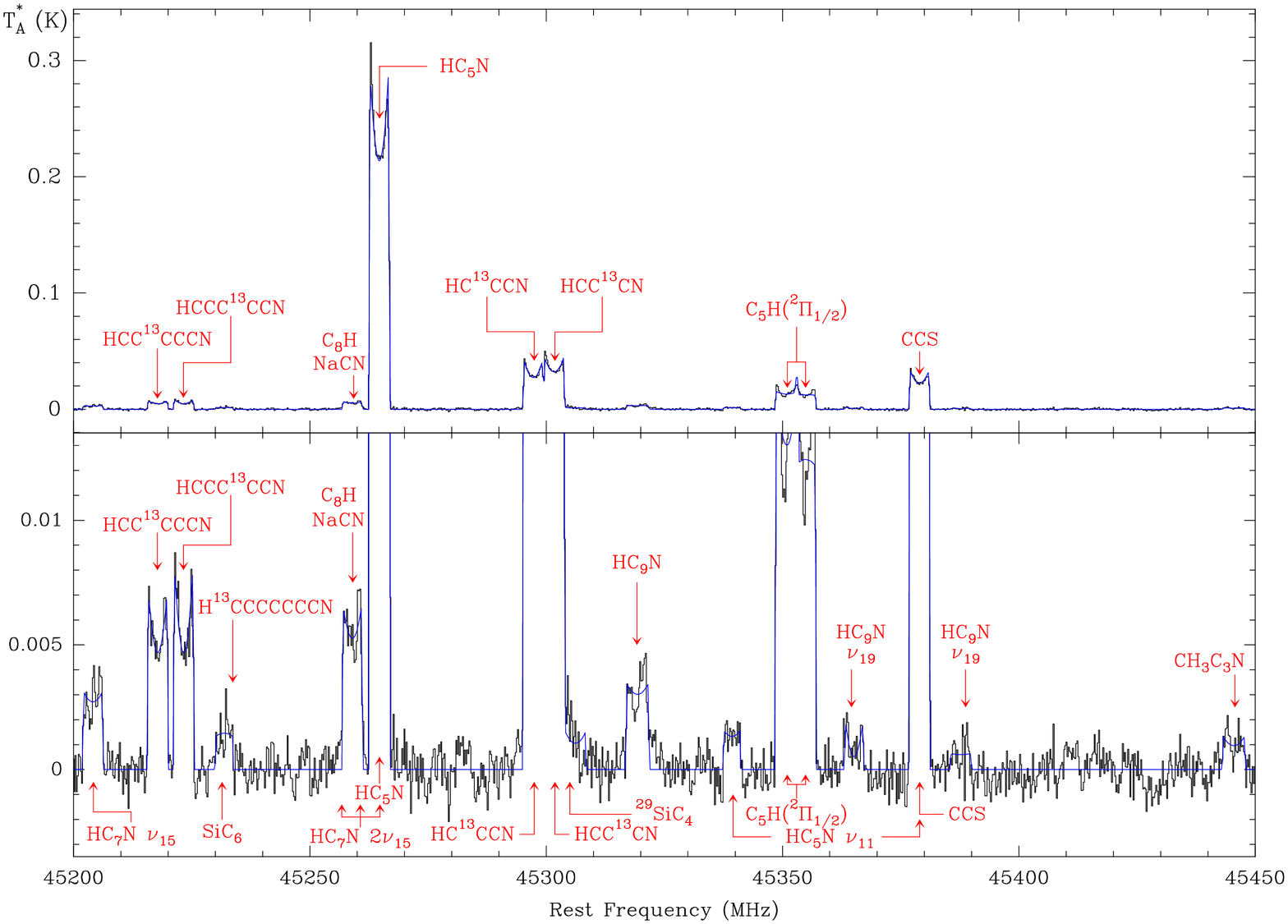}
\caption{IRC+10216 YEBES 40m data, line fits and labels from 
44950 to 45450 GHz.}
\label{fig29}
\end{figure*}

\begin{figure*}                                              
\includegraphics[width=0.93\textwidth]                       {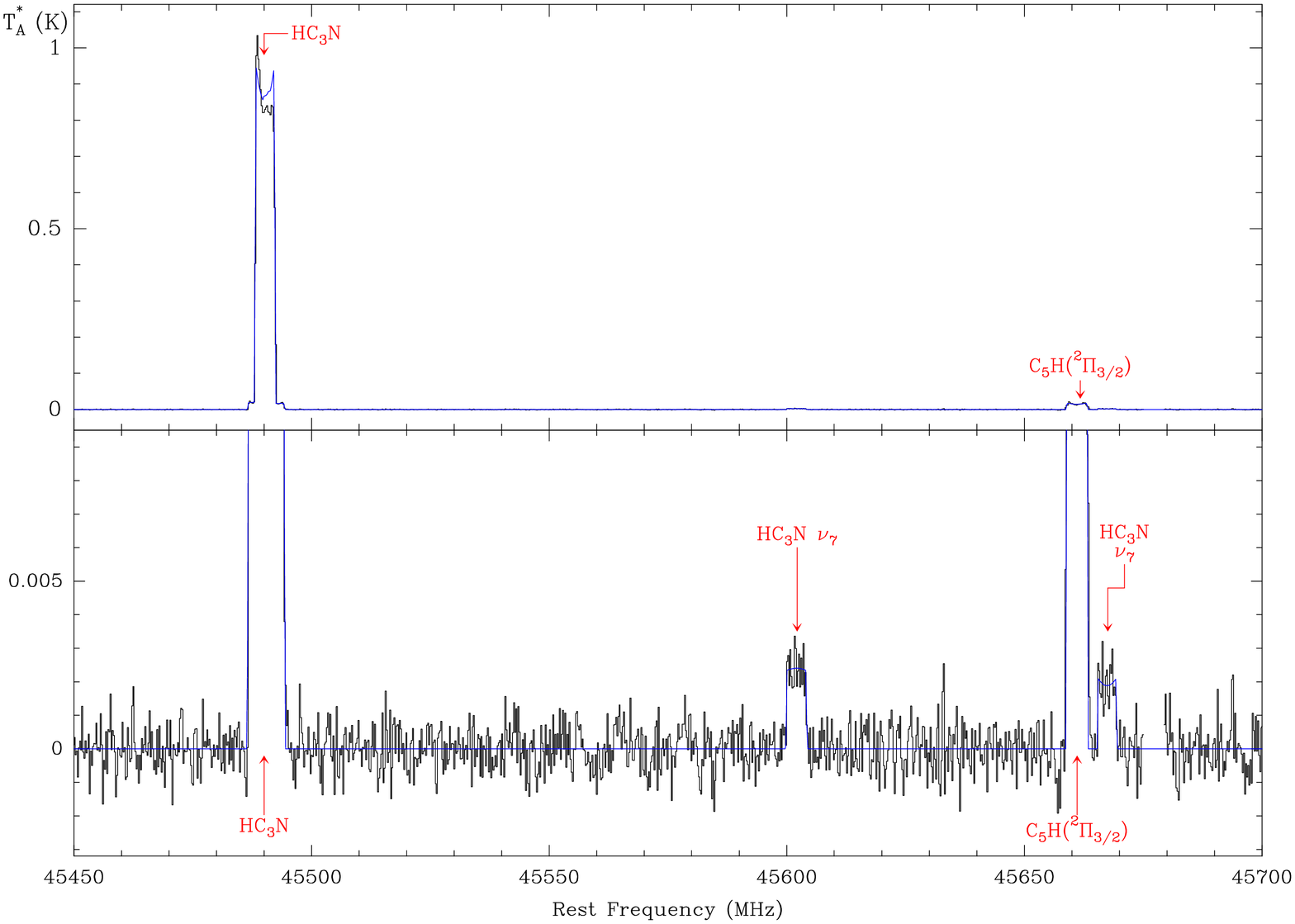}\\
\includegraphics[width=0.93\textwidth]                       {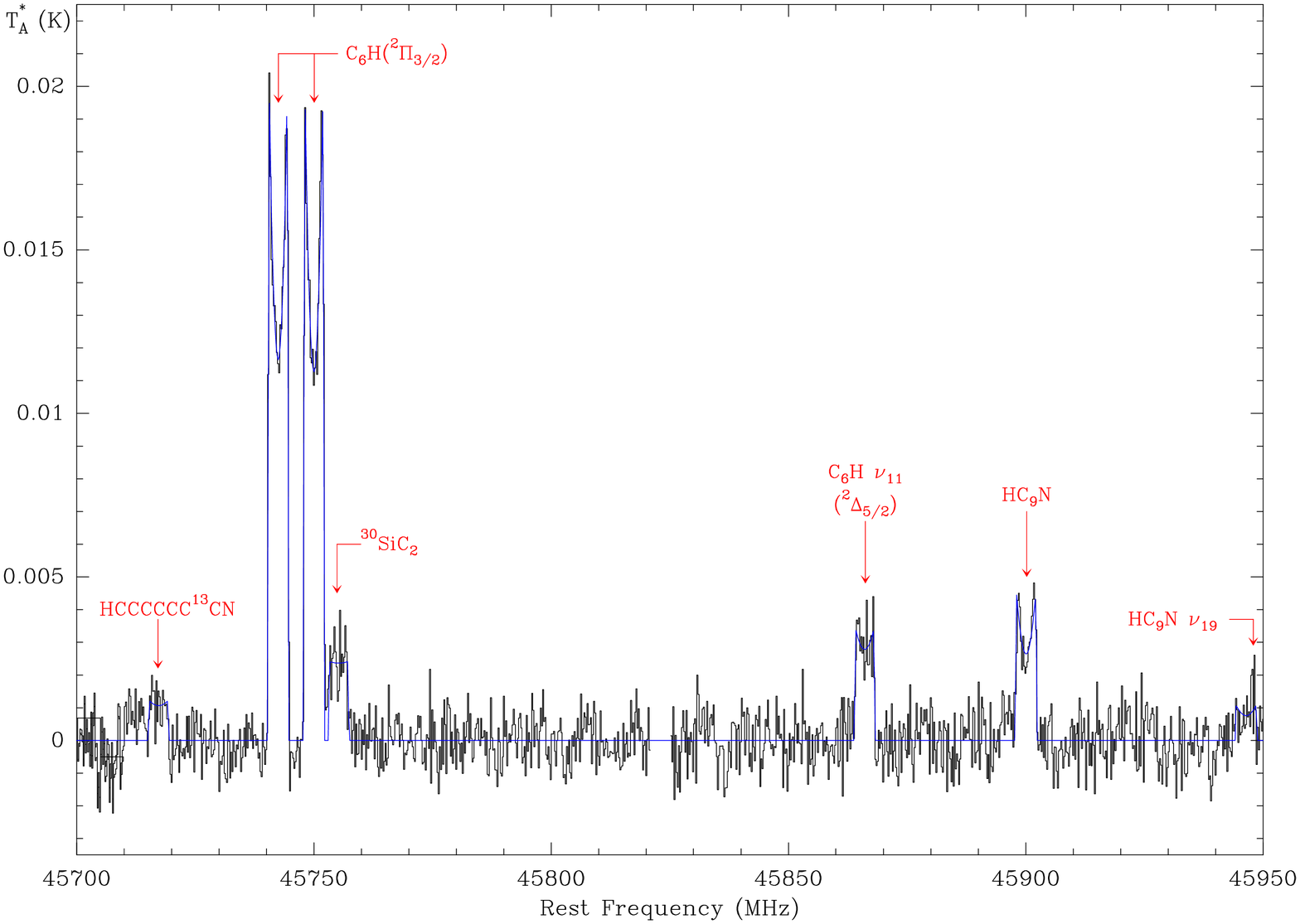}
\caption{IRC+10216 YEBES 40m data, line fits and labels from 
45450 to 45950 GHz.}
\label{fig30}
\end{figure*}

\begin{figure*}                                              
\includegraphics[width=0.93\textwidth]                       {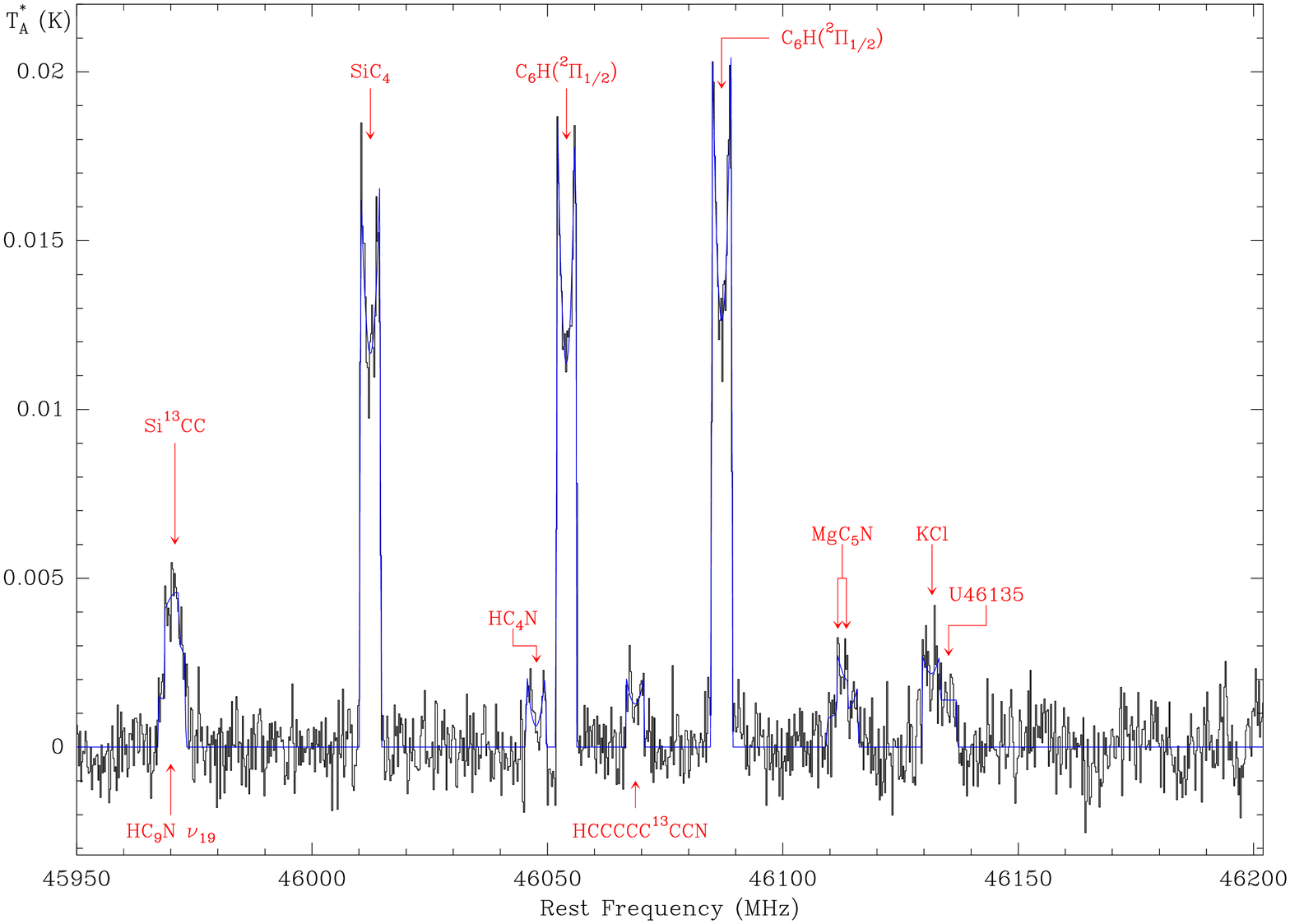}\\
\includegraphics[width=0.93\textwidth]                       {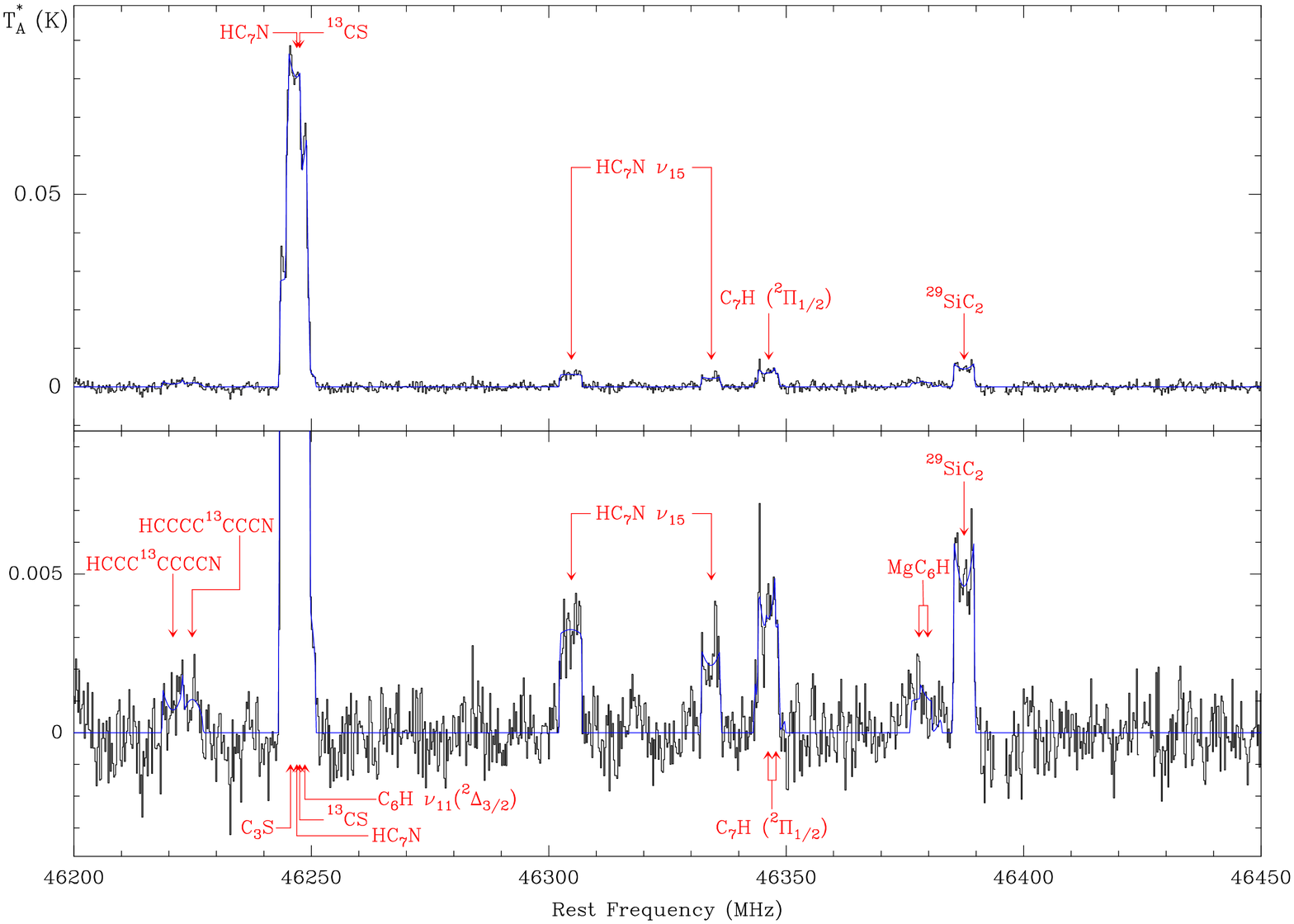}
\caption{IRC+10216 YEBES 40m data, line fits and labels from 
45950 to 46450 GHz.}
\label{fig31}
\end{figure*}

\begin{figure*}                                              
\includegraphics[width=0.93\textwidth]                       {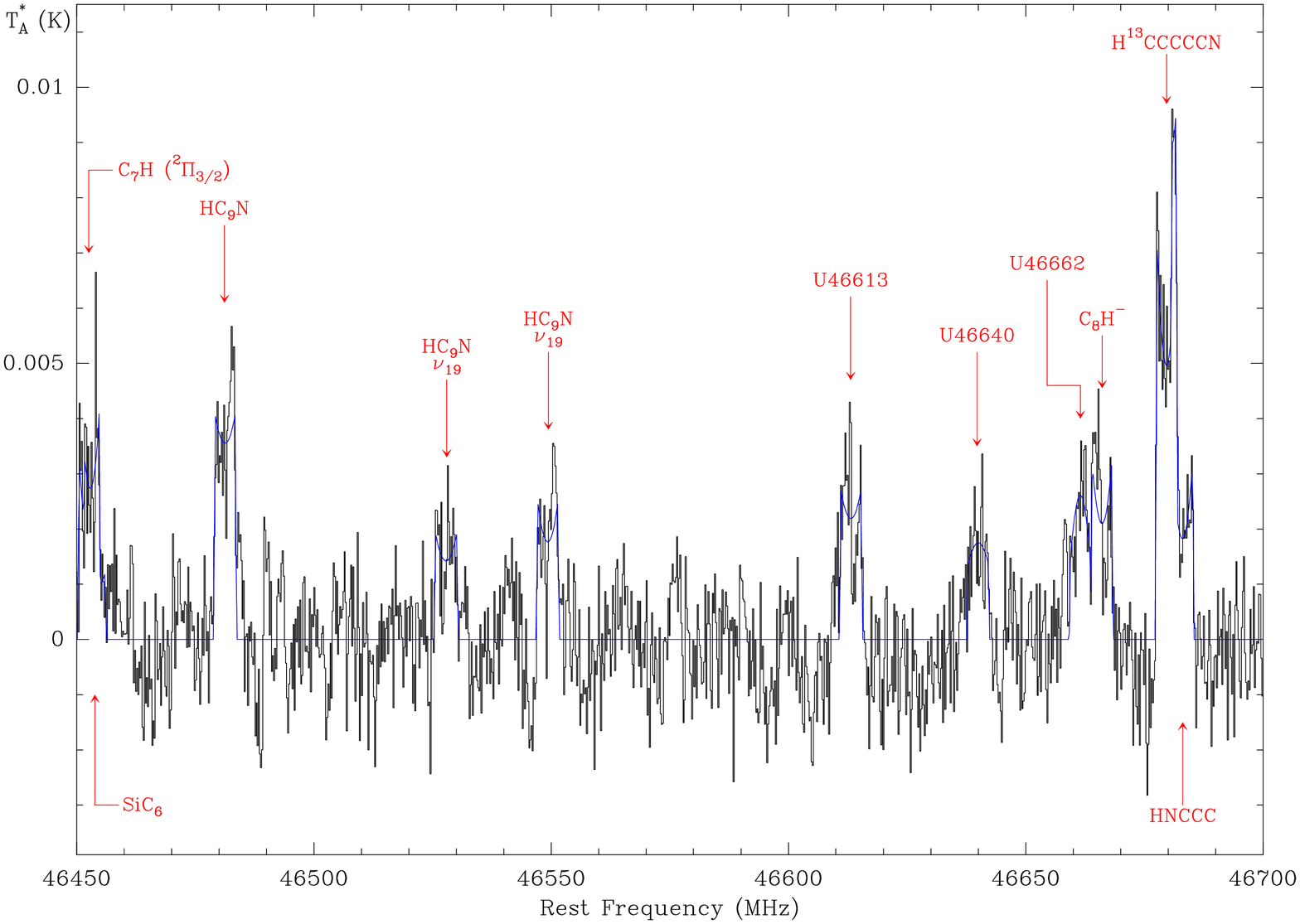}\\
\includegraphics[width=0.93\textwidth]                       {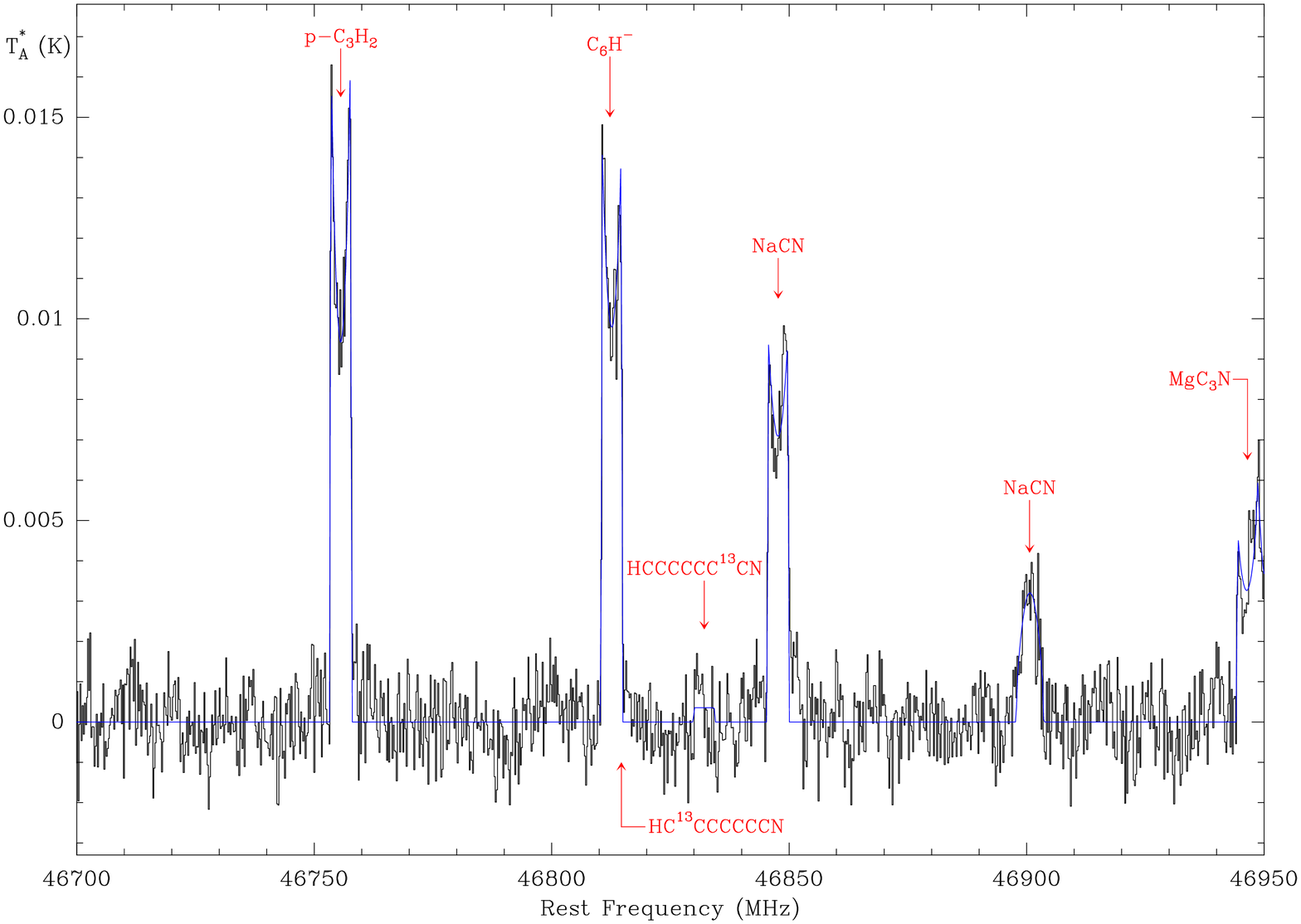}
\caption{IRC+10216 YEBES 40m data, line fits and labels from 
46450 to 46950 GHz.}
\label{fig32}
\end{figure*}

\begin{figure*}                                              
\includegraphics[width=0.93\textwidth]                       {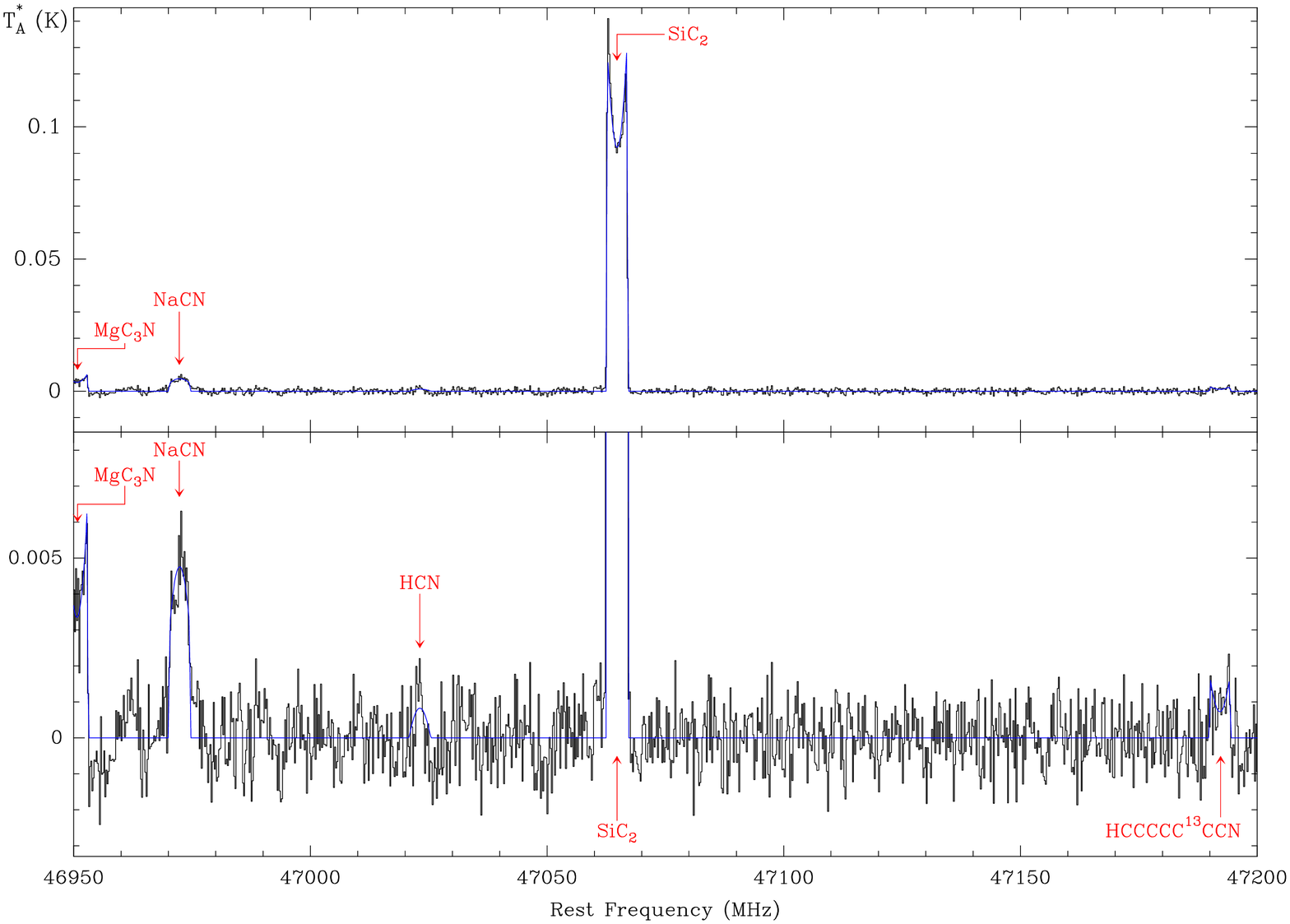}\\
\includegraphics[width=0.93\textwidth]                       {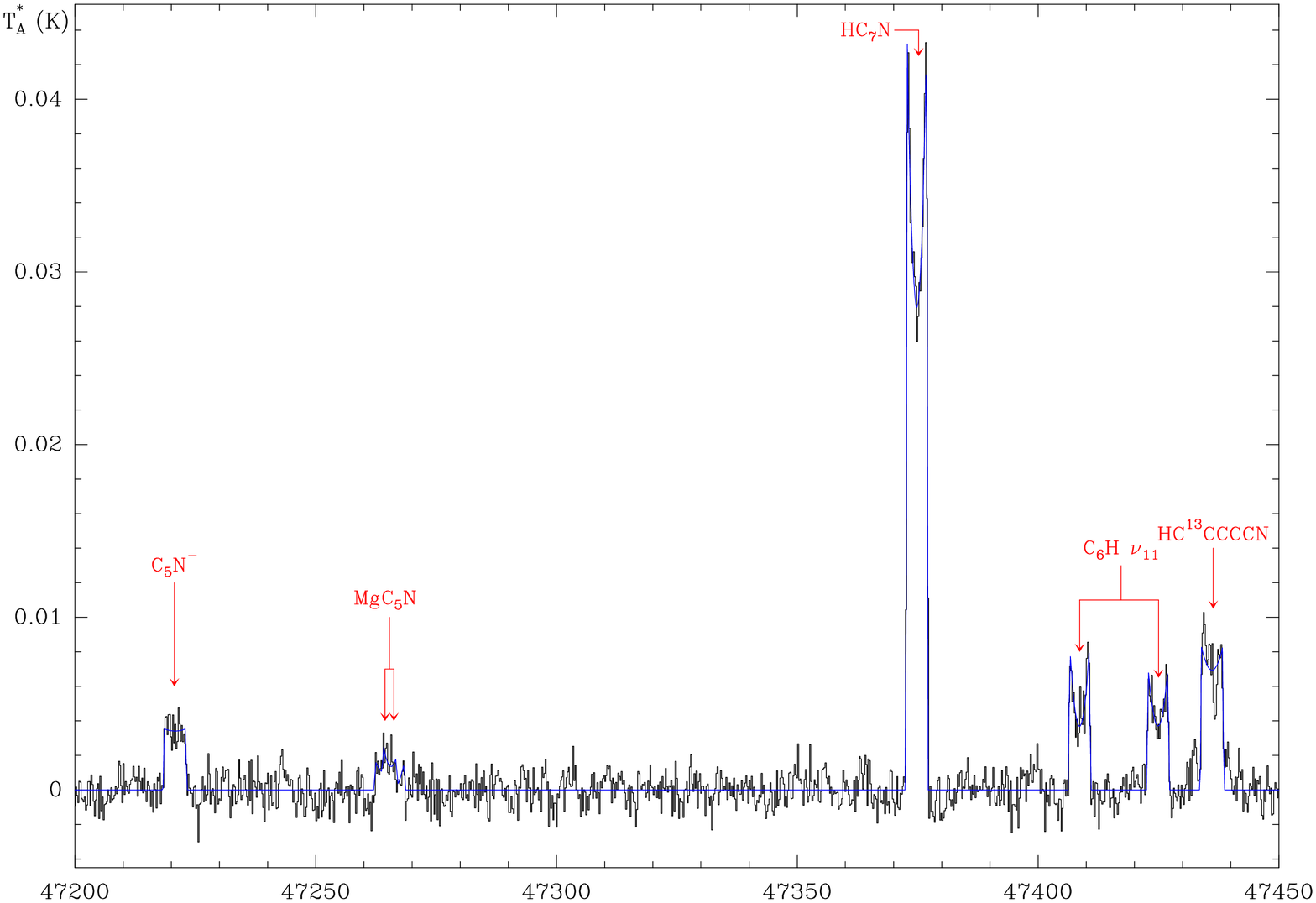}
\caption{IRC+10216 YEBES 40m data, line fits and labels from 
46950 to 47450 GHz.}
\label{fig33}
\end{figure*}

\begin{figure*}                                              
\includegraphics[width=0.93\textwidth]                       {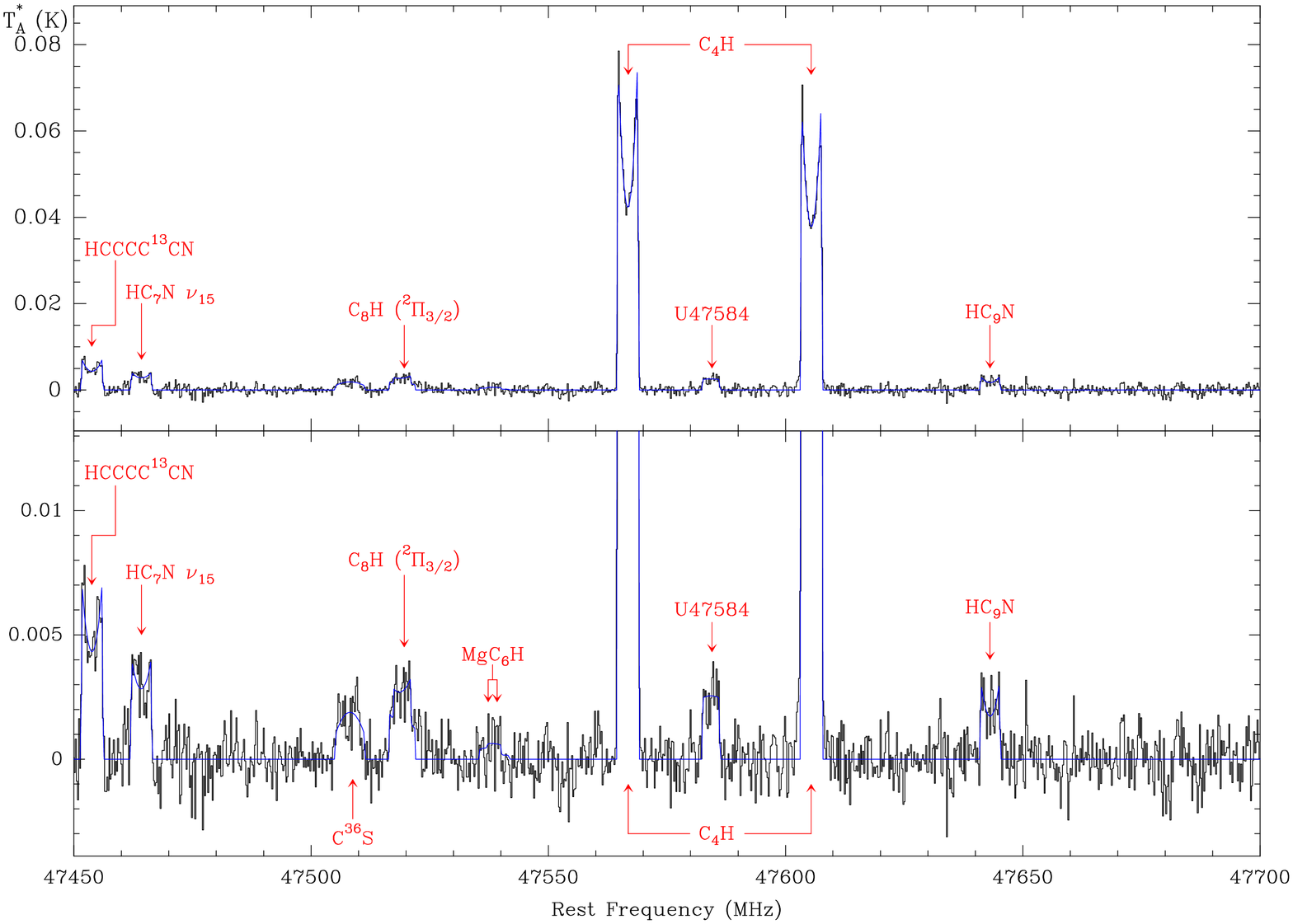}\\
\includegraphics[width=0.93\textwidth]                       {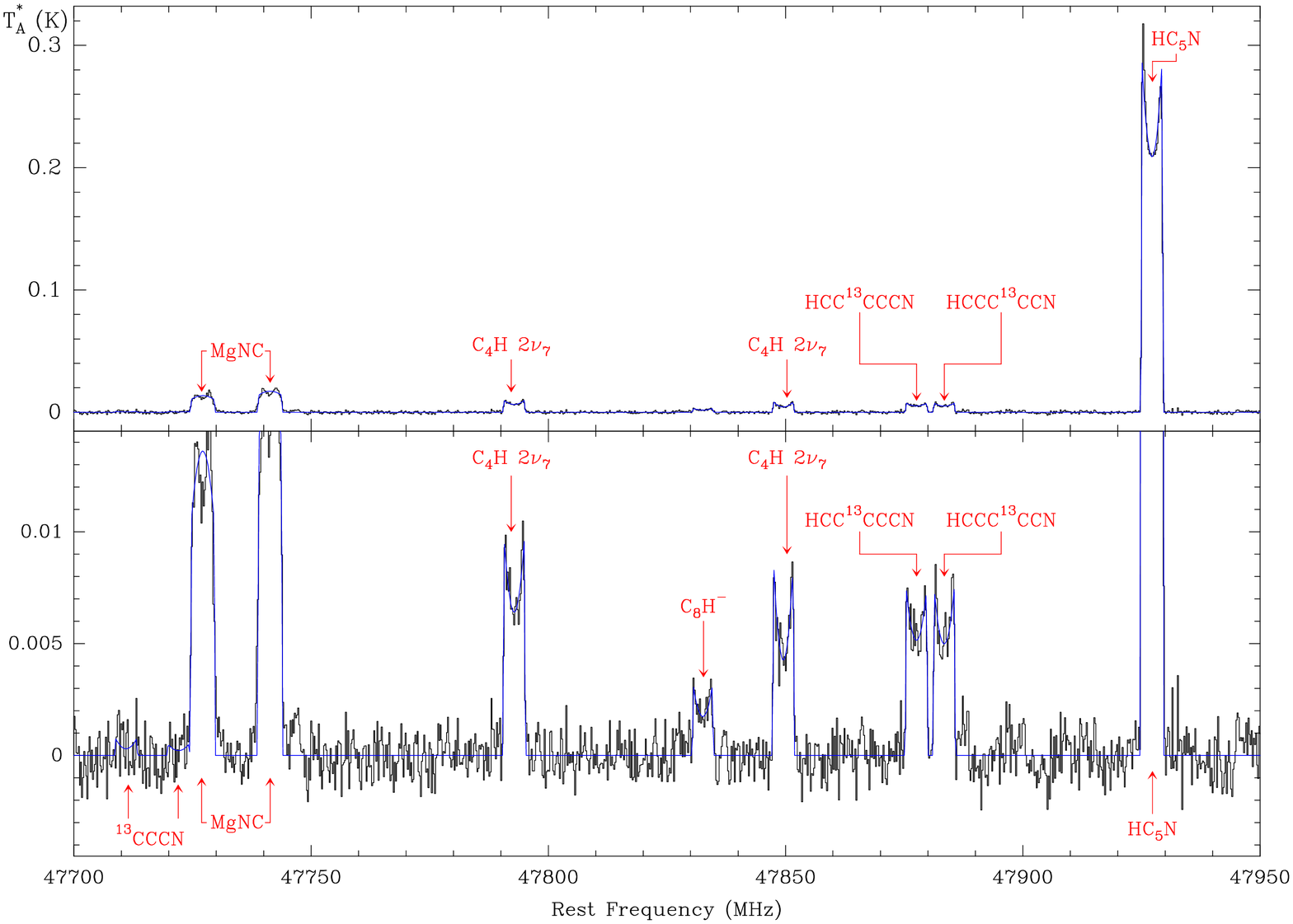}
\caption{IRC+10216 YEBES 40m data, line fits and labels from 
47450 to 47950 GHz.}
\label{fig34}
\end{figure*}

\begin{figure*}                                              
\includegraphics[width=0.93\textwidth]                       {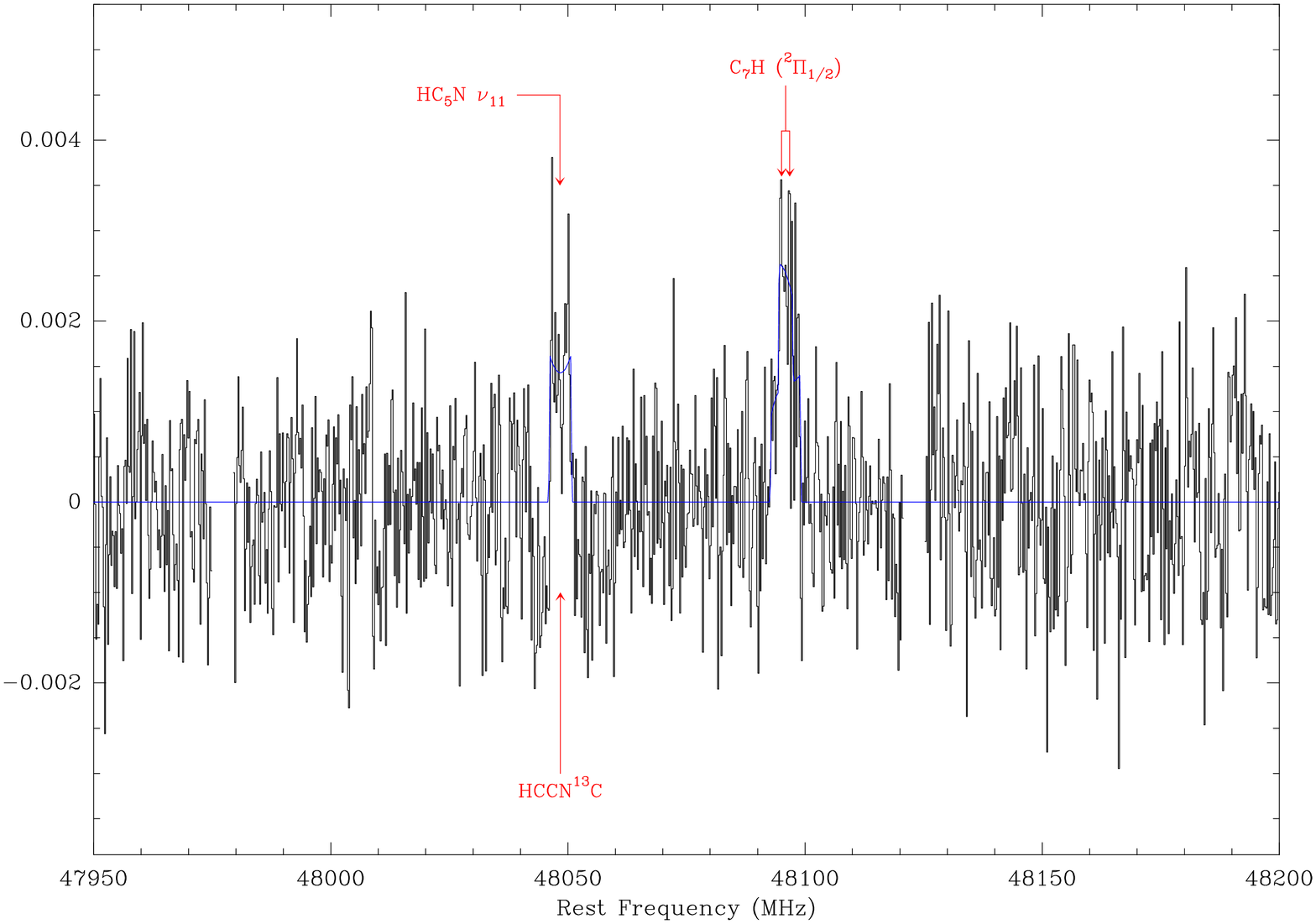}\\
\includegraphics[width=0.93\textwidth]                       {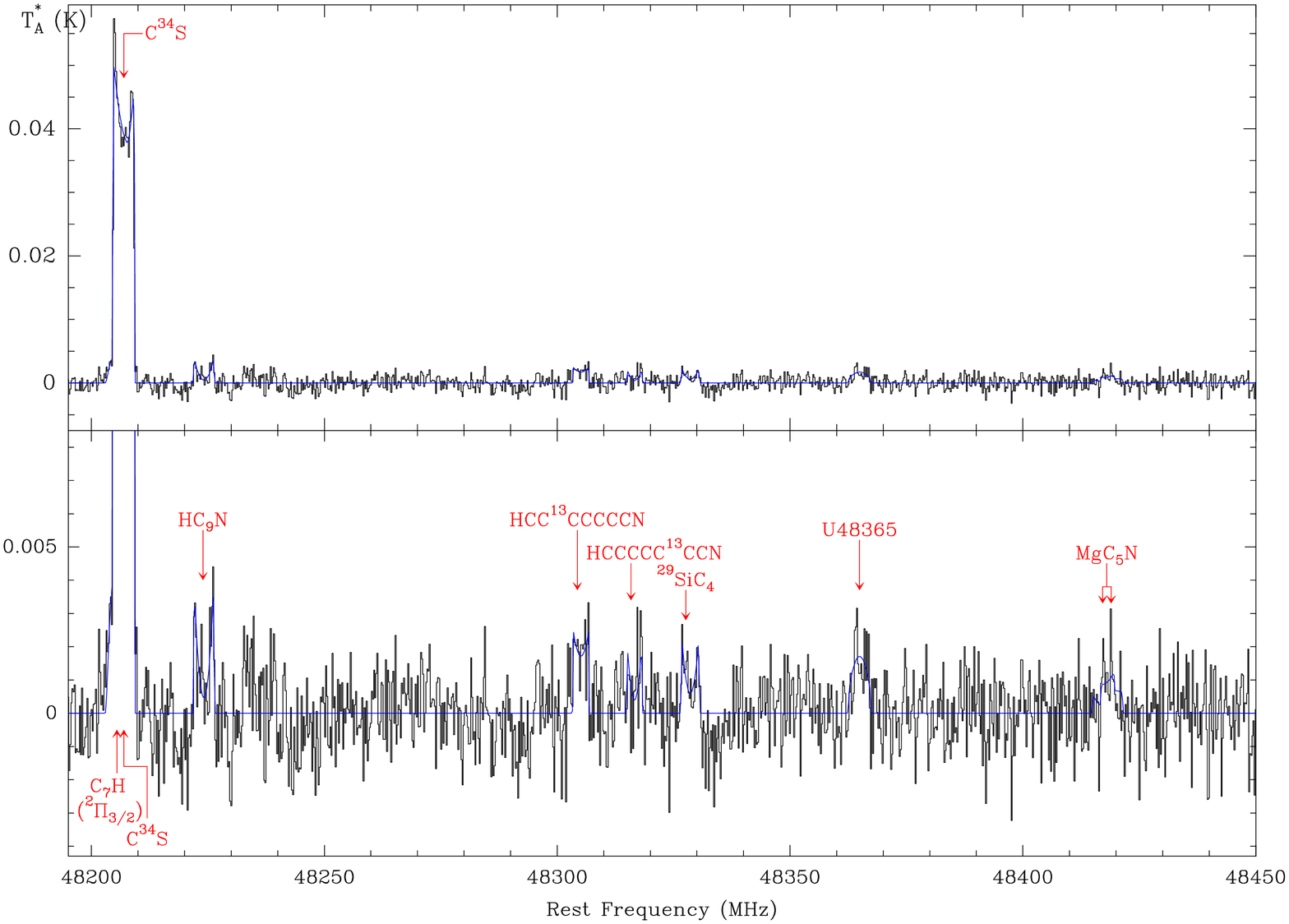}
\caption{IRC+10216 YEBES 40m data, line fits and labels from 
47950 to 48450 GHz.}
\label{fig35}
\end{figure*}

\begin{figure*}                                              
\includegraphics[width=0.93\textwidth]                       {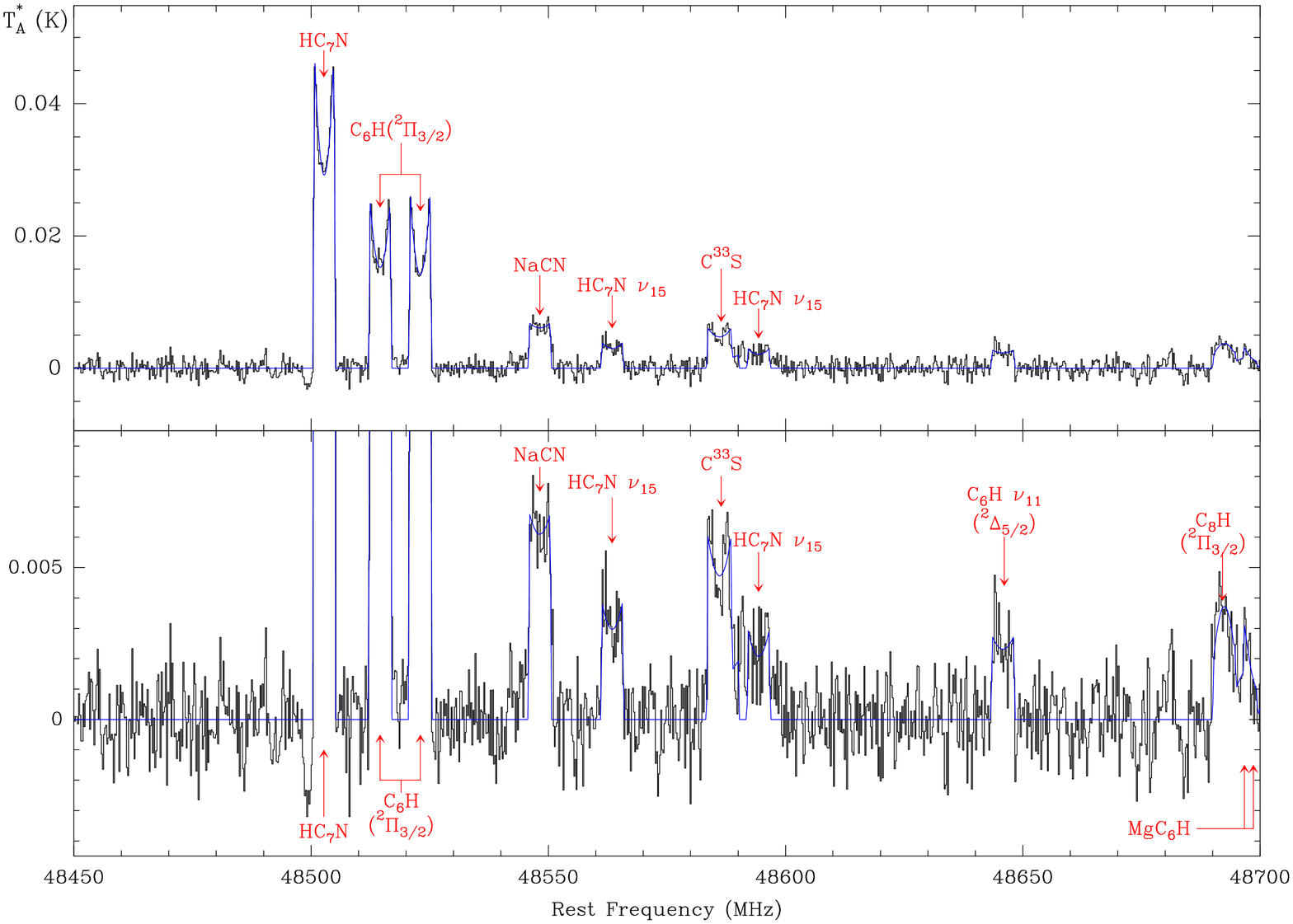}\\
\includegraphics[width=0.93\textwidth]                       {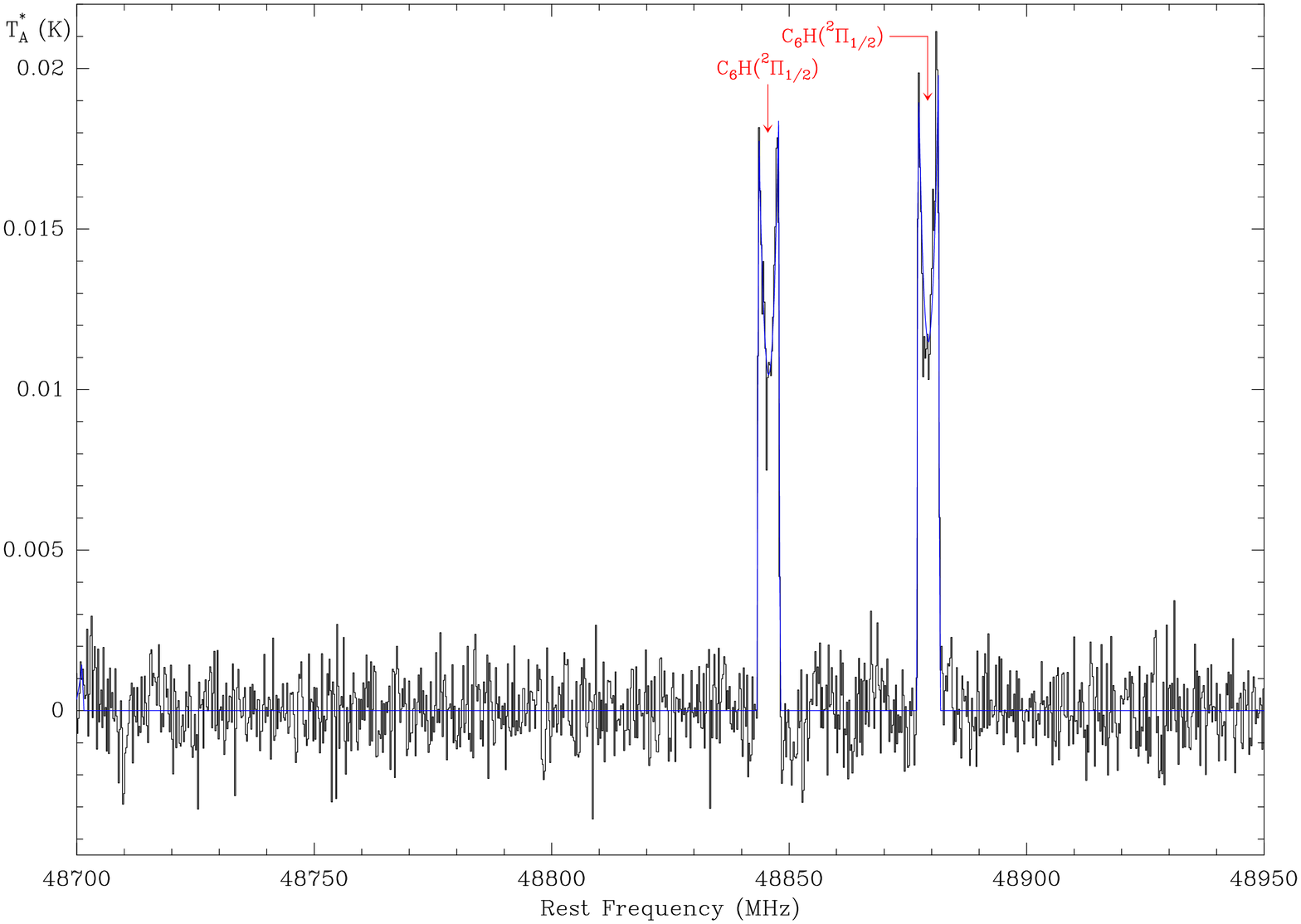}
\caption{IRC+10216 YEBES 40m data, line fits and labels from 
48450 to 48950 GHz.}
\label{fig36}
\end{figure*}

\begin{figure*}                                              
\includegraphics[width=0.93\textwidth]                       {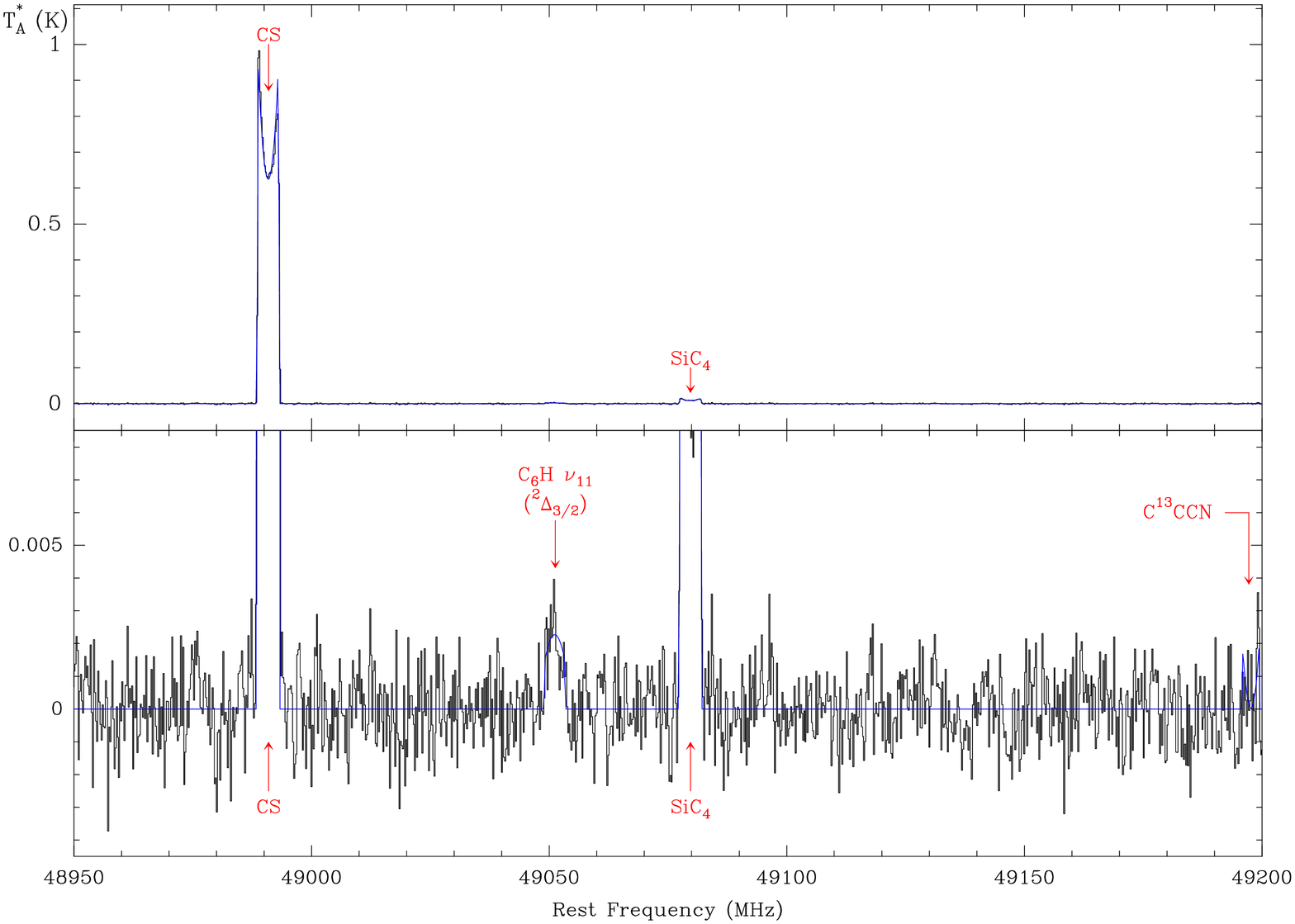}\\
\includegraphics[width=0.93\textwidth]                       {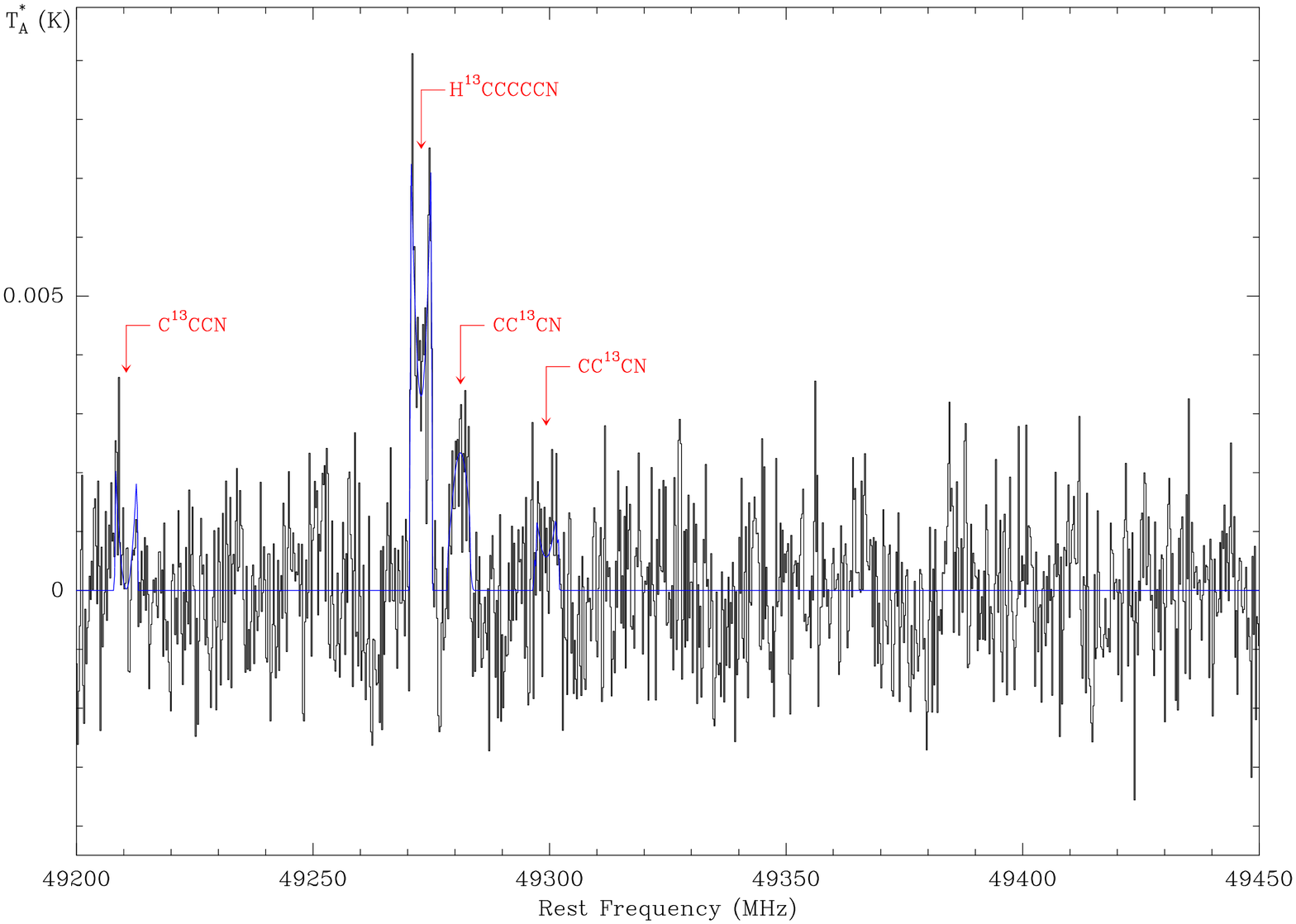}
\caption{IRC+10216 YEBES 40m data, line fits and labels from 
48950 to 49450 GHz.}
\label{fig37}
\end{figure*}

\begin{figure*}                                              
\includegraphics[width=0.93\textwidth]                       {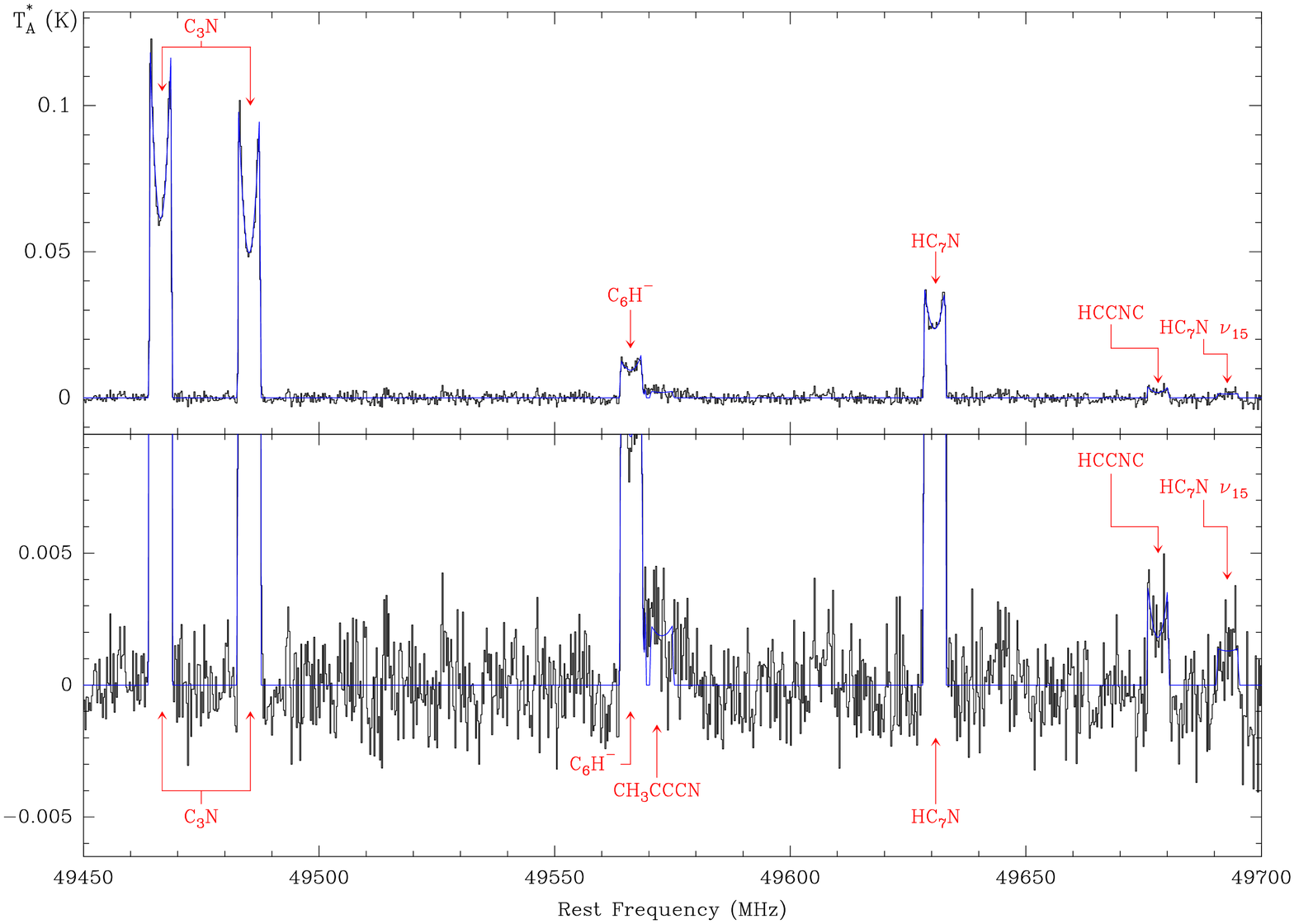}\\
\includegraphics[width=0.93\textwidth]                       {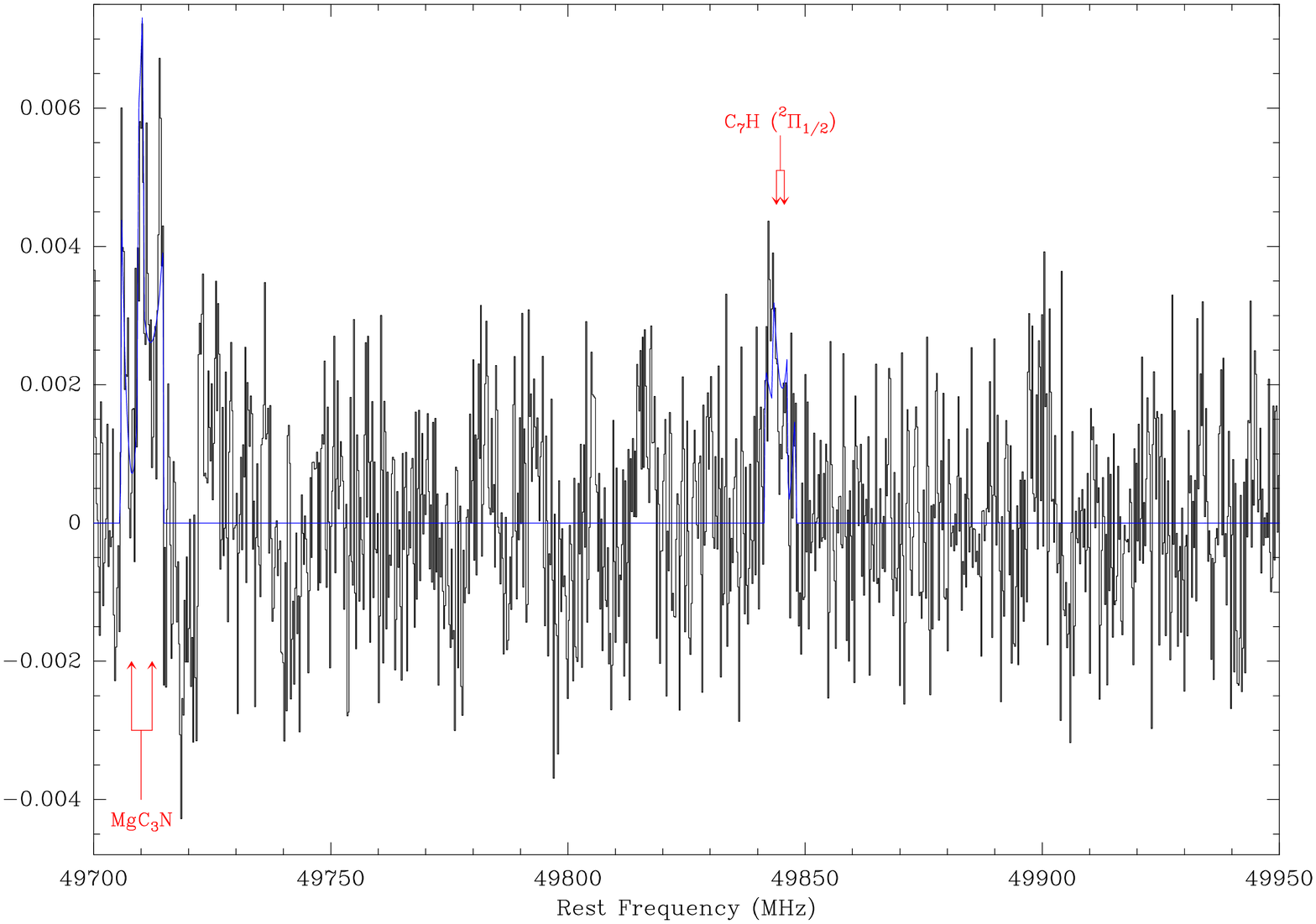}
\caption{IRC+10216 YEBES 40m data, line fits and labels from 
49450 to 49950 GHz.}
\label{fig38}
\end{figure*}

\begin{figure*}                                              
\includegraphics[width=0.93\textwidth]                       {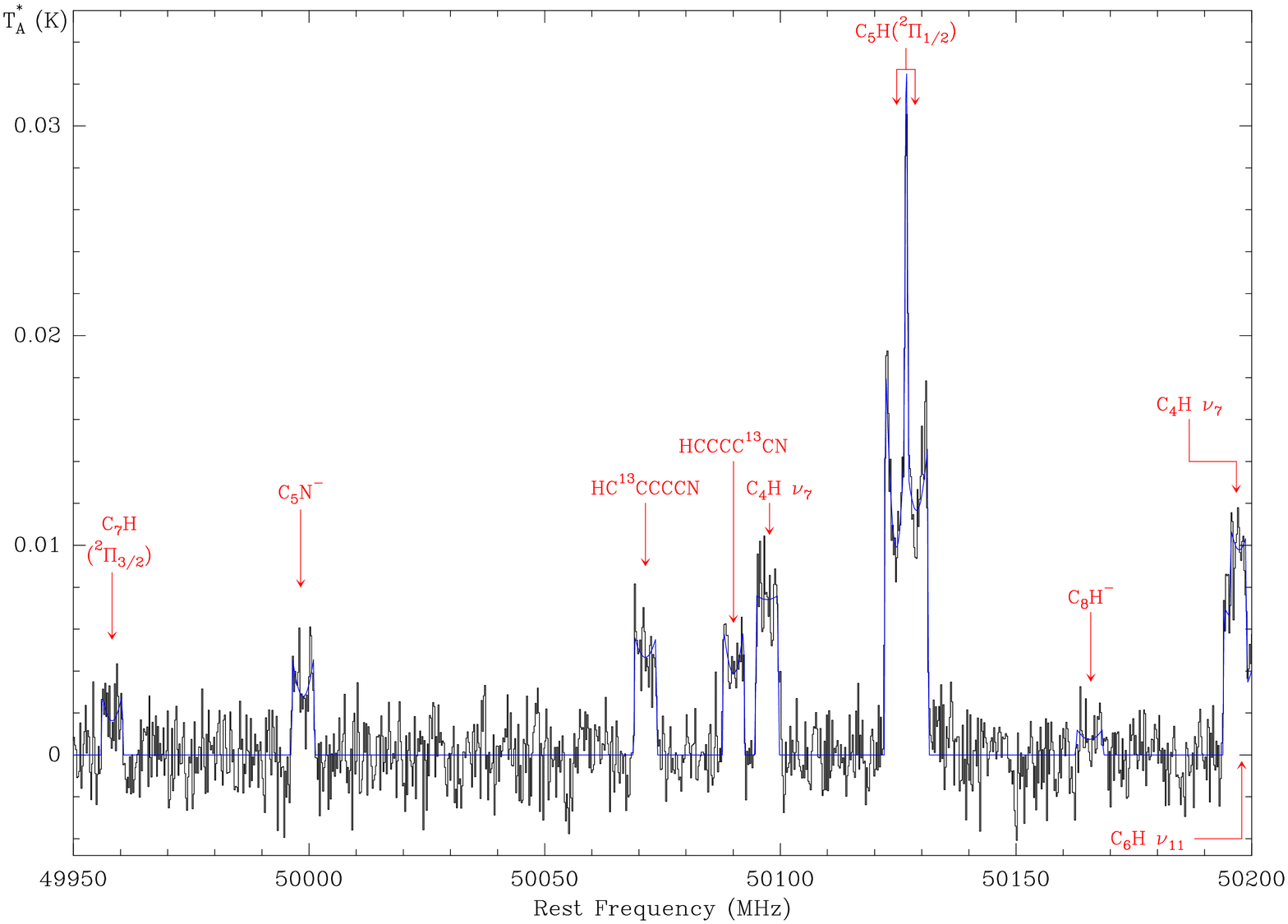}\\
\includegraphics[width=0.93\textwidth]                       {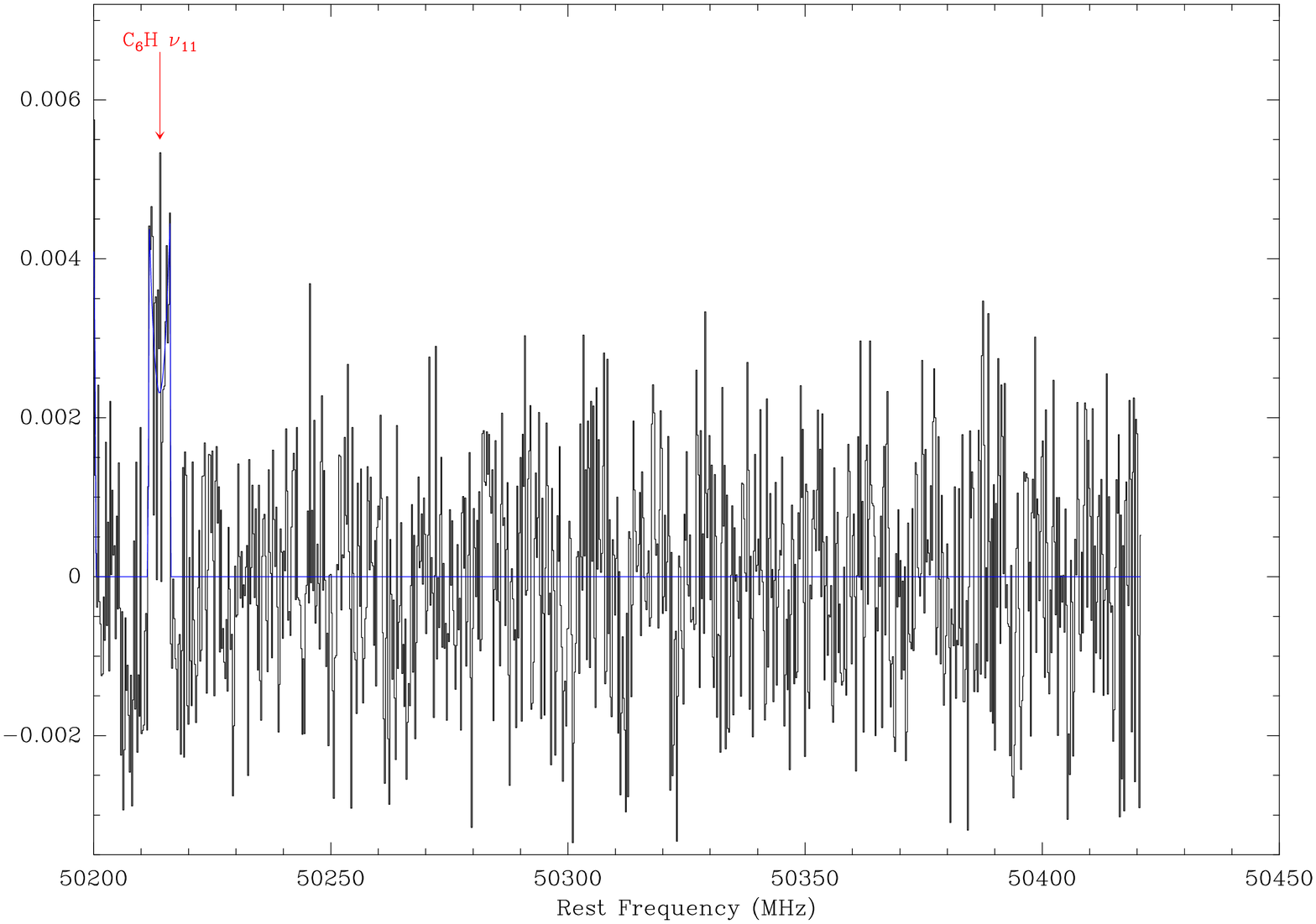}
\caption{IRC+10216 YEBES 40m data, line fits and labels from 
49950 to 50450 GHz.}
\label{fig39}
\end{figure*}

This appendix is devoted to provide the exhaustive           
information of our IRC+10216 Q-band survey with the          
Yebes 40m telescope, in two forms. First, the most           
significant line parameters (theoretical + observed)         
of all spectral lines that can be assigned to an             
identified molecular species are given in Table              
\ref{tb:id_qband}. The accumulated number of distinguishable 
 lines for each species is provided in the last column of    
 Table \ref{tb:id_qband}. This table is the basis to         
build Figures \ref{fg:freqdiag} and                          
\ref{fg:rotdiag}. The observed values in this table,         
 including errors in parenthesis affecting the last digits   
 of the observed frequency and the expansion velocity,       
 come from the line fits performed over the                  
 data with the Gildas SHELL method as shown in               
 Figures \ref{fig01} to \ref{fig39}, that can be found in    
 this appendix after the tables. An error of 0 or 0.0 means  
 the given observed frequency ($\nu_{obs}$) and/or           
 expansion velocity (Vexp) has been forced in the            
 fit. The figures themselves                                 
have labels to identify the positions of all the             
 lines that appear in Table \ref{tb:id_qband}, and           
 they also label the U-lines listed in Table                 
 \ref{ulines}. The frequency scale in all tables and figures 
 assumes a velocity of the source with respect to the local  
 standard of rest of -26.5 kms$^{-1}$.                 The   
 changing telescope beam across the Q-band is taken into     
 account in the calculation of W=$\int$T$_{MB}$dv.

%\onecolumn                                                  
\scriptsize   \begin{center}      \LTcapwidth=\textwidth     
\begin{longtable}{cccccccccp{2.5mm}c}                        
\caption{Line identification and associated parameters       
from our 31.0-50.3 GHz observations presented in this work.  
% Notes:                                                     
%$<$+$>$ Frequency of the strongest hyperfine component.     
%$<$$>$ $\nu_{rest}$ corresponds to average                  
%hyperfine components.                                       
%%+ Assumed width at half power of 14.5 km/s.                
%+ Rest Frequencies determined in \citep{Pardo2020}.         
%++ Rest Frequencies determined in \citep{Pardo2021}.        
%%\$ Blended with another line.                              
%% Line separation allows an                                 
%%individual fit for each line.                              
%% \& Average of two of more                                 
%% hyperfine lines.                                          
%\% Marginal detection and/or assignment.                    
%$[]$ Line too narrow for the expected width                 
%in IRC+10216.                                               
%$()$  Line too broad for the expected width                 
%in IRC+10216, possible blend of several lines.              
 } \label{tb:id_qband}\\    %\hline                          
Molecular & Quantum numbers & $\nu_{rest}$ &                 
$\nu_{obs}$&E$_{up}$& S$_{ul}$ & W=$\int$T$_{MB}$dv&         
$ln\frac{3\;k_B\;W}{8\;\pi^3\;\nu_{rest}\;S_{ul}\;\mu^2}$    
 & Vexp        & Fig. & Line \\ species & of the transition &
(MHz) & (MHz) & (K) & & (mK$\cdot$kms$^{-1}$) &(Rot. Diagram)
& (kms$^{-1}$)& &   count \\ \hline &&&&&&&&&\\[-1em]        
\endfirsthead  %\hline                                       
Molecular & Quantum numbers & $\nu_{rest}$ &                 
$\nu_{obs}$&E$_{up}$& S$_{ul}$ & W=$\int$T$_{MB}$dv&         
$ln\frac{3\;k_B\;W}{8\;\pi^3\;\nu_{rest}\;S_{ul}\;\mu^2}$    
 & Vexp        & Fig. & Line \\ species & of the transition &
(MHz) & (MHz) & (K) & & (mK$\cdot$kms$^{-1}$) &(Rot. Diagram)
& (kms$^{-1}$)& &   count \\ \hline &&&&&&&&&\\[-1em]        
\endhead                                                     
\multicolumn{11}{l}{Continues on next page.}                 
\endfoot   %\multicolumn{11}{l}{}                            
\endlastfoot      \renewcommand{\arraystretch}{2}            
C$_8$H                & $^2\Pi_{\frac{3}{2}}$ J=$\frac{53}{2}\rightarrow\frac{51}{2}$ a                &  31093.03 &  31093.03( 0) &    20.4 &    52.80 &    509 &  20.926 &     14.5(0.0) & \ref{fig01}a &   1   \\ 
%\hline \rule{0pt}{12pt}       
C$_8$H                & $^2\Pi_{\frac{3}{2}}$ J=$\frac{53}{2}\rightarrow\frac{51}{2}$ b                &  31093.42 &  31093.42( 0) &    20.4 &    52.80 &    339 &  20.521 &     14.5(0.0) & \ref{fig01}a &   1   \\ 
%\hline \rule{0pt}{12pt}       
H$^{13}$CCCCCN        & J=12$\rightarrow$11                                                            &  31120.03 &  31119.94(23) &     9.7 &    12.00 &    761 &  23.622 &     13.5(2.2) & \ref{fig01}a &   1   \\ 
%\hline \rule{0pt}{12pt}       
C$_8$H                & $^2\Pi_{\frac{1}{2}}$ J=$\frac{53}{2}\rightarrow\frac{51}{2}$ a                &  31151.11 &  31150.84(23) &    48.3 &    53.00 &    169 &  19.816 &     14.0(2.2) & \ref{fig01}a &   2   \\ 
%\hline \rule{0pt}{12pt}       
C$_8$H                & $^2\Pi_{\frac{1}{2}}$ J=$\frac{53}{2}\rightarrow\frac{51}{2}$ b                &  31160.98 &  31161.29(19) &    48.4 &    53.00 &    337 &  20.508 &     16.3(2.4) & \ref{fig01}a &   3   \\ 
%\hline \rule{0pt}{12pt}       
C$_5$H                & $^2\Pi_{\frac{3}{2}}$ J=$\frac{13}{2}\rightarrow\frac{11}{2}$ a                &  31241.77 &  31241.77( 0) &    40.1 &    11.30 &   1421 &  24.063 &     14.5(0.0) & \ref{fig01}b &   1   \\ 
%\hline \rule{0pt}{12pt}       
C$_5$H                & $^2\Pi_{\frac{3}{2}}$ J=$\frac{13}{2}\rightarrow\frac{11}{2}$ b                &  31242.33 &  31242.33( 0) &    40.1 &    13.20 &   1254 &  23.782 &     14.5(0.0) & \ref{fig01}b &   1   \\ 
%\hline \rule{0pt}{12pt}       
NaCN                  & J$_{Ka,Kc}$=2$_{0,2}$$\rightarrow$1$_{0,1}$                                    &  31262.33 &  31262.36(23) &     2.3 &     2.00 &    751 &  23.966 &     14.5(2.2) & \ref{fig01}b &   1   \\
\end{longtable}   \end{center}                             
% \twocolumn                                                 

 %\scriptsize
 \begin{table*}[h]
 \caption{Observational parameters of the remaining   
 unidentified lines in this work.}
 \setlength\extrarowheight{-2.0pt}
 \label{ulines}
 \begin{tabular}{|cccc|cccc|cccc|}
 \hline 
  {\scriptsize Label} & {\scriptsize $\nu_{obs}$}&
  {\scriptsize $\int$T$_{A}^*$dv} & {\scriptsize Vexp}& 
  {\scriptsize Label} & {\scriptsize $\nu_{obs}$}&
  {\scriptsize $\int$T$_{A}^*$dv} & {\scriptsize Vexp}& 
  {\scriptsize Label} & {\scriptsize $\nu_{obs}$}&
  {\scriptsize $\int$T$_{A}^*$dv} &{\scriptsize Vexp}\\
  & {\scriptsize (MHz)}  & {\scriptsize (mK$\cdot$kms$^{-1}$)} & {\scriptsize (kms$^{-1}$)} &  
  & {\scriptsize (MHz)}  & {\scriptsize (mK$\cdot$kms$^{-1}$)} & {\scriptsize (kms$^{-1}$)} &  
  & {\scriptsize (MHz)}  & {\scriptsize (mK$\cdot$kms$^{-1}$)} & {\scriptsize (kms$^{-1}$)} \\ 
 \hline
  \scriptsize{U31869              }   &   \scriptsize{ 31868.98( 6)}  &   \scriptsize{ 50}  &     \scriptsize{    14.6(0.6)}    &  
  \scriptsize{U32036              }   &   \scriptsize{ 32035.92( 8)}  &   \scriptsize{ 49}  &     \scriptsize{    13.7(0.9)}    &  
  \scriptsize{U32040              }   &   \scriptsize{ 32039.77(11)}  &   \scriptsize{ 49}  &     \scriptsize{    20.1(0.6)}    \\ 
  \scriptsize{U32079              }   &   \scriptsize{ 32078.51( 4)}  &   \scriptsize{ 49}  &     \scriptsize{    12.1(0.4)}    &  
  \scriptsize{U32119              }   &   \scriptsize{ 32118.86( 4)}  &   \scriptsize{ 25}  &     \scriptsize{    12.2(0.4)}    &  
  \scriptsize{U32201              }   &   \scriptsize{ 32200.71( 0)}  &   \scriptsize{ 24}  &     \scriptsize{    11.2(0.0)}    \\ 
  \scriptsize{U32322              }   &   \scriptsize{ 32322.47( 8)}  &   \scriptsize{ 49}  &     \scriptsize{    14.5(0.0)}    &  
  \scriptsize{U32971              }   &   \scriptsize{ 32970.52(10)}  &   \scriptsize{ 35}  &     \scriptsize{    13.0(0.9)}    &  
  \scriptsize{U33174              }   &   \scriptsize{ 33174.03(19)}  &   \scriptsize{ 35}  &     \scriptsize{    15.6(1.4)}    \\
\hline
 \end{tabular}
 \end{table*}
 %\normalsize

\clearpage                                                   
\clearpage                                                   
\clearpage                                                   
\clearpage                                                   
\clearpage                                                   
\clearpage                                                   
\clearpage                                                   
\clearpage                                                   
\clearpage                                                   
\clearpage                                                   
\clearpage                                                   
\clearpage                                                   
\clearpage                                                   
\clearpage                                                   
\clearpage                                                   
\clearpage                                                   
\clearpage                                                   
\clearpage                                                   
\clearpage                                                   
\clearpage                                                   
\clearpage                                                   
\clearpage                                                   
\clearpage                                                   
\clearpage                                                   
\clearpage                                                   
\clearpage                                                   
\clearpage                                                   
\clearpage                                                   
\clearpage                                                   
\clearpage                                                   
\clearpage                                                   
\clearpage                                                   
\clearpage                                                   
\clearpage                                                   
\clearpage                                                   
\clearpage                                                   
\clearpage                                                   
\clearpage                                                   
\clearpage

\end{appendix}                                               

\begin{thebibliography}{}                                    
\bibitem[Alexander et al. (1976)]{Kroto1976}Alexander,       
A.~J., Kroto, H.~W., \& Walton, D.~R.~M., 1976,              
J. Mol. Spectrosc., 62, 175.                                 
\bibitem[Allen et al. (2001)]{Allen2001}                     
Allen, M.D., Evenson, K.M., Brown, J.M., et al., 2001,       
J. Mol. Spectrosc., 209, 143                                 
\bibitem[Anderson et al. (1994)]{An94}Anderson, M.~A.,       
Steimle, T.~C., \& Ziurys, L.~M., 1994, ApJ, 429, L41        
\bibitem[Anttila et al. (1993)]{A93}                         
Anttila, R., Horneman, V.M., Koivusaari, M., et al., 1993,   
J. Mol. Spectrosc., 157, 198                                 
\bibitem[Aoki et al. (1993)]{Aoki1993}                       
Aoki, K., and Ikuta, S., 1993, J. Chem. Phys., 99, 3809.     
\bibitem[Bester et al. (1983)]{Bes1983}                      
Bester, M., Tanimoto, M., Vowinkel, B., et al. 1983,         
Z. Naturforschung., 38a, 64                                  
\bibitem[Bester et al. (1984)]{Bes1984}                      
Bester, M., Yamada, K., Winnewisser, G., et al. 1984,        
A\&A, 137, L20                                               
\bibitem[Bizzocchi et al. (2004)]{Bizzo2004a}                
Bizzocchi, L., Degli Esposti, C., Botschwina, P., 2004,      
J. Mol. Spectrosc., 225, 145                                 
\bibitem[Blanksby et al. (2001)]{Blanksby0000}Blanksby,      
S.~J., McAnoy, A.~M., Dua, S., et al., 2001, MNRAS, 328, 89  
\bibitem[Bogey et al. (1986)]{Bo86}Bogey, M., Demuynck, C.,  
\& Destombes, J.~L., 1986, Chem. Phys. Letters, 125, 383     
\bibitem[Botschwina et al. (1992)]{Bots1992}                 
Botschwina, P., Horn, M., Seeger, S., et al., 1992,          
Chem. Phys. Letters, 195, 427.                               
\bibitem[Botschwina (1996)]{Bots1996}                        
Botschwina, P., 1996, Chem. Phys. Lett., 259, 627            
\bibitem[Botschwina et al. (1997)]{Bots1997}                 
Botschwina, P., Horn, M., Markey, K., et al., 1997,          
Mol. Phys. 92, 381.                                          
\bibitem[Caris et al. (2002)]{Caris2002}                     
Caris, M., Lewen, F., \& Winnewisser, G., 2002,              
Z. Naturforschung, 57a, 663                                  
\bibitem[Caris et al. (2004)]{Caris2004}                     
Caris, M., Lewen, F., Muller, H.S.P., Winnewisser, G., 2004, 
J. Mol. Struct., 2004, 695, 243                              
\bibitem[Cazzoli and Puzzarini (2006)]{C06}                  
Cazzoli, G., \& Puzzarini, C., 2006,                         
J. Mol. Spectrosc., 240, 153                                 
\bibitem[Cernicharo(1985)]{Cernicharo1985} Cernicharo, J.    
 1985, Internal IRAM report (Granada: IRAM)                  
\bibitem[Cernicharo et al. (1991)]{Cer1991}                  
Cernicharo, J., Gu\'elin, M., Kahane, K., et al. 1991, A\&A, 
246, 213                                                     
\bibitem[Cernicharo et al. (2000)]{Cer00} Cernicharo, J.,    
Gu\'elin, M., Kahane, C., 2000, A\&AS, 142, 181              
\bibitem[Cernicharo et al. (2008)]{Cerni2008}                
 Cernicharo, J., Gu\'elin, M., Ag\'undez, M., et al. 2008,   
 ApJ, 688, L83                                               
\bibitem[Cernicharo et al. (2011)]{Ce11}                     
 Cernicharo, J.,                                             
 Ag\'undez, M., Kahane, C., Gu\'elin, M., Goicoechea, J.R.,  
Marcelino, N., de Beck, E.,\& Decin, L., 2011, A\&A, 529, L3 
\bibitem[Cernicharo (2012)]{Cer12} Cernicharo, J., 2012, in  
 ECLA 2011: Proc. of the European Conference on Laboratory   
 Astrophysics, EAS Publications Series, 2012, Ed.: C. Stehl, 
 C. Joblin, \& L. d'Hendecourt (Cambridge: Cambridge Univ.   
 Press), 251;                                                
 \texttt{https://nanocosmos.iff.csic.es/?page$\_$id=1619}    
\bibitem[Cernicharo et al. (2015)]{Cer15} Cernicharo, J.,    
Marcelino, N., Ag\'undez, M., et al. 2015, A\&A, 575, A91    
\bibitem[Cernicharo et al. (2019)]{Cerni2019}                
Cernicharo, J., Cabezas, C., Pardo, J. R., et al. 2019       
A\&A, 63, L2                                                 
 \bibitem[Cernicharo et al. (2021a)]{Ce21a}                  
 Cernicharo, J., Ag\'undez, Cabezas, C.,                     
 Tercero, B., Marcelino, M., Pardo, J.R., \& de Vicente, P., 
 2021, A\&A, 649, L15                                        
 \bibitem[Cernicharo et al. (2021b)]{Ce21b}                  
 Cernicharo, J., Ag\'undez, Kaiser, R.~I., Cabezas, C.,      
 Tercero, B., Marcelino, M., Pardo, J.R., \& de Vicente, P., 
 2021, A\&A, 652, L9                                         
\bibitem[Chen al. (1995)]{Chen95} Chen, W., Novick, S.~E.,   
McCarthy, M.~C., Gottlieb, C.~A., \& Thaddeus, P., 1995,     
J. Chem. Phys. 103, 7828                                     
\bibitem[Cho and Saito (1998)]{C98}                          
Cho, S.-H., \& Saito, S., 1998, ApJ, 496, L51.               
\bibitem[Clouser and Gordy (1964)]{Clo}                      
Clouser, P.L., \& Gordy, W., Phys. Rev. 1964, 134A, 863      
\bibitem[Creswell et al. (1977)]{Creswell1977}Creswell,      
R.~A., Winnewisser, G., \& Gerry, M.~C.~L., 1977,            
J. Mol. Spectrosc. 65, 420                                   
\bibitem[Degli Esposti et al. (2004)]{Bizzo2004b}            
Degli Esposti, C., and Bizzocchi, L., 2004, ApJ., 614, 518   
\bibitem[Ebenstein and Muenter (1984)]{Ebenstein1984}        
Ebenstein, W.L., \& Muenter, J.S., 1984, J. Chem. Phys.,     
80, 3989                                                     
\bibitem[Gripp et al. (2000)]{Gripp2000}Gripp, J., Guarnieri,
 A., Stahl, W., et al., 2000, J. Mol. Struct., 526, 81       
\bibitem[Gordon et al. (2000)]{Gor2000}                      
Gordon, V.D., Nathan, E.S., Apponi, A.J., et al., 2000,      
J. Chem. Phys., 113, 5311                                    
\bibitem[Gottlieb al. (1983)]{Gott83} Gottlieb, C.~A.,       
Gottlieb,  E.~W., Thaddeus, P., \& Kawamura, H. 1983, ApJ    
275, 916                                                     
\bibitem[Gotlieb et al. (1985)]{Gott85}                      
Gottlieb, C. A., Vrtilek, J. M., Gottlieb, E. W., Thaddeus,  
P., \& Hjalmarson, A., 1985, ApJ, 294, L55                   
\bibitem[Gotlieb et al. (1989)]{Got1989}Gottlieb, C.A.,      
 Vrtilek, J.~M., \& Thaddeus, P., 1989, ApJ, 343, L29        
\bibitem[Gottlieb et al. (2003)]{Got2003}Gottlieb, C.,       
 Myers, P.C., \& Thaddeus, P., 2003, ApJ, 588, 655           
\bibitem[Groenewegen et al. (2012)]{Gro2012}Groenewegen,     
 M. A. T., Barlow, M. J., Blommaert, J. A. D. L., et al.     
 2012, A\&A, 543, L8                                         
\bibitem[Guarnieri et al. (1992)]{Guarnieri1992}Guarnieri,   
A., Hinze, R., Kr\"uger, M., et al., 1992,                   
J. Mol. Spectrosc., 156, 39                                  
\bibitem[Gu\'elin et al. (1982)]{Gue1982} Gu\'elin, M.,      
P. Friberg, and A. Mezaoui, 1982, A\&A, 109, 23              
\bibitem[Gu\'elin et al. (2018)]{Gue2018}Gu\'elin, M., Patel,
 N. A., Bremer, M., et al., 2018, A\&A, 610, A4              
\bibitem[Gupta et al. (2007)]{Gupta2007}Gupta, H.,           
 Br\"unken, S., Tamassia, F., et al., 2007, ApJ, 655, L57    
\bibitem[Halfen \& Ziurys (2011)]{HZ11}                      
Halfen, D.T., \& Ziurys, L.M., 2011, ApJ, 730, 107           
\bibitem[Hensel et al. (1993)]{He93}Hensel, K.D., Styger,    
C., J\"ager, W., et al., 1993, J. Chem. Phys., 99, 3320      
\bibitem[Hirahara et al. (1993)]{Hirahara1993}               
Hirahara, Y., Ohshima, Y., \& Endo, Y., 1993, ApJ, 403, L83  
\bibitem[Hirano et al. (2002)]{Hi02}Hirano, T., Ishii, K.,   
Odaka, T.E., et al., 2002, J. Mol. Spectrosc., 215, 42       
\bibitem[Ikuta et al. (2000)]{Ikuta2000} Ikuta, S. Tsuboi,   
T., \& Aoki, K., 2000, J. Mol. Struct., 528, 297             
\bibitem[Kahane et al. (1992)]{Kah1992}Kahane, C.,           
Cernicharo, J., G\'omez-Gonz\'alez, J., \& Gu\'elin, M. 1992,
A\&A, 256, 235                                               
\bibitem[Kaifu et al. (2004)]{Kaif04}                        
Kaifu, N., Ohishi, M., Kawaguchi, K., Saito, S., Yamamoto,   
S., Miyaji, T., Miyazawa, K.,  Ishikawa, S.~I., Noumara, C., 
Harasawa, S., Okuda, M., \& Suzuki, H., 2004, PASJ, 56, 69   
\bibitem[Kanata et al. (1987)]{K7}Kanata, H., Yamamoto, S.,  
\& Saito, S., 1987, Chem. Phys. Letters, 140, 221            
\bibitem[Kasai et al. (1997)]{Kasai1997} Kasai, Y.,          
Sumiyoshi, Y., Endo, Y., 1997, ApJ, 477, L65                 
\bibitem[Kawaguchi et al. (1992)]{Kawaguchi1992}Kawaguchi,   
K., Takano, S., Ohishi, M., et al., 1992, ApJ, 396, L49      
\bibitem[Kawaguchi et al. (1993)]{Ka93}Kawaguchi, K.,        
Kagi, E., Hirano, T., et al., 1993, ApJ, 406, L39            
\bibitem[Kawaguchi et al. (1995)]{Kaw95}Kawaguchi,           
K., Kasai, Y., Ishikawa, S., \& Kaifu, N., 1995, PASJ, 47,   
 853                                                         
\bibitem[Killian et al.(1990)]{Killian1990} Killian, T.~C.,  
Vrtilek, T.~C., Gottlieb, J.~M., 1990 , ApJ, 365, L89        
\bibitem[Kokkin et al. (2011)]{Kok1991}Kokkin, D.~L.,        
Br\"unken, S., Young, K.~H., et al., 2011, ApJS, 196, 17     
\bibitem[Lee (1997)]{Lee97}                                  
Lee, S., 1997, Chem. Phys. Letters, 268, 69.                 
\bibitem[Lide (1965)]{Li65}                                  
Lide, D.~R., 1965, J. Chem. Phys., 42, 1013.                 
\bibitem[Linnartz et al.(1999)]{Linn1999} Linnartz, H.,      
Motylewski, T., Vaizert, O., Maier, J.~P., Apponi, A.~J.,    
McCarthy, M.~C., Gottlieb, C.~A., \& Thaddeus, P. 1999,      
J. Mol. Spectrosc. 197, 1                                    
\bibitem[Lovas et al. (1992)]{Lo92}Lovas, F.~J., Suenram,    
R.~D., Ogata, T., \& Yamamoto, S., 1992, ApJ, 399, 325       
%\bibitem[Maiwald et al.(2000)]{Maiwal2000}Maiwald, F.,Lewen,
%F., Ahrens, V., et al., 2000, J. Mol. Spectrosc., 202, 166  
\bibitem[Mallinson \& de Zafra (1978)]{Mallins1978}          
Mallinson, P.~D., \& de Zafra, R.~L., 1978,                  
Mol. Phys. 36, 827.                                          
\bibitem[Manson et al.(1977)]{M77}Manson Jr, E.~L.,          
Clark, W.~W., DeLucia F.~C., \& Gordy, W.,                   
1977, Phys. Rev. A, 15, 223.                                 
\bibitem[Mbosei et al.(2000)]{Mbosei2000}Mbosei, L., Fayt,   
A., Dr\'ean, P., \& Cosl\'eou, J., 2000 , J. Mol. Struc.,    
517, 271.                                                    
\bibitem[McCarthy et al. (1995)]{McCa1995}                   
McCarthy, M.~C., Gottlieb, C.~A., Thaddeus, P., Horn, M.,    
\& Botschwina, P., 1995, J. Chem. Phys., 103, 7820           
\bibitem[McCarthy et al. (1996)]{McCa1996}                   
McCarthy, M.~C., Travers, M.~J., Kov\'acs, A., Gottlieb,     
C.~A., \& Thaddeus, P., 1996, A\&A, 309, L31                 
\bibitem[McCarthy et al.(1997)]{McCa1997}                    
McCarthy, M.~C., Travers, M.~J., Kov\'acs, A., Gottlieb,     
C.~A., \& Thaddeus, P., 1997, ApJ, 113, 105                  
\bibitem[McCarthy et al.(1999)]{McCa1999}  McCarthy, M.~C.   
Chen, W., Apponi, A.~J., Gottlieb, C.~A., \&                 
Thaddeus, P., 1999, ApJ, 520, 158                            
\bibitem[McCarthy et al.(2000)]{McCarthy2000}McCarthy, M.~C.,
 Levine, E.~S., Apponi, A.~J., \& Thaddeus, P., 2000,        
J. Mol. Spectrosc. 203, 75                                   
\bibitem[McCarthy et al. (2003)]{McCa2003}                   
McCarthy, M.~C., Fuchs, G.~W., Kucera, J., Winnewisser, G.,  
\& Thaddeus, P., 2003, J. Chem. Phys., 118, 3549             
\bibitem[McCarthy et al.(2007)]{McCarthy2007}McCarthy, M.~C.,
Gottlieb, C.~A., Gupta, H., \& Thaddeus, P., 2006,           
ApJ, 652, L141.                                              
\bibitem[Moises et al. (1982)]{Moi1982}Moises, A., Boucher,  
D., Burie, J., et al., 1982, J. Mol. Spectrosc., 92, 497     
\bibitem[Mollaaghababa et al. (1991)]{M91}                   
Mollaaghababa, R., Gottlieb C.~A., Vrtilek J.~M.,            
\& Thaddeus, P., 1991, ApJ, 368, L19                         
\bibitem[M\"uller et al.(2007)]{Mueller2007}M\"uller         
H.~S.~P., McCarthy M.~C., Bizzocchi L., et al., 2007,        
Phys. Chem. Chem. Phys., 9, 1579                             
\bibitem[M\"uller et al. (2009)]{M09}M\"uller, H.~S.~P.,     
Drouin, B.~J., \& Pearson, J.~C., 2009, A\&A, 506, 1487      
\bibitem[Ohishi et al. (1989)]{Ohi1989}Ohishi, M.,           
Kaifu, N., Kawaguchi, K., et al. 1989, ApJ, 345, L83         
\bibitem[Ohshima \& Endo (1992)]{O92}                        
Ohshima, Y, \& Endo, Y., 1992, J. Mol. Spectrosc., 153, 627  
\bibitem[Oswald and Botschwina (1995)]{Oswald1995}Oswald, M.,
\& Botschwina, P., 1995, J. Mol. Spectrosc. 169, 181         
\bibitem[Oyama et al. (2020)]{Oyam2020}Oyama et al., 2020,   
ApJ, 890, A39                                                
\bibitem[Pardo et al.(2001)]{Pardo2001} Pardo, J.~R.,        
Cernicharo, J., Serabyn, E. 2001, IEEE Trans. Antennas and   
Propagation, 49, 12                                          
\bibitem[Pardo et al.(2020)]{Pardo2020} Pardo, J.~R.,        
 Berm\'udez, C., Cabezas, C., Ag\'undez, M., Gallego, J.~D., 
 Fonfr\'{\i}a, J.~P., Velilla-Prieto, L., Quintana-Lacaci,   
 G., Tercero, B., Gu\'elin, M., \& Cernicharo, J.,           
 2020, A\&A, 640, L13                                        
\bibitem[Pardo et al.(2021)]{Pardo2021} Pardo, J.~R.,        
 Cabezas, C., Fonfr\'{\i}a, J.~P., Ag\'undez, M., Tercero,   
B., de Vicente, P., Gu\'elin, M., \& Cernicharo, J.,         
 2021, A\&A, 652, L13                                        
\bibitem[Pavone et al. (1990)]{Pav90}Pavone, F.~S., Zink,    
L.~R., Prevedelli, M., et al., 1990,                         
J. Mol. Spectrosc., 144, 45                                  
\bibitem[Raymonda et al. (1970)]{R70}                        
Raymonda, J.~W., Muenter, J.~S., \& Kemplerer W.~A., 1970,   
J. Chem. Phys., 52, 3458                                     
\bibitem[Saito et al. (1987)]{Sai1987}Saito, S., Kawaguchi,  
K., Yamamoto, S., et al., 1987, ApJ, 317, L115               
\bibitem[Sanz et al.(2003)]{Sanz03} Sanz, M.~E., McCarthy,   
M.~C., Thaddeus, P., 2003, J. Chem. Phys., 119, 11715        
\bibitem[Simeckova et al. (2004)]{S04}Simeckova, M., Urban,  
S., Fuchs, U., et al., 2004, J. Mol. Spectrosc., 226, 123    
\bibitem[Spezzano et al.(2012)]{Sp12}Spezzano, S.,           
 Tamassia, F., Thorwirth, S., et al. 2012, ApJS, 200, 1      
\bibitem[Steimle and Bousquet (2001)]{SB01}Steimle, T.~C.,   
\& Bousquet, R.~R., 2001, J. Chem. Phys., 115, 5203          
\bibitem[Suenram et al. (1989)]{Sue1989}Suenram, R.~D.,      
Lovas, F.~J., \& Matsumura, K., 1989, ApJ, 342, L103         
\bibitem[Suenram \& Lovas (1994)]{S94}                       
Suenram, R.~D., \& Lovas, F.~J., 1994, ApJ, 429, L89         
\bibitem[Tang et al. (1995)]{Ta95}                           
Tang, J.~A., Saito, S., 1995, J. Mol. Spectrosc., 169, 92    
\bibitem[Tang et al.(1999)]{Tang1999} Tang, J.~A., Sumiyoshi,
 Y., \& Endo, Y., 1999 , Chem. Phys. Letters, 315, 69        
\bibitem[Tercero et al. (2021)]{Tercero2021}Tercero, F.,     
L\'opez-P\'erez, J.~A., Gallego, J.~D., et al., 2021, A\&A,  
645, A37                                                     
\bibitem[Thaddeus et al.(1985)]{Th85}Thaddeus, P.,           
Vrtilek, J.~M., \& Gottlieb, C.~A., 1985, ApJ, 299, L63      
 \bibitem[Thaddeus et al.(2008)]{Tha2008}                    
Thaddeus, P.,                                                
 Gottlieb, C.~A., Gupta, H., Br\"unken, S.,                  
 McCarthy, M.~C., Ag\'undez, M., Gu\`elin, M.,               
 \& Cernicharo, J., 2008, 677, 1139                          
\bibitem[Thorwirth et al.(2000)]{Thorwirth2000} Thorwirth,   
S., M\"uller, H.~S.~P., \& Winnewisser, G., 2000,            
J. Mol. Spectrosc. 204, 133                                  
\bibitem[Thorwirth et al.(2001)]{Thorwirth2001}              
Thorwirth, S., M\"uller, H.~S.~P., \& Winnewisser, G.,  2001,
 Phys. Chem. Chem. Phys., 3, 1236                            
\bibitem[Tiemann et al.(1972)]{Tiemann1972}                  
Tiemann, E., Renwanz, E., Hoeft, J., et al., 1972,           
Z. Naturforschung, 27a, 1566                                 
\bibitem[Timp et al. (2012)]{Timp2012}Timp, B.~A., Doran,    
 J.~L., Iyer, S., et al., 2012, J. Mol. Spectrosc., 271, 20  
\bibitem[Van Vaals (1984)]{VV84}Van Vaals, J.~J.,            
Meerts, W.~L., \& Dymanus, A., 1984, Chem. Phys., 86, 147    
\bibitem[Vrtilek et al.(1987)]{Vr87}Vrtilek, J.~M.,          
Gottlieb, C.~A., \& Thaddeus, P., 1987, ApJ, 314, 716        
\bibitem[Vrtilek et al.(1990)]{Vr90}Vrtilek, J.~M.,          
Gottlieb, C.~A., Gottlieb, E.~W., Killian, T.~C. \&          
Thaddeus, P., 1990, ApJ, 364, L53                            
\bibitem[Winnewisser \& Cook (1968)]{W68} Winnewisser, G.,   
\& Cook, R.~L., 1968, J. Mol. Spectrosc. 28, 266.            
\bibitem[Woon (1995)]{Woon1995}                              
Woon, D.~E., 1995, Chem. Phys. Lett. 244, 45                 
\bibitem[Wyse \& Gordy (1972)]{WG72}                         
Wyse, F.~C., \& Gordy, W., 1972, J. Chem. Phys., 56, 2130.   
\bibitem[Yamamoto et al. (1987)]{Ya87}Yamamoto, S., Saito,   
S., Kawaguchi, K., et al., 1987, ApJ, 317, L119              
\bibitem[Yamamoto et al. (1990a)]{Yama90}                    
Yamamoto, S. Saito, S., Suzuki, H., Deguchi, S., Kaifu, N.,  
Ishikawa, S.~I., \& Ohishi, M., 1990, ApJ, 348, 363          
\bibitem[Yamamoto et al. (1990b)]{Yama1990}Yamamoto, S.,     
Saito, S., \& Kawaguchi, K., 1990, ApJ, 361, 318             
 \bibitem[Zelinger et al. (2003)]{Zel2003}                   
Zelinger, Z.,                                                
 Amano, T., Ahrens, V., et al., 2003, J. Mol. Spectrosc.,    
 220, 233                                                    
\end{thebibliography}
 \end{document}